%% file: sage.tex
\documentclass[a4paper,fleqn,usenatbib]{mnras}

\usepackage{caption}
\usepackage{graphicx}

\usepackage{amsmath}
\usepackage{amsfonts}
\usepackage{amssymb}
\usepackage{morefloats}
\usepackage{tikz}
\usepackage{hyperref}
\hypersetup{colorlinks,citecolor=blue,filecolor=black,linkcolor=blue, urlcolor=blue}

\newcommand{\st}{^{\circ}}

\title{Shaping Asteroid Models Using Genetic Evolution (SAGE)}
\author[P. Bartczak, G. Dudzi\'{n}ski]
{
	P. Bartczak, G. Dudzi\'{n}ski
	\\
	Astronomical Observatory Institute, Faculty of Physics, Adam Mickiewicz
	University, S{\l}oneczna 36, 60-286 Pozna{\'n}, Poland
}

\begin{document}

\include{main_text}

\clearpage
\appendix
\include{appendices}

%

\end{document}

%% file: main_text.tex

\maketitle

\label{firstpage}

\begin{abstract}

	In this work we present SAGE (Shaping Asteroid models using Genetic
	Evolution) asteroid modelling algorithm based solely on photometric
	lightcurve data.  It produces non-convex shapes, rotation axes orientations
	and rotational periods of asteroids. The main concept behind a genetic
	evolution algorithm is to produce random populations of shapes and spin axis
	orientations by mutating a seed shape and iterating the process until it
	converges to a stable global minimum. To test SAGE we have performed tests
	on five artificial shapes. We have also modelled (433) Eros and (9) Metis
	asteroids, as ground truth observations for them exist, allowing us to
	validate the models. We have compared derived Eros shape with NEAR Shoemaker
	model and Metis shape with adaptive optics and stellar occultation
	observations as with other available Metis models from various inversion
	methods.


\end{abstract}

\begin{keywords}
	Minor planets, asteroids, Methods: numerical, Techniques: photometric,
	lightcurve inversion
\end{keywords}

\section{Introduction}

Asteroid shapes were unknown until 1991.
That year
Galileo spacecraft took flyby photos of (951) Gaspra, and (243) Ida two years
later. Images revealed shapes and topographic features of these two objects --
far from spherical and dotted with impact craters. One cannot say though that
insight into asteroids' physical properties prior to Galileo mission had been
void, as some methods of acquiring information about Solar System's small bodies
from ground-based data were already being developed.

Untill this day photometry remains the main source of information about shapes,
rotational states and physical properties of asteroids. Studying their
lightcurves can lead to the creation of models that  explain observations, at
least to some extend and with some assumptions.

First thorough analysis of the lightcurve inversion problem was done by
\cite{Russell}. His conclusions were rather pessimistic -- from lightcurves one
can only deduce a spin axis orientation; unambiguous determination of shape is
beyond the grasp of analytical methods and any shape can be mimicked by albedo
variations on a body's surface. However, Russell's study dealt only with
zero-phase angles and geometric ''scattering law''.

Nonetheless, some methods were introduced later, mainly to calculate spin axes
of asteroids. The magnitude-amplitude and epoch approach
(\cite{asteroids2magnusson}, \cite{michalowski93})  assumed body's homogeneous
albedo and triaxial ellipsoid shape. Given many apparitions providing different
aspect angles it is possible to determine the spin axis orientation and triaxial
ellipsoid's axis ratios $a/b$ and $b/c$.

Increased computer power available to researchers led to the development of
numerical methods (e.g. \cite{Uchida87}, \cite{Karttunen89}) whose biggest
achievement was representing asteroids' models by small surface elements,
enabling the application of arbitrary scattering laws. The spin axis orientation
and ellipsoid axes were iteratively changed to provide the best fit to available
lightcurves.  Some numerical methods went beyond the simple triaxial ellipsoid
model \citep{Cellino87} merging eight different ellipsoids into one shape.

A new lightcurve inversion method was introduced by \cite{Kaasalainen01I} and
\cite{Kaasalainen01II}. It allows to produce asteroids' shape models,
rotational periods and spin axis orientations, the only constrain on asteroid's
shape being its convexity. The resulting model is a convex hull containing shape
of an asteroid. Lightcurves must be obtained form multiple apparitions in order
to provide a unique solution.

Nonetheless spacecraft missions revealed that asteroids' shapes are far more complex than
simple geometric shapes and are non-convex in general. A growing number of
adaptive optics and stellar occultation observations encourages to include
them in the modelling process.
An attempt to combine lightcurve inversion with Adaptive Optics and stellar
occultations KOALA (Knitted Occultation, Adaptive optics and Lightcurve
Analysis) was presented in \cite{Carry2010}, where it was used to model
(2) Pallas. This method has been developed further to include more observational
techniques \citep{Kaasalainen2011, Carry12}.
Another method combining various types of data, ADAM (All-Data Asteroid
Modelling algorithm), was described in \cite{ADAM15}.

In this work we introduce SAGE (Shaping Asteroids with Genetic Evolution) method
of modelling asteroids' shapes, spin axis orientations and period determination.
In this approach only lightcurve data is used to produce non-convex shapes of
asteroids, assuming a homogeneous albedo and mass distribution of a body and
rotation about a single axis. Each model produced by SAGE is a physical one,
i.e. the rotation axis always lies along model's greatest moment of inertia axis
and goes through its centre of mass.

In section \ref{sec:method} we explain the concept behind a genetic evolution
algorithm and describe the modelling process. In order to validate the method we
have performed numerical tests on artificial test models
(section~\ref{sec:num_tests}). We have also successfully modelled
(433) Eros and (9) Metis (section~\ref{sec:models_of_asteroids}) for which
\textit{in situ} observations (for Eros), adaptive optics and stellar
occultation (for Metis) exist allowing us to validate the results.

\section{Method}
\label{sec:method}

\subsection{Shape representation}

An asteroid shape model is represented by a mesh of vertices in 3D space with
triangular faces. Each face is defined as a list of three vertices in the
counter-clockwise order which defines a surface normal vector direction.
In SAGE 242 vertices are used to describe a shape and their positions in
space are free parameters in the inversion process. To lower the degrees of
freedom, every vertex lies along a ray oriented in a fixed direction.
Rays are evenly distributed on a sphere. By allowing a vertex to move only along
a ray it's position is reduced to a single variable and a redefinition of
faces is not needed when changes to the shape are made.

\begin{figure}
	\begin{center}
		\includegraphics[width=8cm]{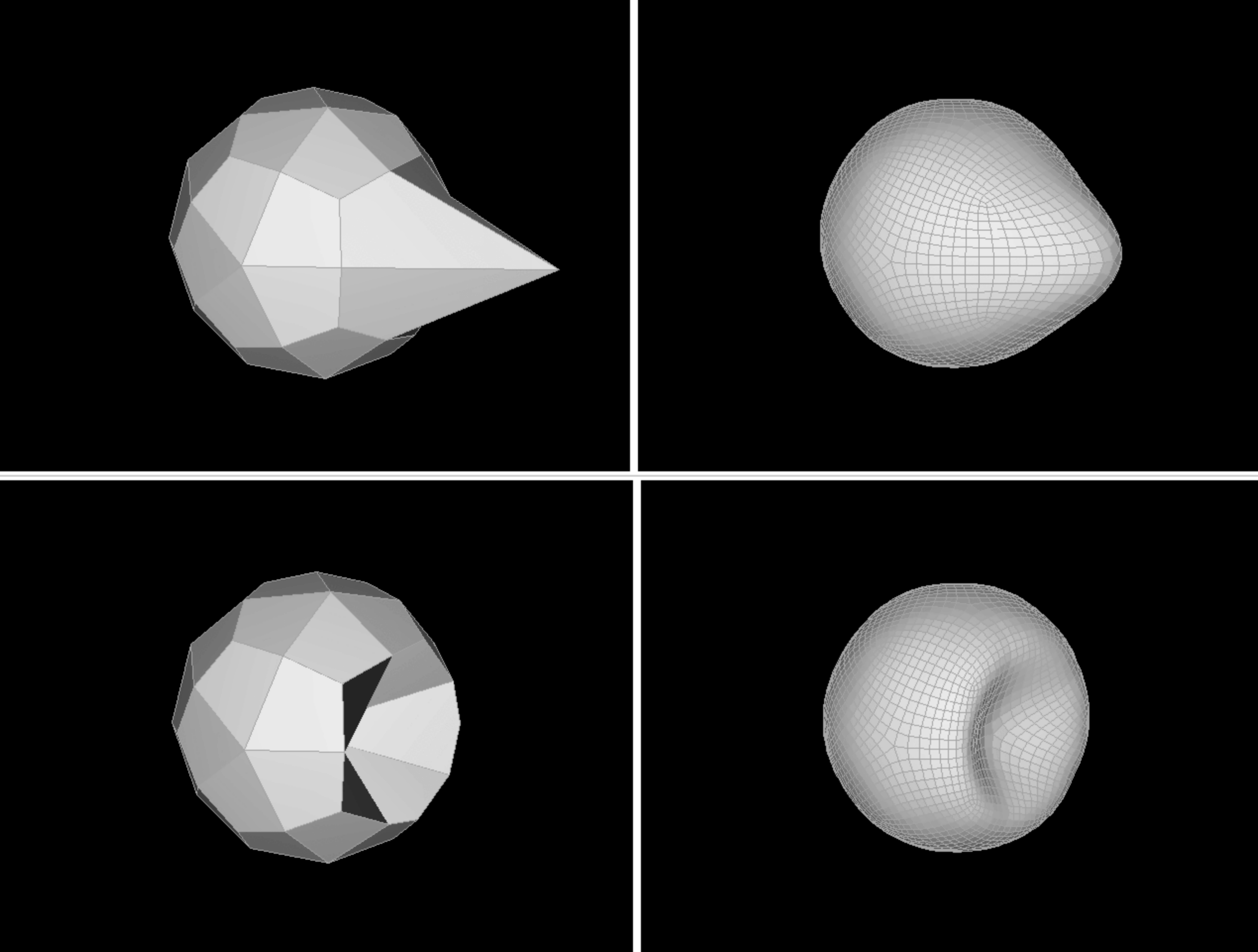}
	\end{center}
	\caption{Example of Catmull-Clark algorithm applied to a mesh. Images on the
	left show initial, rough body, on the right the same body after a
surface subdivision. The new mesh is smoother and has more uniformly
distributed vertices and faces.}
	\label{fig:Catmull}
\end{figure}

A more detailed model is used for lightcurve calculation. 242 parameter mesh is
refined by Catmull-Clark surface subdivision algorithm \citep{Catmull-Clark} for
surfaces smoothening (Fig. \ref{fig:Catmull}). The resulting asteroid shape model
consists of 3842 vertices and 7680 faces. Only the smoothened model is used for
lightcurve generation and is considered an asteroid's shape.

The choice of 242 shape parameters is dictated by several factors. We took the
largest possible platonic solid, an icosahedron, with 20 congruent regular
triangular faces and with each of 12 vertices being a meeting point for the same
number (i.e. 5) of facets. Then, we applied consecutive Catmull-Clark surface
subdivisions to get shapes with a larger number of vertices and with a
distribution of face sizes as uniform as possible. The resulting solids have 62,
242, 962, 3842, \dots, vertices. The 242 version offers a sufficient number of
parameters to produce detailed-enough models (with 3842 vertices) with respect
to the amount of information present in lightcurves. On the other hand, the
greater the number of parameters in the modelling, the more computing power,
database load and time needed for the inversion process. The number of
parameters is therefore a trade-off between models' detail level, computing
power needed and time available.  

\subsection{Orientation in space}
\label{sec:orientation}

Observed photometric lightcurves are compared to models' ones throughout the
modelling process.
To compute model's lightcurves asteroid's orientation and position
in space, as well as the Sun's and the Earth's need to be
replicated for the times of the observations.
Vectors in the ecliptic reference
frame (originating in the Sun's centre) are obtained from NASA JPL's
HORIZONS\footnote{http://ssd.jpl.nasa.gov/horizons.cgi}  service.


Shape model is defined in its own reference frame, where $z$ axis lies along
the model's largest inertia vector (see section \ref{sec:inertia}) and
constitutes an axis of rotation. The centre of a reference frame is always the
centre of mass of a model, assuming even distribution of mass. To
orient the model's spin axis Euler angles $\alpha$, $\beta$ and $\gamma$ are
used with 3-1-3 rotations about $z$, $x$ and $z$ axes.

In order to combine translation (orbital position) and spin axis orientation a
4x4 matrix $M$ of a form
\begin{equation}
	\label{eq:modelMatrix}
	M = T~ R_z(\alpha)~ R_x(\beta)~ R_z(\gamma)
\end{equation}
is used, where $R$ stands for a rotation matrix about the axis indicated by the
subscript and  $T$ stands for translation matrix.  Dimensions of $M$ are such
that the translation and rotations operations can be represented by a single
matrix. Positioning of a model in space, that changes the reference frame form
model-centred to the orbital one, is accomplished by multiplying vertices by
matrix $M$.

\subsection{Centre of mass and moments of inertia }
\label{sec:inertia}

Every asteroid model created in every sub-step of SAGE method is a physical one,
assuming homogeneous density distribution. In order to calculate model's
principal axes and define model-centred reference frame the centre of mass needs
to be found. Methods described in depth in \citep{Dobrovolskis} are employed for
a centre of mass and moments of inertia calculations.

For an arbitrary polyhedron composed of tetrahedrons, a centre of mass vector
$\mathbf{R}$ is given by the formula

\begin{equation}
	\bmath{R} =  \sum \frac{\Delta V \Delta \bmath{R}}{V},
\end{equation}
where $\Delta V$ and $\Delta\mathbf{R}$ are a tetrahedron volume and centroid, $V$ is a
polyhedron volume.

Next, an inertia tensor $\mathbf{I}$ of a form
\begin{equation}
	\mathbf{I} =
\begin{bmatrix}
I_{xx} & I_{xy} & I_{xz} \\
I_{xy} & I_{yy} & I_{yz} \\
I_{xz} & I_{yz} & I_{zz}
\end{bmatrix},
\end{equation}
is calculated.  Using parallel axis theorem it is possible to compute an inertia
tensor relative to the centre of mass

\begin{equation}
	\mathbf{I}' = \mathbf{I} - \mathcal{M}
\begin{bmatrix}
	Y^2 + Z^2 & -XY & -XZ \\
	-XY & X^2 + Z^2 & -YZ\\
	-XZ & -YZ & X^2 + Y^2
\end{bmatrix},
\end{equation}
where $\mathcal{M}$ is total mass of a body and $X$, $Y$ and $Z$ are
Cartesian components of a centre of mass vector $\mathbf{R}$.

To create body's own reference frame where $z$ axis is the axis of the largest
inertia, $\mathbf{I'}$ has to be rotated into new coordinate
system in which it becomes diagonal. It is done by finding inertia
tensor's eigenvalues.

\subsection{Synthetic lightcurve generation}
\label{sec:light-curveGeneration}

Generating lightcurves of asteroids' models is a computationally expensive
process. A shape is described by 3842 vertices on which 7680 faces are
defined. SAGE, in general, produces non-convex shapes so shadowing effects have
to be taken into account complicating computations even more.

During the modelling process with thousands of iterations, millions of
lightcurves are generated; to run SAGE efficiently parallelization of some parts
of the calculations is more then necessary. To accelerate the modelling process
graphics cards (GPU) and OpenGL\footnote{https://www.opengl.org} graphics
libraries are used. According to our internal tests, the process of
rasterization executed on a GPU offers about 100 times speed-up compared to an
equivalent code run on a CPU.



To mimic light reflected from a surface, a linear combination of Lambert
and Lommel-Seeliger scattering law of a form

\begin{equation}
	S = (1-c)\frac{\mu \mu_0}{\mu + \mu_0} + c\mu\mu_0
	\label{eq:scatteringLaw}
\end{equation}
is used \citep{Kaasalainen01I}, where $\mu$ and $\mu_0$ denote cosines of the
angles between surface normal and the direction to an observer and direction to
the Sun respectively; a linear factor $c$ equals $0.1$.

\begin{figure}
	\centering
	\includegraphics[width=8cm]{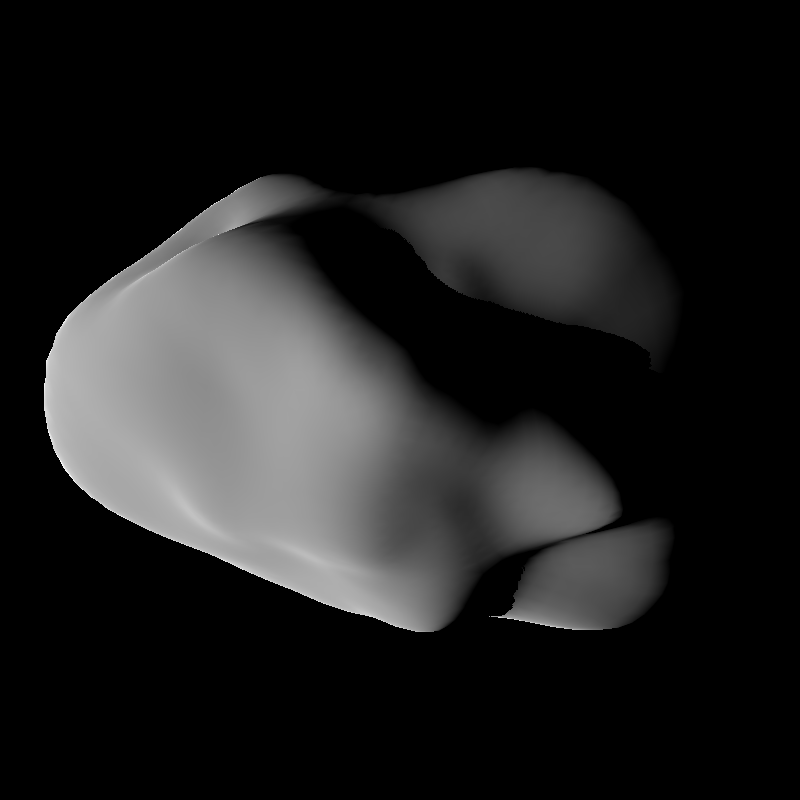}
	\caption{An example of an image generated for lightcurve data point
	calculation at high phase angle showing self shadowing effect.}
	\label{fig:view}
\end{figure}

In the process of lightcurve generation a 3D scene is composed in an
heliocentric ecliptic reference frame. Camera -- an observer -- is put in the
position of the Earth as it was during the time of the observation. Similarly,
asteroid model is translated to its corresponding orbital position and rotations
are applied to orient the spin axis of the model.  Next, in the process of
rasterization an image is created, in which a model is visible as if it was
observed from the Earth with a telescope of infinite resolution (Fig.
\ref{fig:view}). Background has a value of 0 and every surface element has
its own colour value computed using the scattering law (Eq.
\ref{eq:scatteringLaw}). The sum of the pixels' values of an image is one point
on a lightcurve. To generate the whole lightcurve, a body is gradually rotated
by $\Delta\gamma$ angle and the process ends when a full rotation is performed.

To simulate shadows a scene from the Sun's point of view is generated first. Its
goal is to create a shadow map on a model's surface that is used in generating a
scene from the Earth's point of view to determine whether a surface element is
illuminated or not. This method is fast and produces accurate shadows when only
one light source is present.

Generated lightcurves consist of points of relative fluxes. Each point is
recalculated to give a magnitude in logarithmic scale so the synthetic
lightcurves can be compared with the observed ones. During RMSD calculation a
synthetic lightcurve is allowed to shift vertically to find the best fit.
It is necessary as most of the time observed lightcurves come from relative
photometry, synthetic lightcurve generation does not take into account the
distance form an observer to a model, and size and albedo of a body are
unknown.

\subsection{Period search}
\label{sec:periodSearch}

\begin{figure}
	\includegraphics[width=8cm]{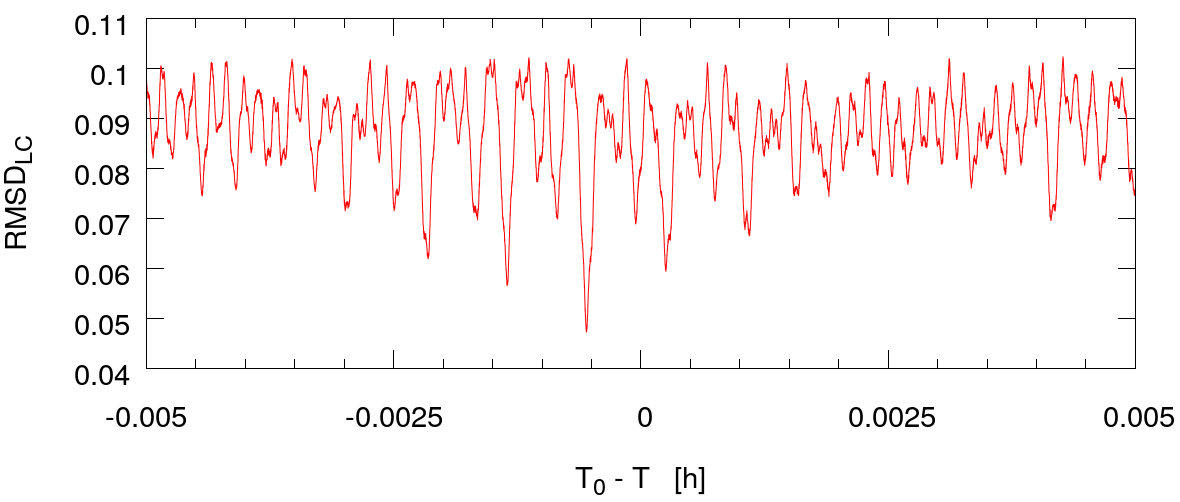}
	\caption{An example periodogram created duging search for the best
		rotational period. Root mean square deviation of all the lightcurves
		RMSD$_{LC}$ is shown against the difference between initial period $T_0$
		and scanned period $T$.}
	\label{fig:periodogram_example}
\end{figure}

During a weighting process, search for best rotational period is performed. The
need to repeat the search is justified by changes in shape and pole
orientation after every iteration that results in a new set of lightcurves.
An example periodogram can be seen in Fig.~\ref{fig:periodogram_example}.

The step of the rotation period search is $10^{-6}h$, which is sufficient
considering that only small changes to the shape are applied and lightcurves do
not change dramatically from one iteration to another.

The whole scope is scanned to search for the period that results in the best fit
for all lightcurves. Observations are performed over many years (at least 3
apparitions needed) and the bigger the time span, the more dramatic the effect
of rotation period change, making it easier to find the global minimum.
Uncertainty of a period depends on a time range of observations, a period search
step and a number of points on synthetic and observed lightcurves. If
uncertainty based on observations' time range is smaller than a period search
step, the former is used as final period uncertainty.

Synthetic lightcurves are shifted in the time dimension to obtain the best
fit determined by overall RMSD value. Each lightcurve comparison is a separate
problem, so  parallelization can be applied again.
CUDA\footnote{http://www.nvidia.com/object/cuda\_home\_new.html} libraries are
used for a period search to perform necessary computations on GPUs.

Period uncertainty is computed for a final model. It is defined as a range
of period values at a $\sigma$ level calculated as follows:
\begin{equation}
	\sigma = \frac{RMSD}{\sqrt{N -n}}
\end{equation}
where $N$ is a number of points in lightcurves and $n$ is a number of model's
degrees of freedom. A value of period uncertainty is different for each model
and depends on a data set and model itself.

\subsection{Modelling process}
\label{sec:modellingProcess}

A number of free parameters in the modelling process makes brute-force scanning
for global minimum impossible. Therefore genetic evolution algorithm is adopted
to search for an asteroid model that best fits observations. The modelling
process runs on a loop and every iteration generates a new population of random
shapes and pole orientations based on the seed shape.

The modelling process does not make any assumptions about shape and pole
orientation prior to modelling. A sphere with a random pole orientation is
always used as a starting point.

In every iteration a population of shapes is created based on a seed shape by
applying small, random changes to parameters describing the shape, so that
every model in a population resembles the seed model. The centre of mass and
inertia tensor are calculated for every model. Next, a set of random pole
orientations is created to be applied later to the models in a population.

For every combination of shape and pole orientation the synthetic lightcurves
are computed (see section \ref{sec:light-curveGeneration}) in order to compare
them with the observed ones. To decide which model's lightcurves best
resemble the real ones a RMSD (root mean square deviation) value defined as
follows

\begin{equation}
	\text{RMSD} = \sqrt{ \frac{1}{n} \sum_i^n  (\hat{y}_i - y_i)^2}
\end{equation}
is used, where $n$ is a number of points on a lightcurve, $y_i$ is an
observed magnitude and $\hat{y}_i$ is a computed magnitude. The model
with the lowest RMSD value is then chosen as the best one and serves as a seed
for the next iteration of the modelling process.

Before a new iteration begins there is an observation weighting step. Every
observed lightcurve is compared with a corresponding synthetic one, giving
separate RMSD value.  Observed lightcurve with the biggest RMSD is
given the  highest weight directing the flow of models' changes from population
to population so that said lightcurve is reproduced better. After a few
iterations weights can change and stress is placed on a different
lightcurve. This method ensures the modelling process does not fall into a
local minimum and is capable of crating a shape that fits all observations.

SAGE algorithm generates models with high accuracy rotation periods,
depending on the time span of the observations' set. This is achieved through a
rotation period grid search after every iteration during weighting process (see
section \ref{sec:periodSearch}).

After choosing the best model in the population its RMSD value is compared with
the ones from previous iterations. When RMSD becomes stable (does not change
from one iteration to another within a threshold) the modelling process is
stopped giving the final model (see top image in Fig.~\ref{fig:family}).

\begin{figure}
	\includegraphics[width=8cm]{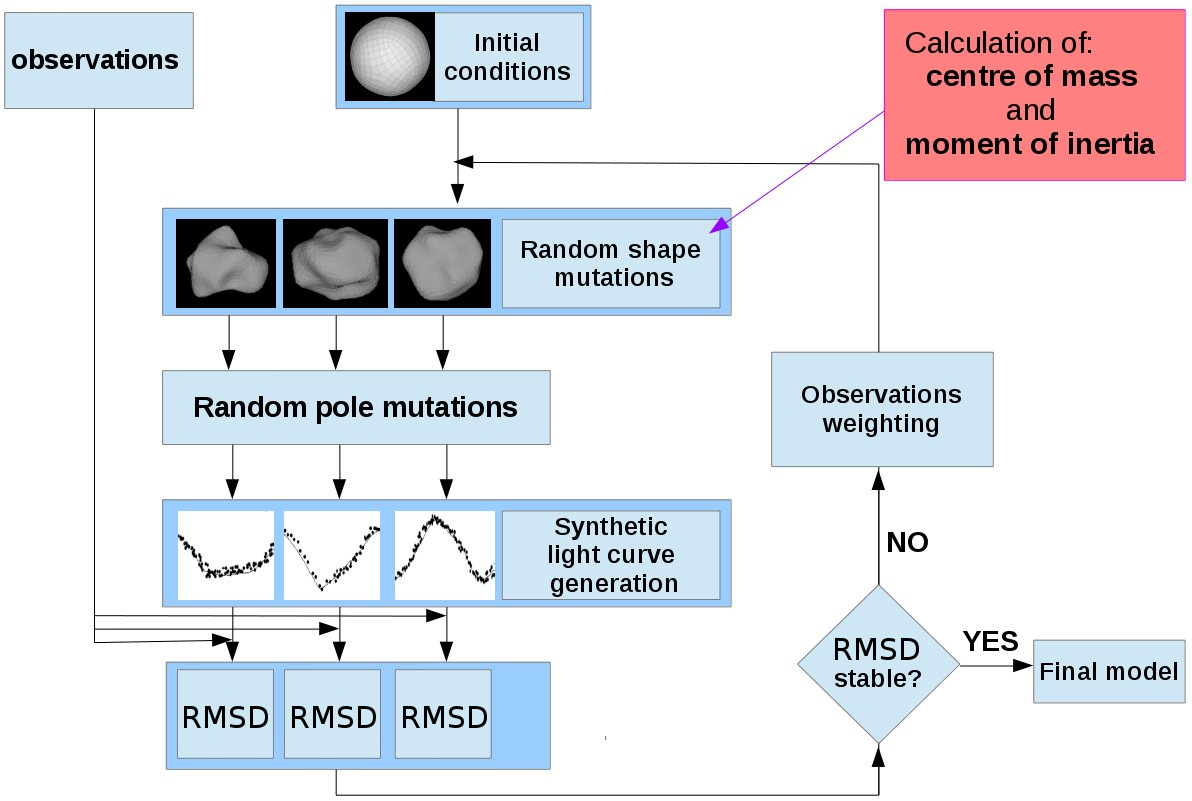}
	\caption{SAGE modelling scheme. See
		sec~\ref{sec:modellingProcess} for description.}
	\label{fig:schemat}
\end{figure}

\subsection{Family of solutions}

\begin{figure}
	\centering
	\includegraphics[width=8cm]{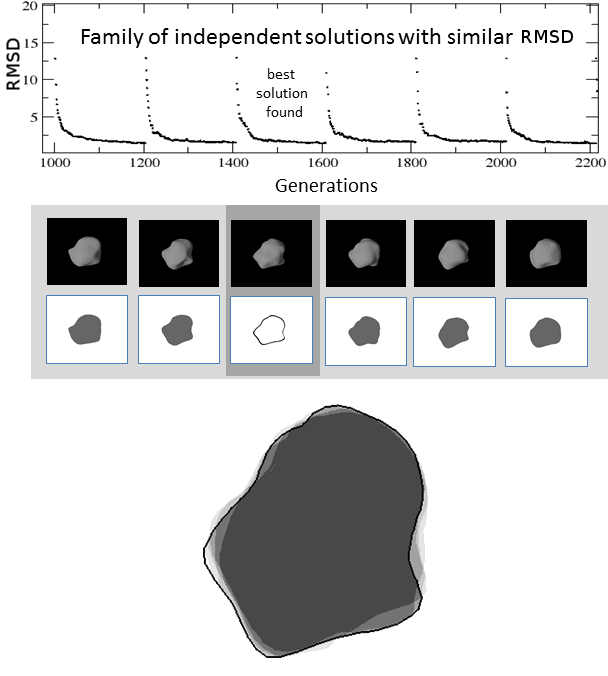}
	\caption{An example of a family of solutions. Top: RMSD shown for multiple
	runs of modelling process. RMSD value drops fast at the beginning of the
modelling gradually plateauing. Middle: different shapes produced by separate
modelling runs. Bottom: composition of family of solutions shapes projections on
the plane of the sky at an arbitrary epoch, showing the differences between
models. The darker the area is the more frequently it appeared in the family. Solid line
represents the best shape model found.}
	\label{fig:family}
\end{figure}

The modelling process described in section \ref{sec:modellingProcess} is run
multiple times and every run, starting form a sphere, follows a different path
and produces a separate model (Fig. \ref{fig:family}). This is typical for
genetic codes to ensure that the global minimum is found. The collection of
models from separate modelling runs is called a family of solutions.  In an
ideal scenario, when the amount of observational data is sufficient and covers
many apparitions of an asteroid (at least 3) all models in a family of
solutions are alike.

Most of the time a family of solutions will consist of two subsets of models,
one for each possible, ambiguous pole orientation. As mentioned, all models
are similar, except  models from one subset are flipped by the $xy$ plane
with the longitude of the pole $\lambda$ differing by $180^{\circ}$. These subsets
represent the two possible -- pro and retrograde -- senses of rotation of the
same body.

The resulting, final models (one for each of the two pole solutions) are chosen
from a family of solutions based on the overall RMSD value describing the fit
to all observed lightcurves.

\subsection{Convergence}
At a first glance, if a small amount of data (e.g. one lightcurve) is
given, a genetic algorithm should find a solution quickly. It would be a local
minimum -- one of many possible shapes able to explain the data.
But it is actually not the case. Observational data determine the fitness
function (namely, global RMSD) for shapes in randomly generated populations.
The smaller the set of data to compare with, the smaller the selective pressure
acting on a population. As a result, we deal with genetic drift rather than
natural selection under the fitness function. The algorithm then behaves more
like a random walk and actual evolution (changes in models leading to a
solution) happens very slowly if it happens at all. It is due to the fact, that
parameters' values alterations are rarely reflected in fitness function
changes. Therefore, the number of populations, which translates directly into
computing time needed, grows dramatically making a genetic algorithm inefficient
in such situations.


To minimize the number of iterations and prevent premature convergence from
happening, sufficient amount of data is required and it depends on factors such
as asteroid's spin axis' orientation, mutual positions of a target and the Earth
or data quality, to name a few. This is target specific and cannot be calculated
precisely. Before the modelling starts, we can assess the amount of information
present in the data by making apparitions' plots (like shown in
Fig.~\ref{fig:Eros_app}) to see available geometries or check the coverage and
quality of the lightcurves.  

\subsection{Model validation}
The success of modelling depends mainly on a data set. There might be parts of
asteroid's surface which are poorly or not at all covered in lightcurves. By
studying a family of solutions we can tell if a data set provides good coverage
of geometries and grants a unique shape solution. The differences between the
models in a family of solutions reveal parts of a shape that were not covered in
data, and, if solutions are not alike, modelling is considered inconclusive.

Formally, the best model found has the lowest RMSD value. It is a very good
criteria to evaluate a model, but one number does not tell the whole
story. What also matters greatly is model's ability to reproduce some
distinctive and unique features present in the observed lightcurves without
producing additional artifacts. However, some observed features might be bogus
making validation process tricky.

During the modelling the best RMSDs found for individual lightcurves are saved.
These are the best fits that occurred in the whole process regardless of
how they fit other lightcurves.
The weighting process is based on these values and disallows a
situation where e.g. one lightcurve is perfectly fitted (assuring low overall
RMSD) with others fitted poorly. Thanks to that, local minima are avoided and
the model explains all the lightcurves at a comparable level.

When dealing with the fact that observed lightcurves are noisy (the noise level
 being often underestimated or unknown) best RMSDs for individual
lightcurves serve as reference points and show what can be achieved based on a
given set of observations; this equips us with some additional information about
the quality of global RMSD.

\section{Numerical tests}
\label{sec:num_tests}

The setup used in tests represents ideal case for the purpose of testing
SAGE's capability of recreating shape, pole orientation and period without any
interference that would come from albedo variations, wrong scattering law or
uneven mass distribution. The lightcurves are similarly not affected by
atmosphere nor by photometric system normally used in asteroid observations.

\subsection{Test bodies}

To create test models we used Gaussian random sphere generation code based on
the algorithm described by Muinonen \citeyearpar{MuinonenGaussian}.  The code
takes spherical harmonics order and surface grid subdivision depth as input
parameters and creates a random body that is later triangulated. Some of the
test bodies were further altered by hand to create more extreme cases.

We moved the center of a body to the computed centre of mass and also computed
an inertia tensor to align spin axis with the axis of greatest inertia. That
simulates homogeneous physical body. We did not introduce any albedo variations
on bodies' surfaces.

The shapes and modelling results can be found in appendices available online;
the models are labeled with Latin alphabet capital letters. Model A was studied
in more detail and is presented below.

\subsection{Orbit}

\begin{figure}
	\begin{center}
		\begin{tikzpicture}
			\def\R{2.5}

			\draw (0,0) circle (\R);
			\node(1) at (0,0) {$\Huge{\lambda}$};
			\draw (0,0) circle (0.06*\R);

			\foreach \x in {0,45,..., 315}
			{
				\pgfmathtruncatemacro\y{\x/45 + 1}
				\fill[fill=white] (\x +90:\R) circle (0.06*\R);
				\draw (\x +90:\R) circle (0.06*\R);
				\node(1) at (\x +90:\R) {\y};
				\node(1) at (\x +90:0.7*\R) {$\x^{\circ}$};
			}
		\end{tikzpicture}

		\includegraphics[width=8cm]{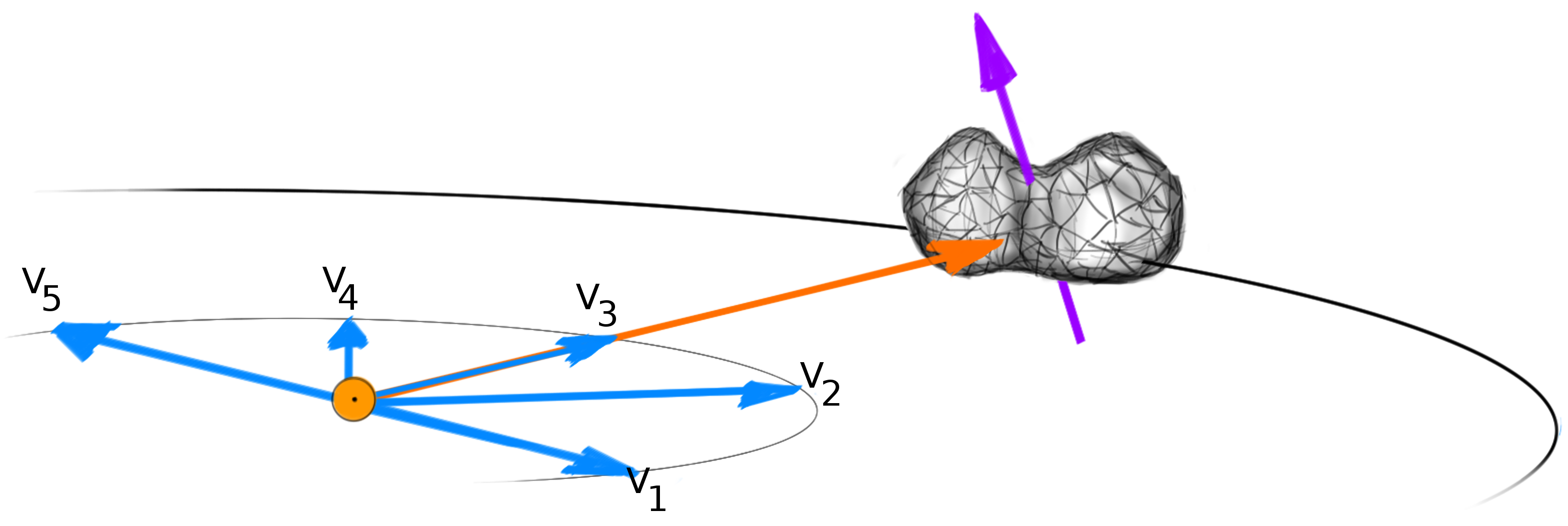}
		\end{center}
		\caption{
		Heliocentric test orbits schema. Top
		image shows the distribution of apparitions (observer is in the center
		of the graph). Bottom image represents a setup for single apparition.
		Vectors $v_1$ through $v_5$ define observer's positions while body's position
		is defined by $av_3$. To create 8 different apparitions the initial vectors
		are rotated about $z$ axis by $45\st$, $90\st$, \dots, $315\st$.
}
	\label{fig:test_orbit}
\end{figure}
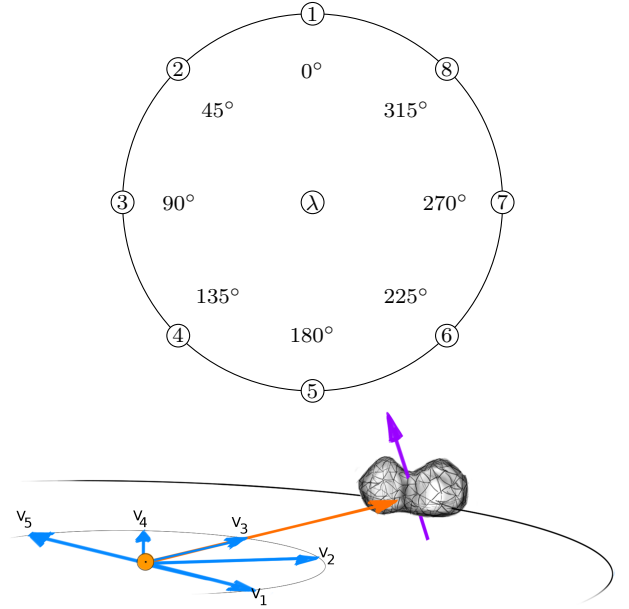

We placed test bodies on an artificial circular orbit around a source of light
with semimajor-axis $a = 3.5AU$ with orbital period of $6.55$ years. Julian day
is used to represent time. An observer is situated on a circular, coplanar,
non-physical orbit at $1AU$. We created $8$ evenly distributed apparitions every
$45^{\circ}$ (Fig.~\ref{fig:test_orbit}) covering one revolution around the
light source.  At every apparition lightcurves were created form $5$ locations
(eq.~\ref{eq:positions}) at the same time, then the whole setup was rotated
$45^{\circ}$ and $1/8^{th}$ of the orbital period was added to the time. The
initial vectors used to place an observer in space were

\begin{equation}
	\label{eq:positions}
\begin{aligned}
	\boldsymbol{v}_1 &= (0, -1, 0)^T \\
	\boldsymbol{v}_2 &= (\frac{1}{\sqrt 2}, -\frac{1}{\sqrt 2}, 0)^T \\
	\boldsymbol{v}_3 &= (1, 0, 0)^T \\
	\boldsymbol{v}_4 &= (\frac{1}{\sqrt 2}, \frac{1}{\sqrt 2}, 0)^T \\
	\boldsymbol{v}_5 &= (0, 1, 0)^T
\end{aligned}
\end{equation}
A test body was put in the position $a\boldsymbol{v}_3$ in every apparition.
As mentioned above, this setup is non-physical (orbital periods of an observer
and a test body are the same despite different semimajor-axes) and this
experiment could not be replicated in reality.
Nonetheless, the relevant aspect is the geometries at which we observe the
bodies could be obtained by extending the time of observations,
i.e. the lightcurves would not be
collected from one revolution but from multiple ones. The actual times of
observations are not important. Choosing one particular real-life orbit  would
produce uneven distribution of apparitions and introduce biases, especially when
we reduced the number of apparitions in the tests.  


\subsection{Lightcurves}

Every lightcurve in every apparition covers a full body rotation and consists of
$180$ evenly distributed points every $2^{\circ}$ of the rotation
phase.

To create lightcurves for test bodies the same code was used as in SAGE
algorithm. This ensures that modeling is affected by the same ''numerical
reality'', with exactly the same scattering law, shadowing effects or numerical
errors (e.g.  floating point rounding). The shape representations however are
different, as spherical harmonics were used to create test models and SAGE uses
a set of direction fixed vectors. The only similarity lies with models being
represented with triangle faces in both cases, but their number and
distribution is different as well.

\subsection{Modelling}

Test bodies were modeled using different sets of data.  We varied the amount of
apparitions and their distribution on the orbit, the orientation of the spin
axis and the phase angle (i.e. the set of observer vectors used in an
apparition).

Test body's locations are numbered $1$ through $8$ and are placed
ecounter-clockwise very $45\st$ (Fig.~\ref{fig:test_orbit}, top graph) starting
from $\boldsymbol{v}^{\text{init}}_{\text{body}} = a (1, 0, 0)^T$.  The
\textit{app.} abbreviation used in results tables enumerates apparitions'
indexes used in the modeling process.



To test the agreement of the models with the test body we constructed topography
maps. Given a direction in space we calculated a mean distance from surface
elements to the origin within $10\st$-wide cone and then subtracted the radius
of a circumsphere. Each vertex was normalized so the length of most distant
vertex from the origin equals $1$, therefore the sphere radius is $1$ as
well.

When we subtract a topography map of a model from a topography map of a test
model we get a map of differences between the two. This allows us to compare
models and interpret results. A differences map can be displayed directly on
the surface of the models
in arbitrary orientation.


\subsection{Results for model A}

\begin{figure}
	\centering
	\includegraphics[width=8cm]{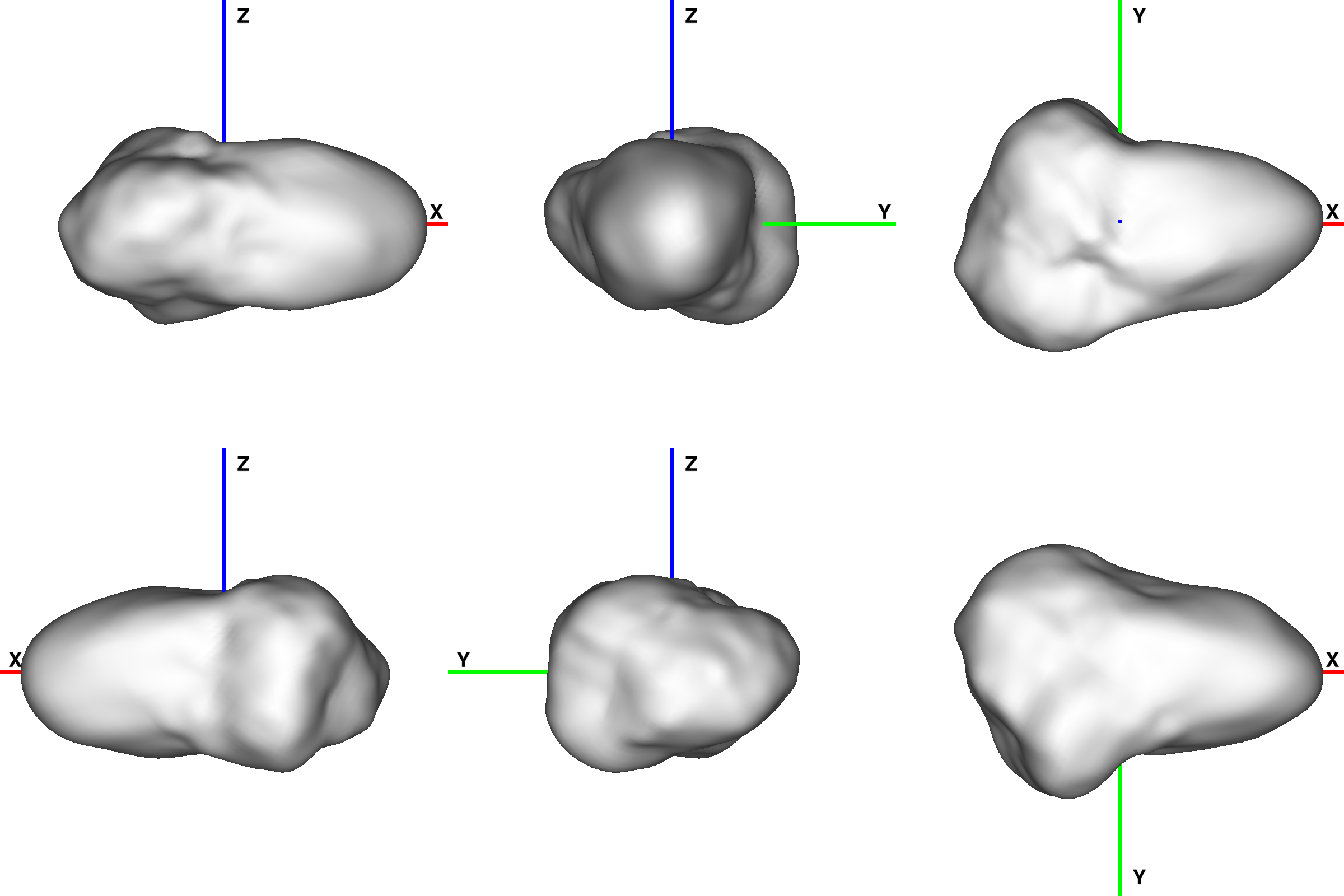}
	\includegraphics[width=8cm]{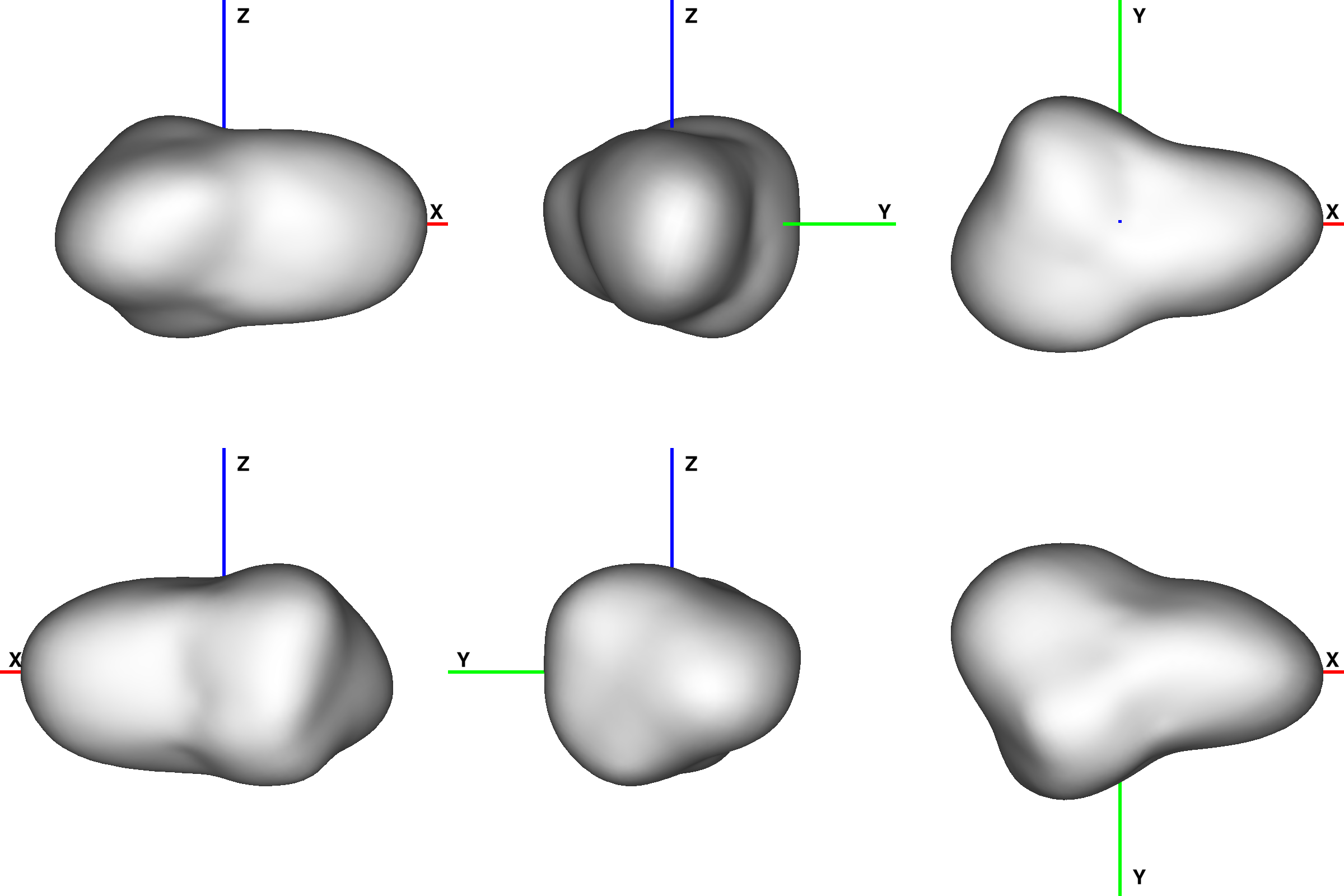}
	\caption{$xz$, $yz$, $xy$, $-xz$, $-yz$ and $-xy$ projections of the test
		model A. First and second rows: test model, third and fourth rows: the
	best model A from inversion.}
	\label{fig:modelA}
\end{figure}

\begin{figure}
	\begin{center}
		\includegraphics[width=8cm]{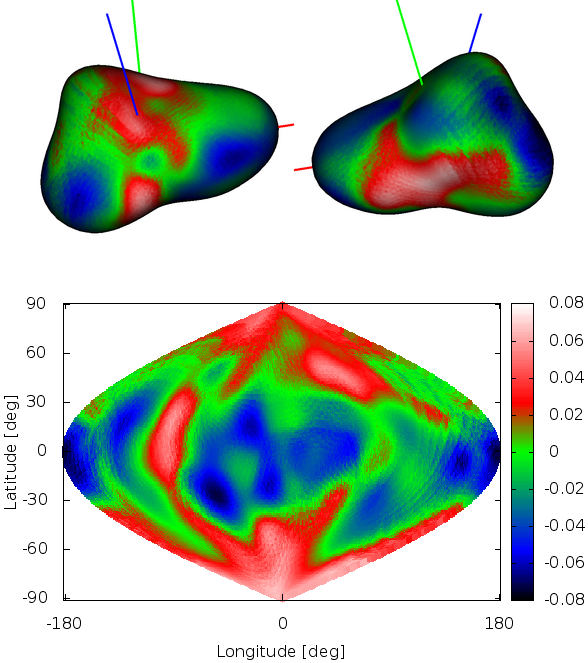}
	\end{center}
	\caption{Topography maps for model A. Colours
	correspond to the difference between test and modelled body in the units of
	a circumsphere radius. Top row shows the topography map on the surface of
	the body on two viewing geometries.}
	\label{fig:topography_modelA}
\end{figure}

\begin{figure}
	\includegraphics[width=8cm]{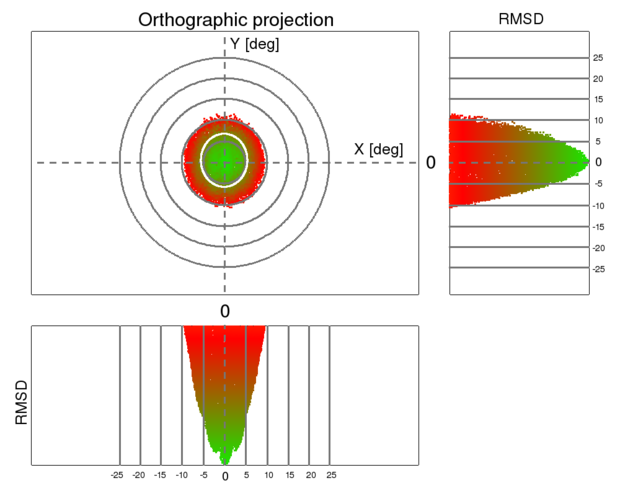}
	\caption{Pole solution map for model A. The map shows
	coordinates of the best pole solutions projected from the sphere onto a plane.
The sphere and plane tangent point is at the position of the test body's correct
pole. Colours correspond to RMSD of the solution, green being the best one.
The white ellipse represents the variance of the pole solution in the $x$ and $y$
plane coordinates.}
	\label{fig:pole_modelA}
\end{figure}

\begin{figure}
	\includegraphics[width=8cm]{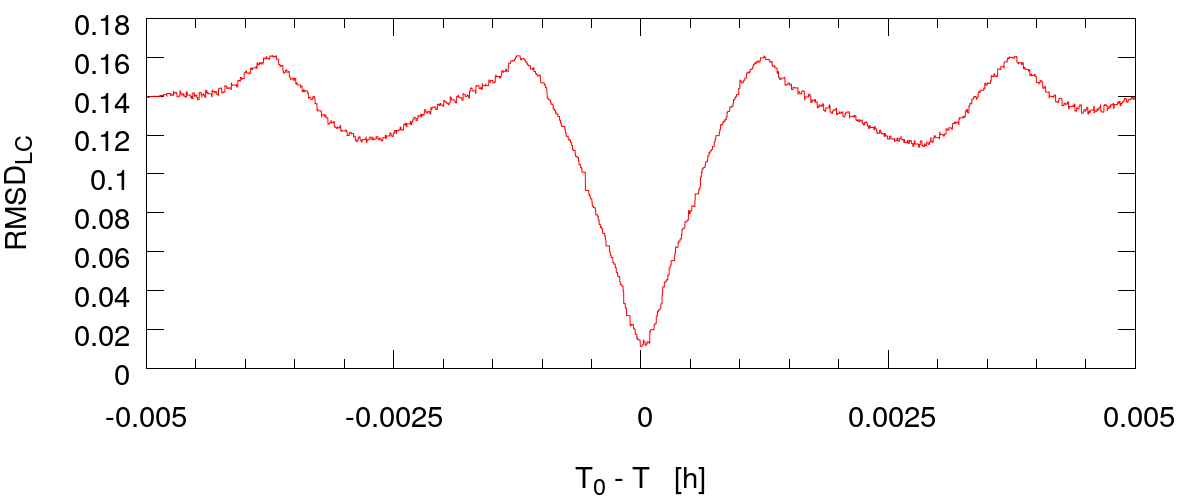}
	\caption{Periodogram for the best result for model A.}
	\label{fig:periodogram_A}
\end{figure}

Table \ref{tab:modelA} presents a summary of all modelling runs for this
model. Model A and its best modelled result can be seen on figure
\ref{fig:modelA} along with topography, pole solution maps and periodogram on
figures~\ref{fig:topography_modelA},  \ref{fig:pole_modelA} and
\ref{fig:periodogram_A}.

\begin{table*}
\centering
\caption{
	Tests results summary for models A (Fig.~\ref{fig:modelA}), B
	(Fig.~\ref{fig:projections_model_B}), C (Fig.~\ref{fig:projections_model_C})
	and D (Fig.~\ref{fig:projections_model_D}).
	$\lambda_{init}$, $\beta_{init}$, $\lambda$ and $\beta$ are the test
	and modelled bodies' spin axis coordinates respectively, $P_{init}$ and $P$
	are rotation periods, \textit{app.} enumerates apparitions used in the modelling
	(e.g. ''1234'' means apparitions 1 through 4 were used), $\theta$ is a phase
	angle. RMSD$_\text{model}$ describes the test and
	modelled shapes' fit, while  RMSD$_\text{LC}$ describes test and modelled
	bodies' lightcurve fit.
}
\label{tab:modelA}
\begin{tabular}{|c|c|c|c|c|c|c|c|c|c|c|c|}
\hline
ID & Model & $\lambda_{\text{init}}[\st]$ & $\beta_{\text{init}}[\st]$ & $\lambda[\st]$ &
$\beta[\st]$ & $P_{\text{init}}[h]$ & $P[h]$ & app. & $\theta[\st]$ &
$RMSD_{\text{model}}$ & $RMSD_{\text{LC}}$  \\ \hline \hline
1 & A & $0$  & $0$ & -- & $0\pm6$ & $12$ & $12.00001\pm10^{-5}$ & all $8$ &0, 14, 16& $0.032662$ & $0.006023$ \\ \hline
2 & A &  $0$ & $90$ & $173\pm12$ & $90\pm12$ & $12$ & $12.00001\pm10^{-5}$ & all $8$ &0, 14, 16& $0.033616$ & $0.010631$ \\ \hline
3 & A &  $90$ & $45$ & $89\pm7$ & $43\pm7$ & $12$ & $12.00001\pm10^{-5}$ & all $8$ &0, 14, 16& $0.023411$ & $0.006601$ \\ \hline
\multicolumn{12}{l}{}\\ \hline
4 & A &  $90$ & $45$ & $88\pm8$ & $44\pm8$ & $12$ & $11.99999\pm4\cdot10^{-5}$ & $1234$ &0, 14, 16& $0.035303$ & $0.008891$ \\ \hline
5 & A &  $90$ & $45$ & $-90\pm9$ & $41\pm9$ & $12$ & $11.99999\pm3\cdot10^{-5}$ & $1256$ &0, 14, 16& $0.032389$ & $0.011149$ \\ \hline
6 & A &  $90$ & $45$ & $93\pm9$ & $44\pm9$ & $12$ & $11.99997\pm3\cdot10^{-5}$ & $1357$ &0, 14, 16& $0.042925$ & $0.010062$ \\ \hline
\multicolumn{12}{l}{}\\ \hline
7 & A &  $90$ & $45$ & $91\pm7$ & $44\pm7$ & $12$ & $12.00001\pm10^{-5}$ & all $8$ &0, 14& $0.030033$ & $0.006312$ \\ \hline
8 & A &  $90$ & $45$ & $90\pm6$ & $45\pm6$ & $12$ & $12.00001\pm10^{-5}$ & all $8$ &0& $0.037269$ & $0.005038$ \\ \hline
\multicolumn{12}{l}{}\\ \hline
9 & B & $290$  & $45$ & $291\pm13$ & $45\pm13$ & $6.75$ & $6.75000\pm10^{-5}$ & all $8$ &0, 14, 16& $0.026721$ & $0.009015$ \\ \hline
10 & C & $330$  & $45$ & $330\pm7$ & $41\pm7$ & $12$ & $11.99999\pm10^{-5}$ & all $8$ &0, 14, 16& $0.014984$ & $0.005942$ \\ \hline
11 & D & $330$  & $45$ & $330\pm7$ & $41\pm7$ & $12$ & $11.99999\pm10^{-5}$ & all $8$ &0, 14, 16& $0.034730$ & $0.014918$ \\ \hline
\end{tabular}
\end{table*}

\subsubsection{Pole orientation}

In each case we received two separate pole solutions with the same $\beta$ and
$\lambda$ differing by $180\st$; shapes for both pole solutions were alike.
Table~\ref{tab:modelA} shows the best model found from both sets
of pole solutions.

We tested three cases for $\beta$: $0\st$, $45\st$ and $90\st$.  Models with
smallest $RMSD_{\text{model}}$ were obtained using all of the apparitions evenly
distributed on an orbit and with $\beta=45\st$. In such geometry the whole body
is seen by an observer throughout one revolution about a light source, therefore
lightcurves may contain information about the whole body. The best fit was
$RMSD_{\text{model}}=0.023411$ for such a case.

For $\beta=90\st$ (i.e an asteroid spin vector pointing north ecliptic pole)
lightcurves form all the apparitions consist of the same information as the
aspect angle does not change with the position on the orbit therefore the worst
fit can be explained. The amount of information is also greatly reduced when
$\beta=0\st$ due to the fact that many lightcurves are almost or completely
flat.

\subsubsection{Apparitions}

Taking into consideration results for eight apparitions with different $\beta$
we used  $\beta=45\st$ for further tests as the best geometry for shape
modelling.

As we decreased the number of apparitions and their distribution we obtained
worse fits. The 1256 case was the best one among them with
$RMSD_{\text{model}}=0.032389$, consisting of two apparitions $45\st$ apart with
additional two corresponding ones on the opposite sides of the orbit. The 1234
case -- apparitions from half orbit -- was slightly worse
($RMSD_{\text{model}}=0.035303$). The cross-like 1357 apparitions distribution
turned out to be significantly worse than the rest with
$RMSD_{\text{model}}=0.042925$.

Removing apparitions depletes the amount of information present in the
lightcurve dataset making it more difficult for the modelling process to derive
shape and pole solutions. Especially in cross-like setup, two of the four
apparitions consist of the same information so the effective amount of
apparitions is then actually smaller.


\subsubsection{Phase angles}

The observer's and model's positions used when constructing the test model's
lightcurves (fig~\ref{fig:test_orbit}) yield $0\st$, $14.2\st$ and $15.95\st$
phase angles. We reduced observer's position vectors to $\boldsymbol{v}_2$,
$\boldsymbol{v}_3$ and $\boldsymbol{v}_4$ giving phase angles $0\st$ and
$14.2\st$ in one case and $0\st$ phase angle exclusively when reduced to
$\boldsymbol{v}_3$ alone. We used all eight apparitions during the modelling.

The $0\st$ and $14\st$ model had a $RMSD_{\text{model}}=0.030033$; the
$0\st$-only case had $RMSD_{\text{model}}=0.037269$, which is still better than
any of the models obtained with smaller number of apparitions. The modelling
algorithm was still able to reproduce major concavities although they were
rather shallow.


\subsection{Other models}

\begin{figure}
	\includegraphics[width=8cm]{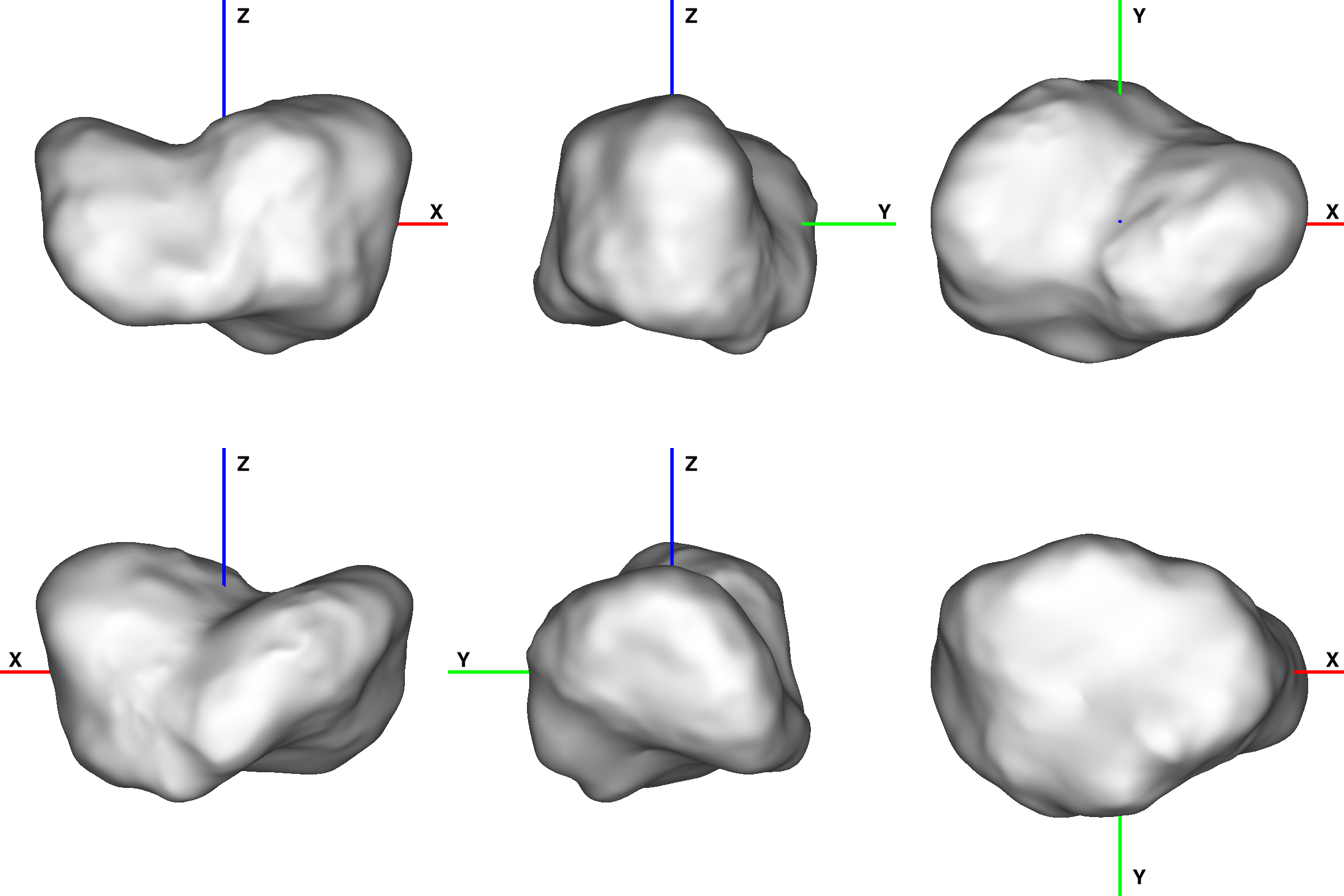}
	\includegraphics[width=8cm]{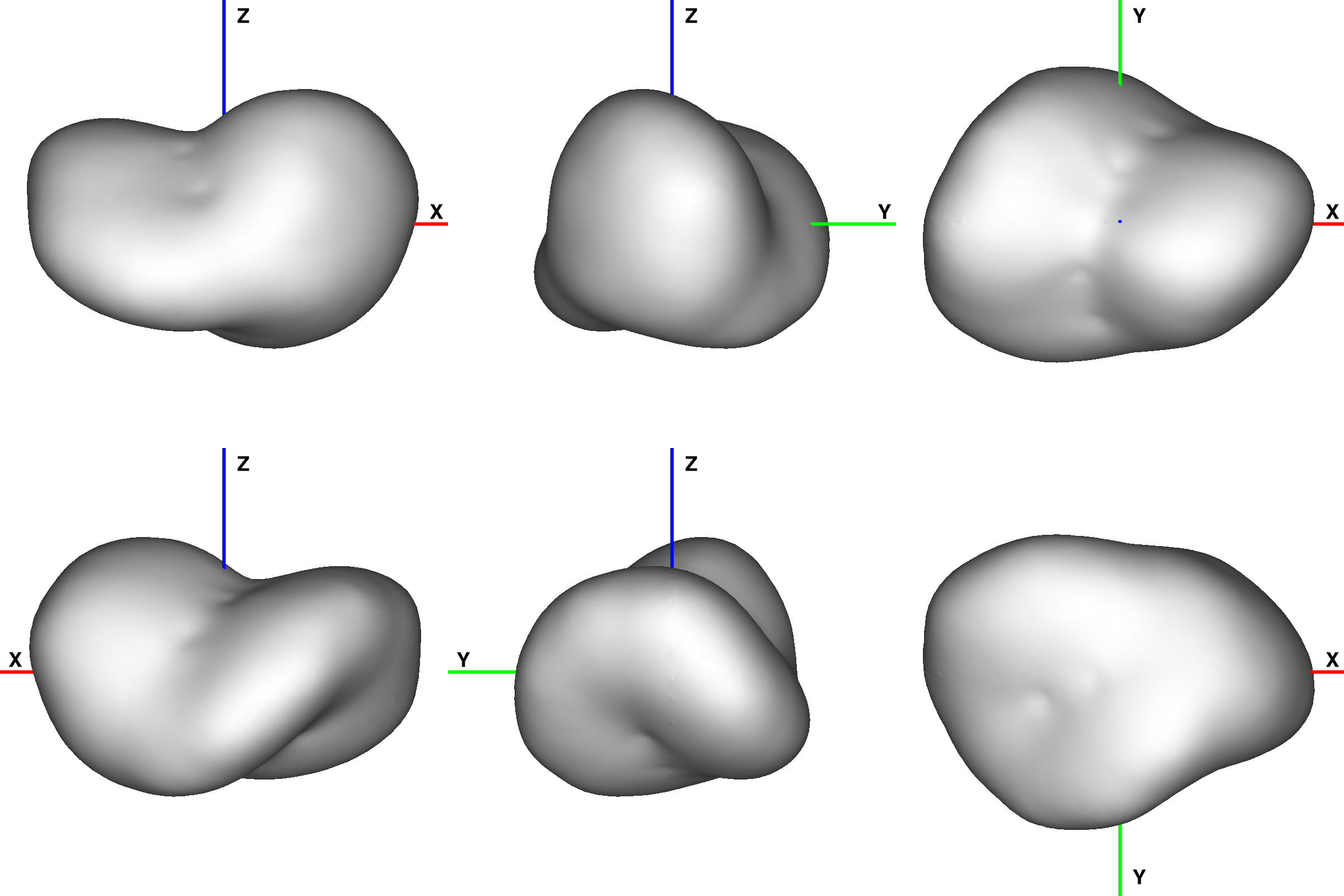}
	\caption{Projections of model B. First and second rows: test model, third and
	fourth rows: the best model B from inversion. }
	\label{fig:projections_model_B}
\end{figure}

\begin{figure}
	\includegraphics[width=8cm]{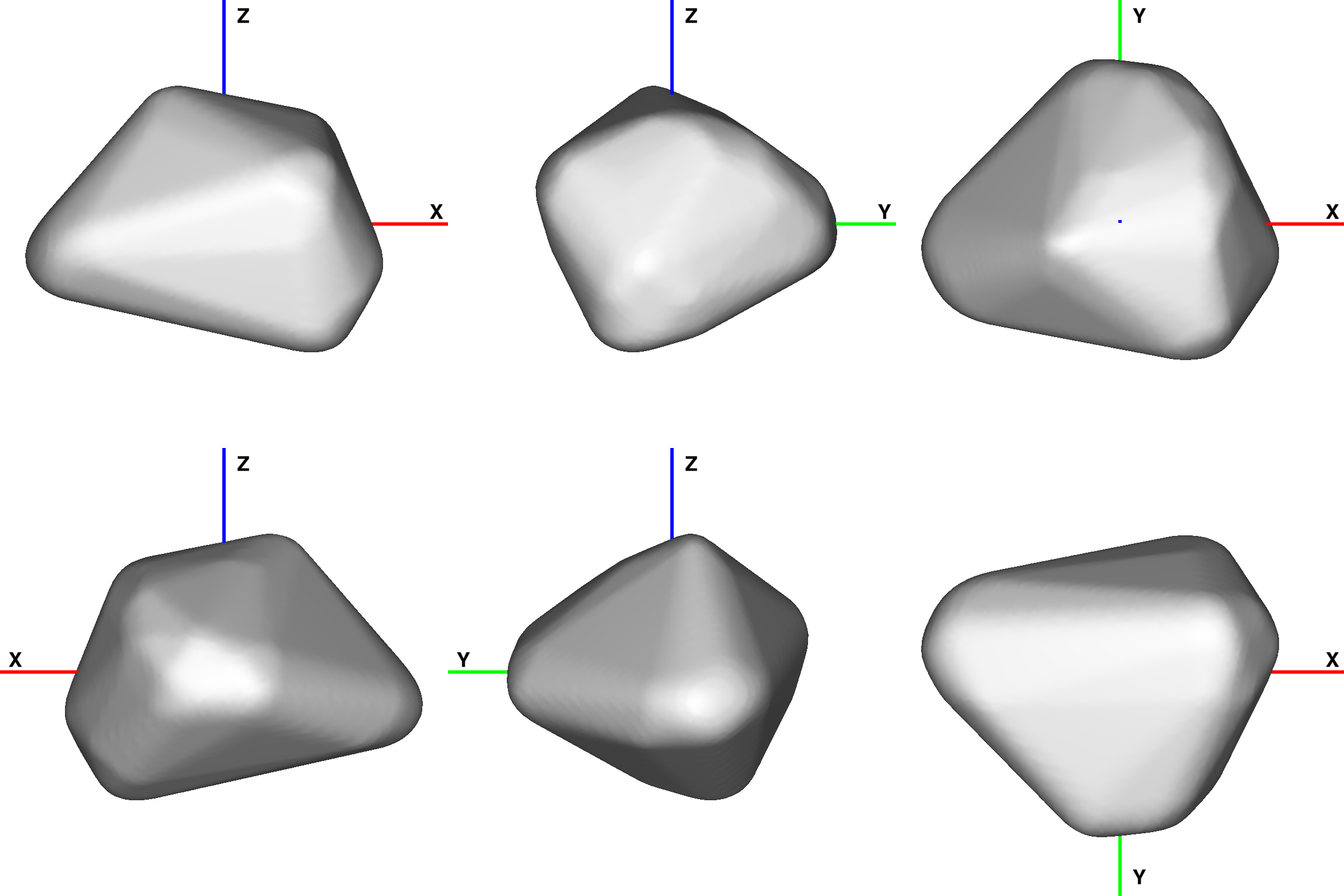}
	\includegraphics[width=8cm]{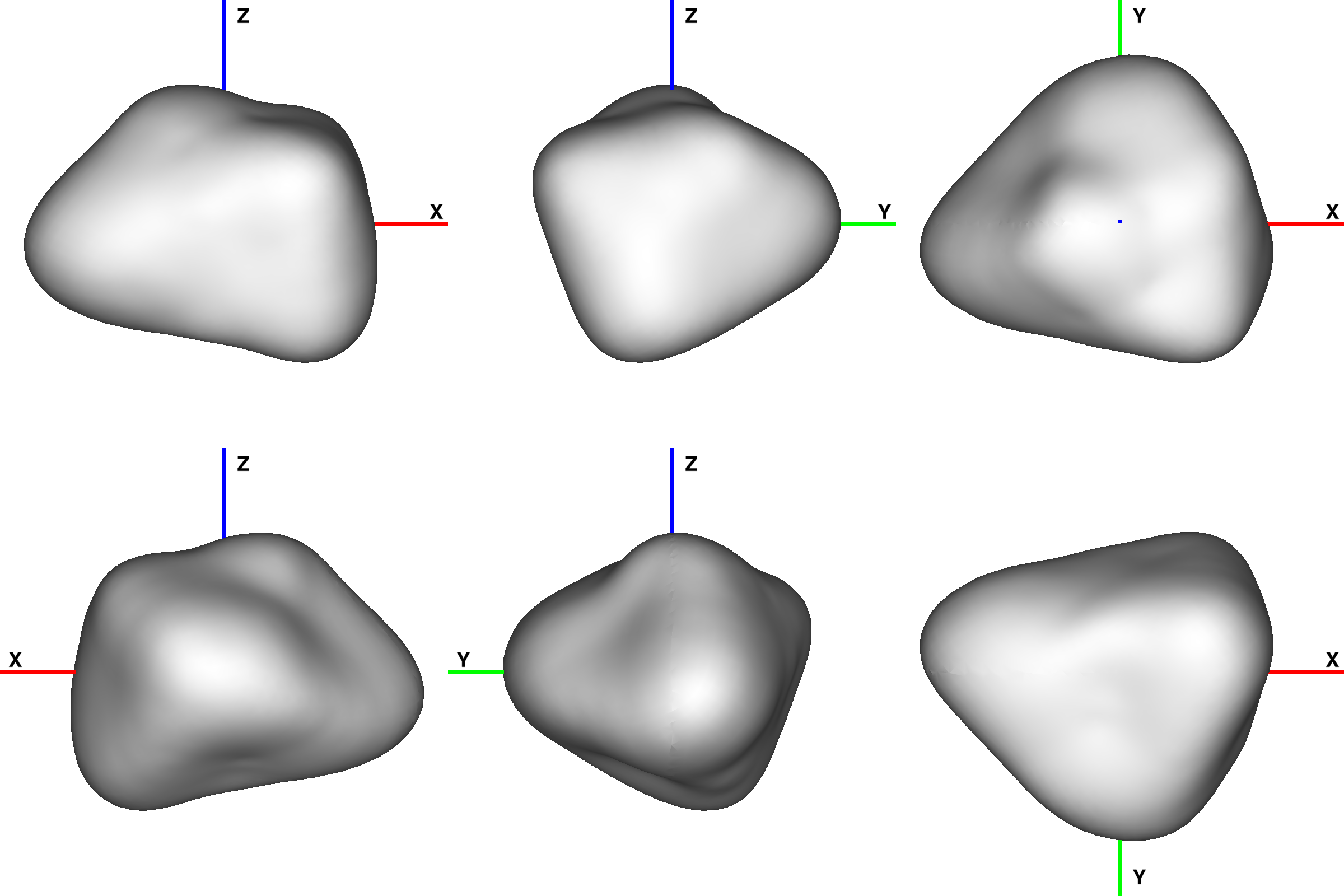}
	\caption{Projections of model C. First and second rows: test model, third and
	fourth rows: the best model C from inversion. }
	\label{fig:projections_model_C}
\end{figure}

\begin{figure}
	\includegraphics[width=8cm]{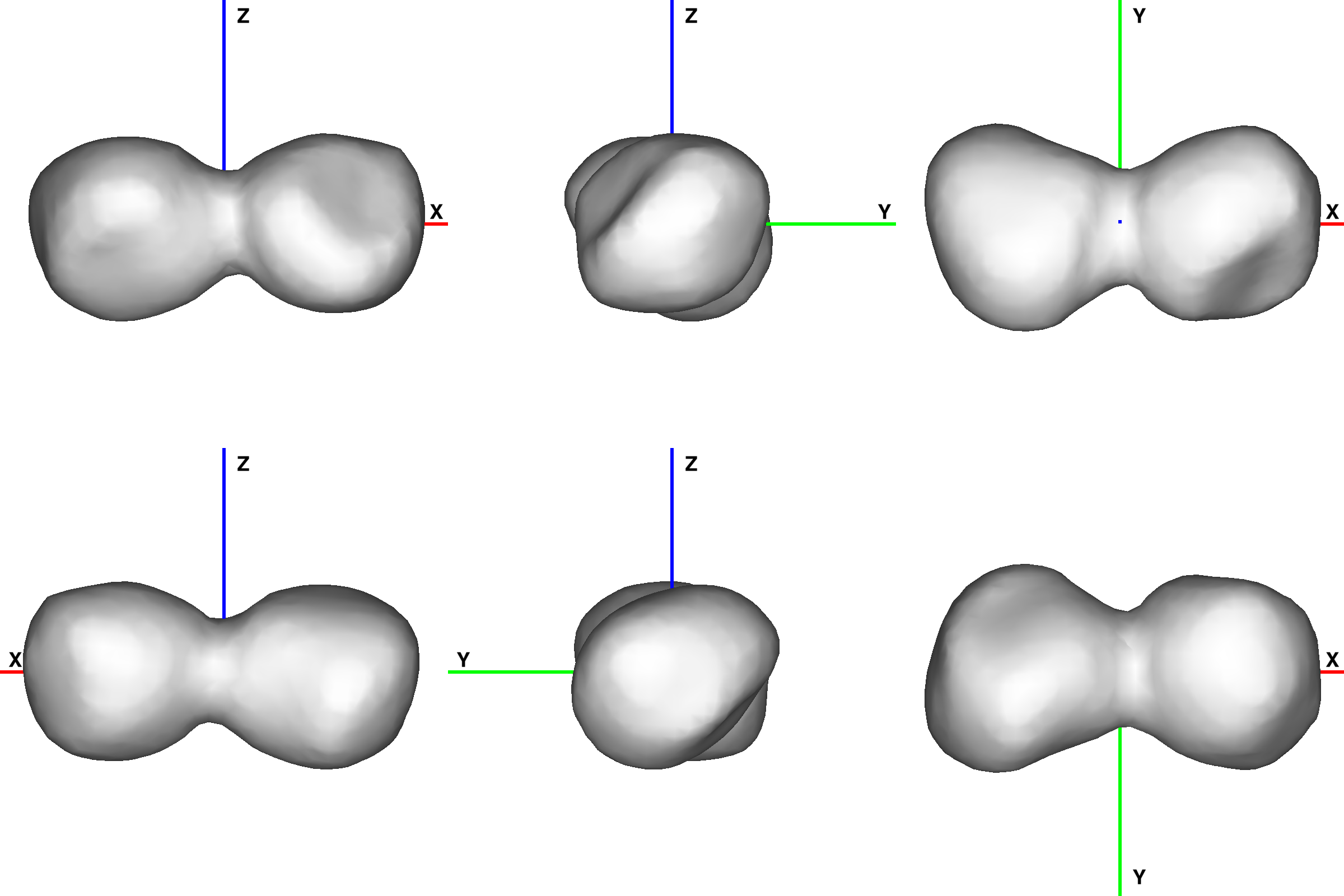}
	\includegraphics[width=8cm]{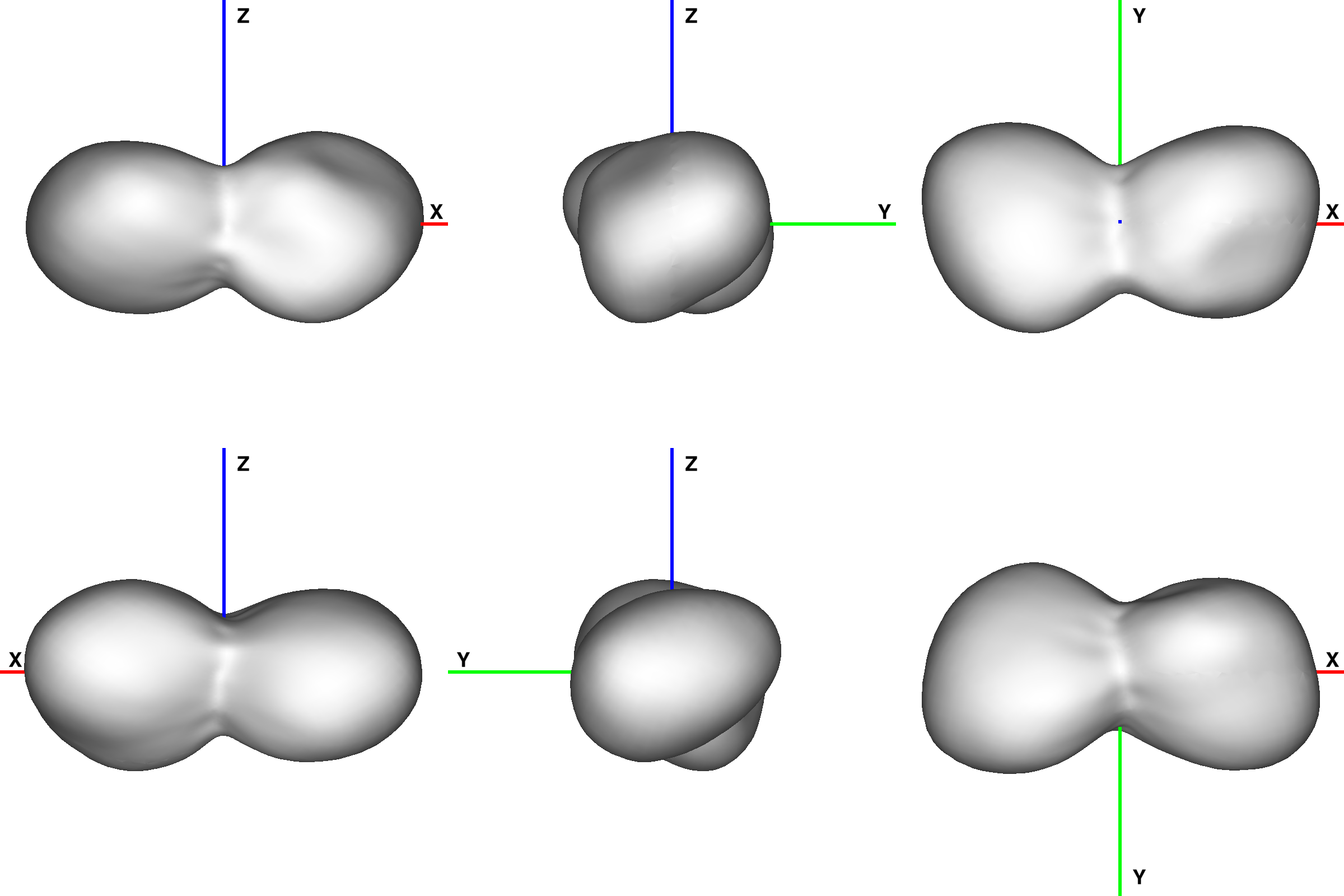}
	\caption{Projections of model D. First and second rows: test model, third and
	fourth rows: the best model D from inversion. }
	\label{fig:projections_model_D}
\end{figure}

We have modelled other test bodies with $\beta=45\st$ to test SAGE's ability
to reconstruct shapes and find pole solutions. The results are summarized in
table~\ref{tab:modelA}. Models are labeled with capital letters, and their
projections can be seen on figures \ref{fig:projections_model_B},
\ref{fig:projections_model_C} and \ref{fig:projections_model_D}.  Diagrams for
pole solutions and periods can be seen on figures \ref{fig:model_B_spin},
\ref{fig:model_C_spin} and \ref{fig:model_D_spin} in appendices available
online.





\section{Models of asteroids}
\label{sec:models_of_asteroids}

\subsection{(433) Eros}

Discovered in 1898, Eros is a well studied S-type NEA.
Thanks to NEAR Shoemaker orbiter probe and its Laser Rangefinder measurements
we have a detailed shape model of this asteroid \citep{Zuber2000}.

With more than one hundred available lightcurves obtained during 5 apparitions
(see Tab.~\ref{tab:Eros} for details) Eros is a very good case to test SAGE
modelling method. Observations of Eros used for modelling were obtained over
large time-span (1951 -- 1993), combined with Eros' orbit giving huge span of
geometries as seen on Fig.~\ref{fig:Eros_app}. Moreover, some lightcurves were
obtained during Eros and the Earth close approaches (1951, 1981, 1974) offering
large and varying phase angles.

\input{img/eros_app.tex}

Eros is a very elongated body which produces lightcurves with amplitudes
exceeding 1 magnitude at equatorial aspects. Moreover, Eros has some
distinctive surface features, such as a giant crater in the middle, making
modelling both interesting and challenging.


Eros model parameters found by SAGE are:
\begin{itemize}
	\item pole coordinates:
		\begin{itemize}
			\item[] $\lambda : 17\st \pm 5\st$
			\item[] $\beta : 8\st \pm 5\st$
		\end{itemize}
	\item rotation period: $5.270256h \pm10^{-6}h$.
\end{itemize}

\begin{figure}
	\includegraphics[width=8cm]{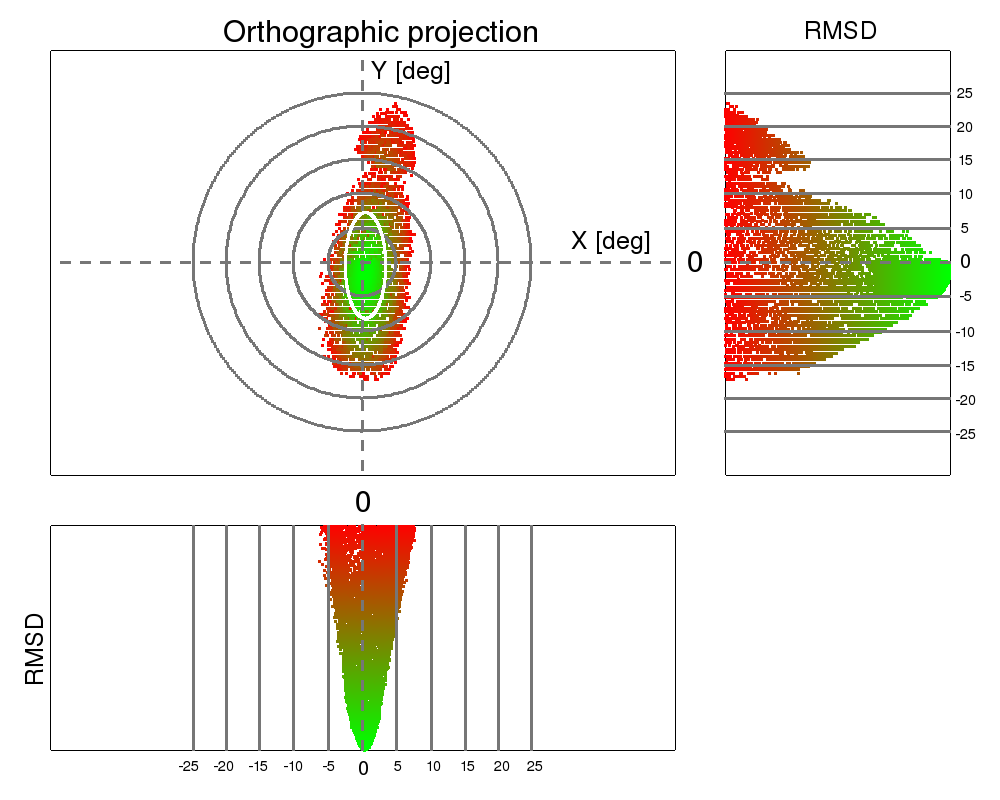}
	\caption{Pole solution map for (433) Eros model.}
	\label{fig:eros_pole}
\end{figure}

\begin{figure}
	\includegraphics[width=8cm]{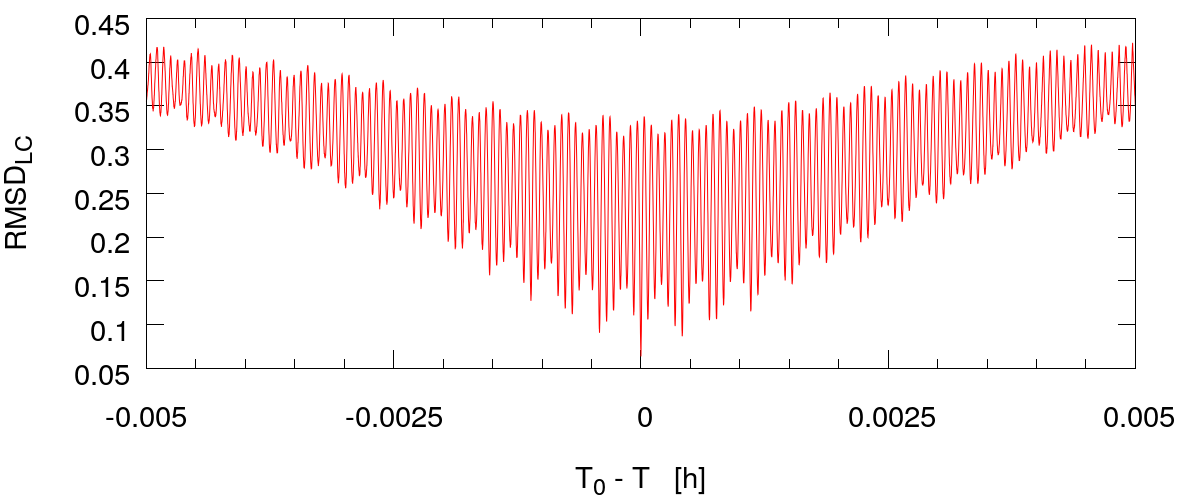}
	\caption{Periodogram for (433) Eros.}
	\label{fig:periodogram_eros}
\end{figure}

These values are in agreement with  the ones found by \cite{Miller02} based
on data from NEAR Shoemaker probe, which
was $\lambda=17.2387\st\pm0.003\st$, $\beta=11.3515\st\pm0.006\st$ and
$P=5.27025547h$.  The pole solution map can be seen on
Fig.~\ref{fig:eros_pole}, while periodogram on Fig.~\ref{fig:periodogram_eros}.

\begin{figure}
	\center
	\includegraphics[width=8cm]{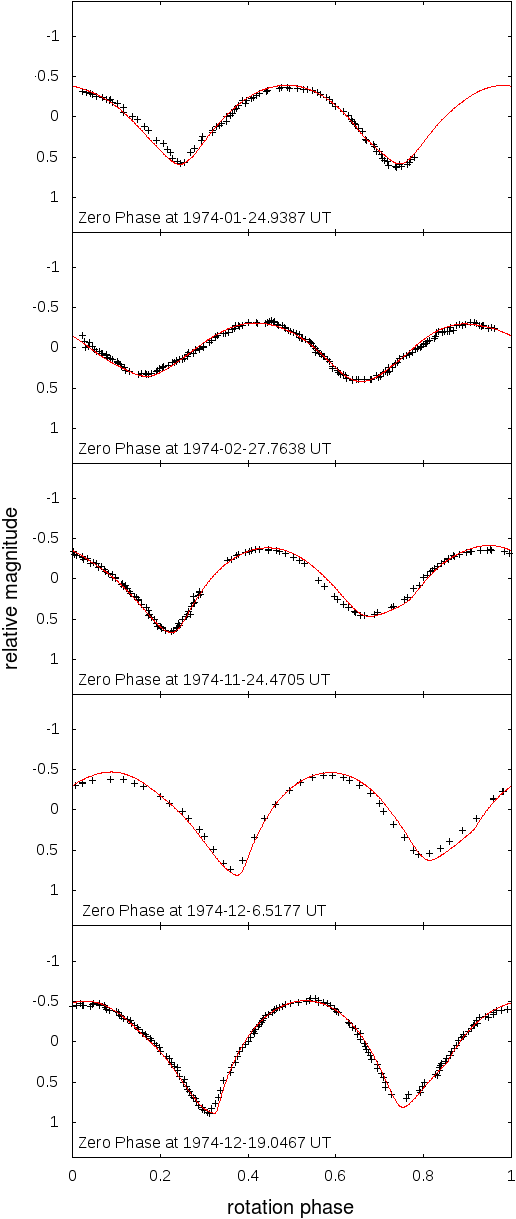}
	\caption{Eros model fit (solid line) to some of the photometric lightcurves
	(dots). The data from top to bottom: Beyer 1953, Dunlap 1976, Cristescu
	1976, Durmmond et al. 1985, Krugly \& Shevchenko 1999.}
	\label{fig:Eros_lc}
\end{figure}

\begin{figure}
	\includegraphics[width=8cm]{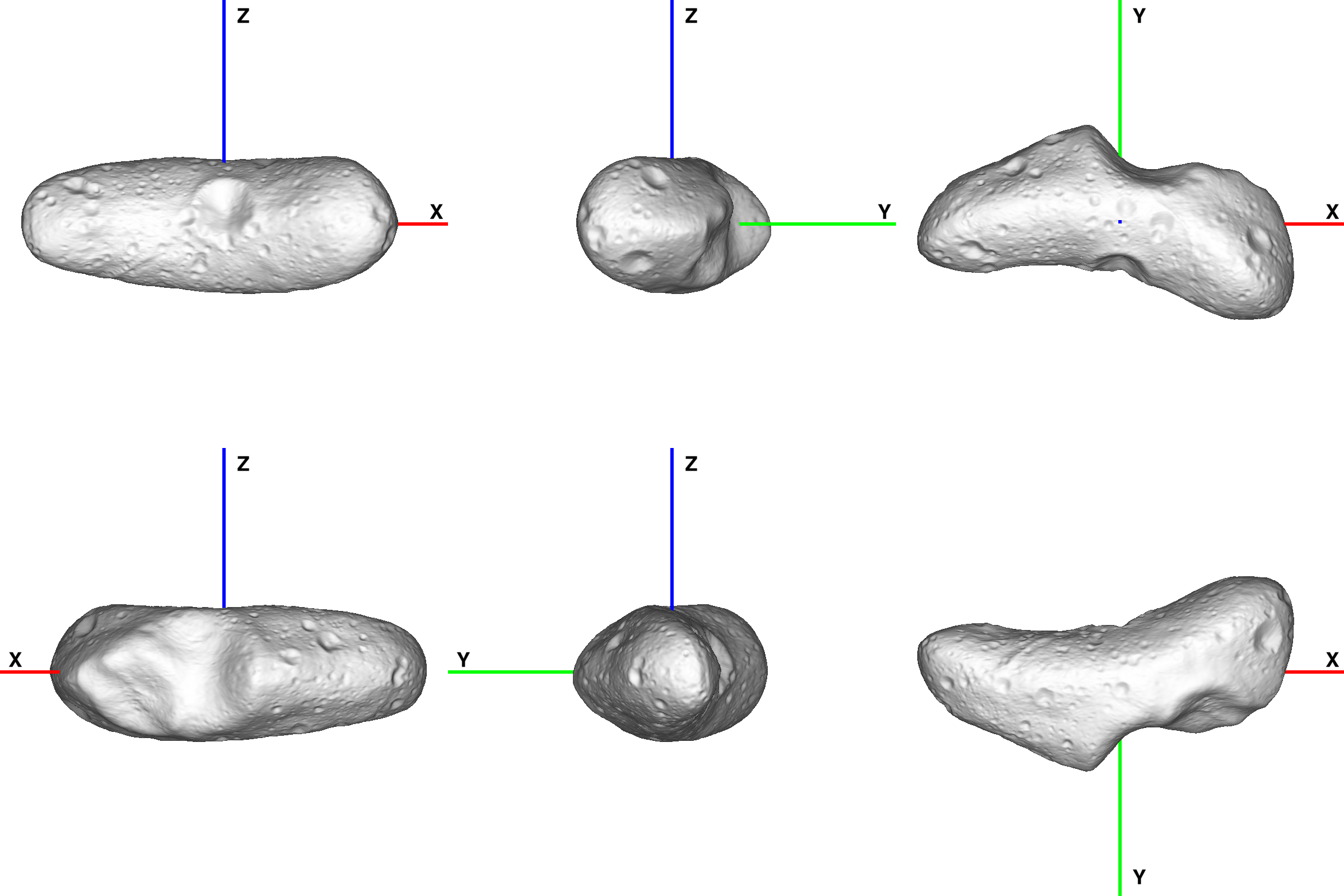}
	\includegraphics[width=8cm]{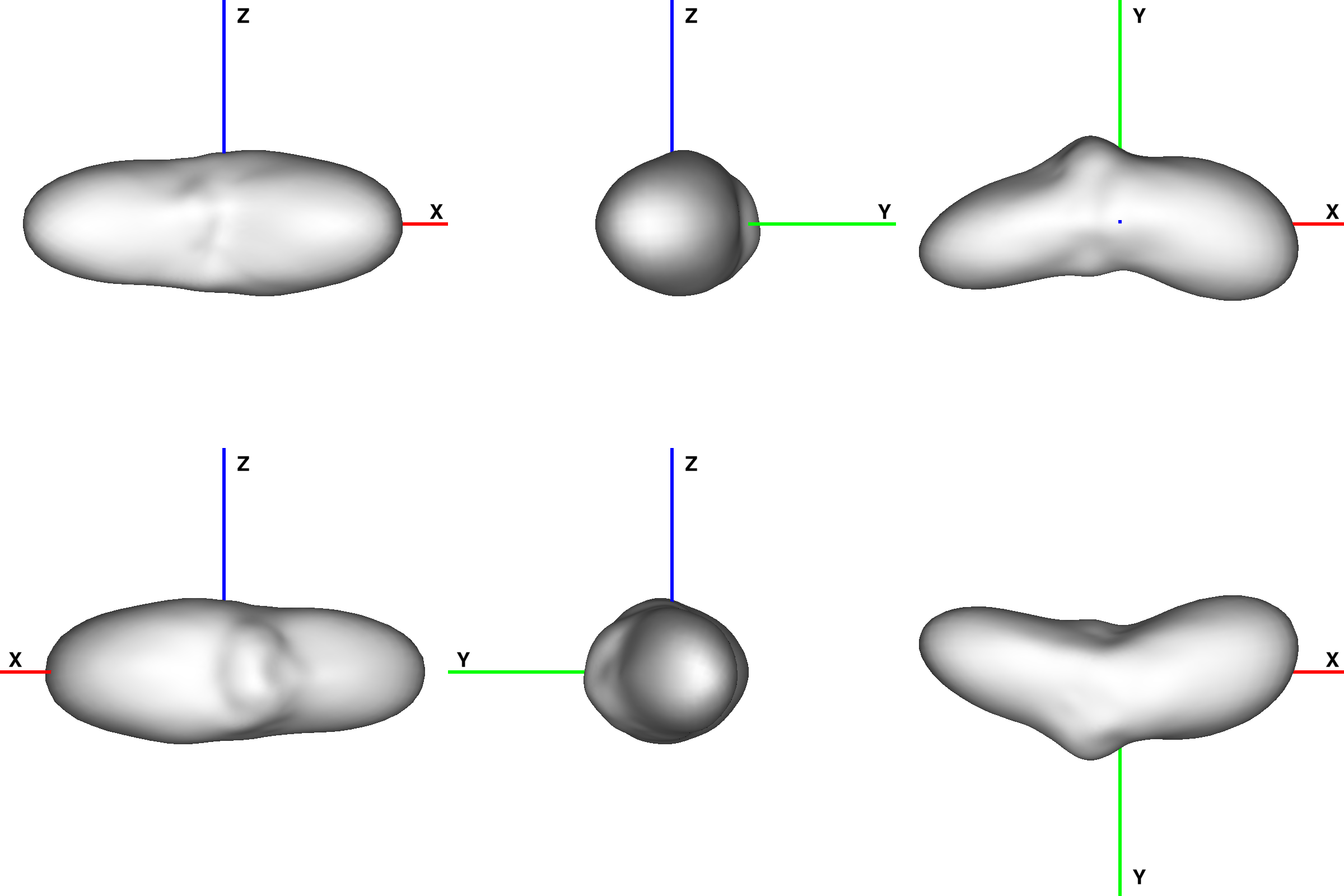}
	\caption{Projections of (433) Eros model obtained from NEAR Shoemaker
		mission (first and second rows), and SAGE model of Eros from photometry
		(third and fourth rows).}
	\label{fig:Eros_projections}
\end{figure}

\begin{figure}
	\includegraphics[width=8cm]{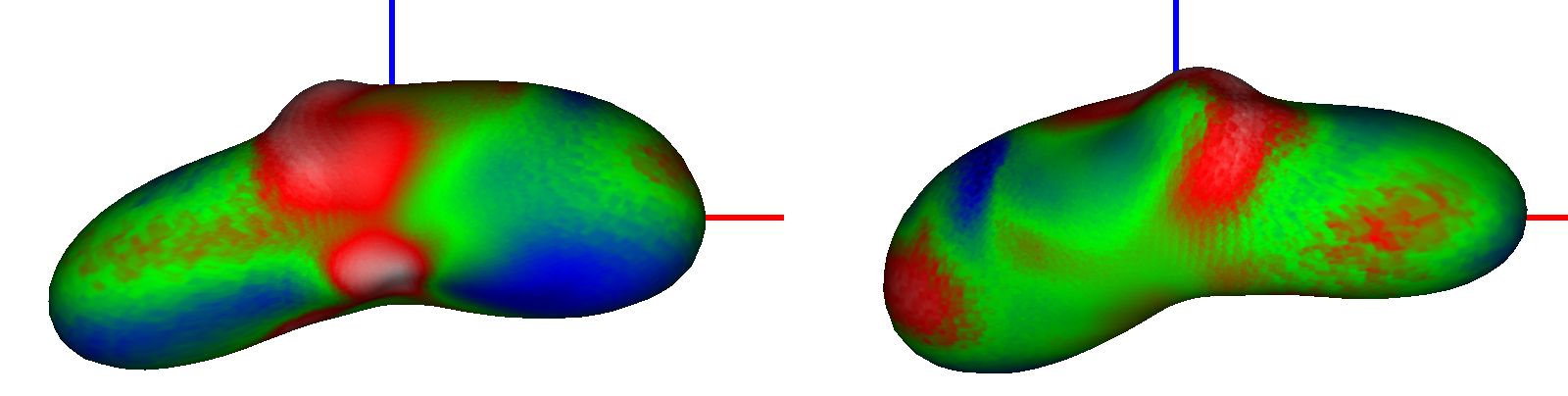}
	\includegraphics[width=8cm]{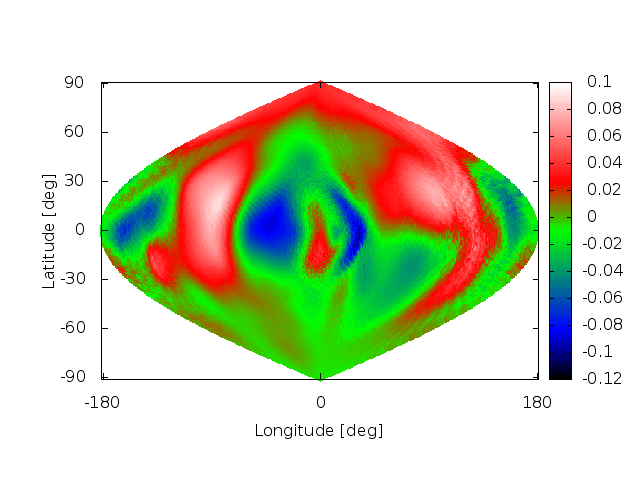}
	\caption{Topography map for (433) Eros SAGE model by comparison with Eros'
		model form NEAR.}
	\label{fig:eros_topography}
\end{figure}

The presented model of Eros is successfully reproducing lightcurves (see
Fig.~\ref{fig:Eros_lc} and Appendix~\ref{app:eros_lc} available online for some
examples), and is in very good visual agreement with the high resolution model
based on the observations obtained during the NEAR Shoemaker rendezvous (see
models' projections in Fig.~\ref{fig:Eros_projections}).  Having a detailed
model from \textit{in situ} measurements allowed us to make a topography map
(Fig.~\ref{fig:eros_topography}) as we did in case of test models. The fit was
at the level of RMSD=$0.025959$ with the largest difference of $0.1R_{max}$.
for targets of unknown size we usually scale the shape model so $R_{max}=1$.  

\begin{table*}
 \caption{Details of the lightcurve data used for 433 Eros modelling. $\alpha$
 -- phase angle, $\lambda$ -- ecliptic longitude, $\beta$ -- ecliptic latitude.}
  \label{tab:Eros}
\begin{tabular}{ccccccc}
\hline
\hline
  Apparition & Year & N$_{lc}$ & $\alpha$ $[^{\circ}]$ & $\lambda$ $[^{\circ}]$ & $\beta$ $[^{\circ}]$ & reference\\
\hline
  1  & 1951/1952 &  28  & 18.6 -- 59.2 & 5.1 -- 118.5 & -10.3 -- 21.5 & \cite{beyer53},\\
  2  &   1972    &   1  &    17.2      &    341.5      &     8.5       & \cite{dunlap76}\\
  3  & 1974/1975 &  68  &  8.6 -- 44.3 & 52.5 -- 158.0 & -31.0 -- 33.8 & \cite{cristescu76}, \cite{dunlap76},\\
							          &&&&&& \cite{millis76},\\
				 			          &&&&&& \cite{miner76}, \cite{pop76},\\
				 			          &&&&&& \cite{scaltriti76}, \cite{tedesco76}\\
 4   & 1981/1982 &  4   & 28.6 -- 53.5 & 42.4 -- 125.6 & -18.6 -- 36.7 & \cite{drummond85}, \cite{harris99}\\
 5   &   1993    &  8   &  1.0 -- 18.1 & 296.1 -- 308.3&  -0.8 -- 3.9  & \cite{krugly99}\\
\hline
\end{tabular}
\end{table*}


\subsection{(9) Metis}

Metis, discovered in 1848, is one of the largest main-belt asteroids. The
available photometric lightcurve data set is rich and spans over many decades
providing good coverage of observing geometries (see Tab.~\ref{tab:Metis_lc} and
Fig.~\ref{fig:Metis_app}).

Metis has also been observed using Adaptive Optics and stellar
occultation events on several occasions. As Metis has not been visited by any
spacecraft to provide \textit{in situ} observations both AO and stellar
occultations may serve as the ground truth to validate SAGE model. Moreover
stellar occultations allowed us to scale the model.

\begin{figure}
	\includegraphics[width=8cm]{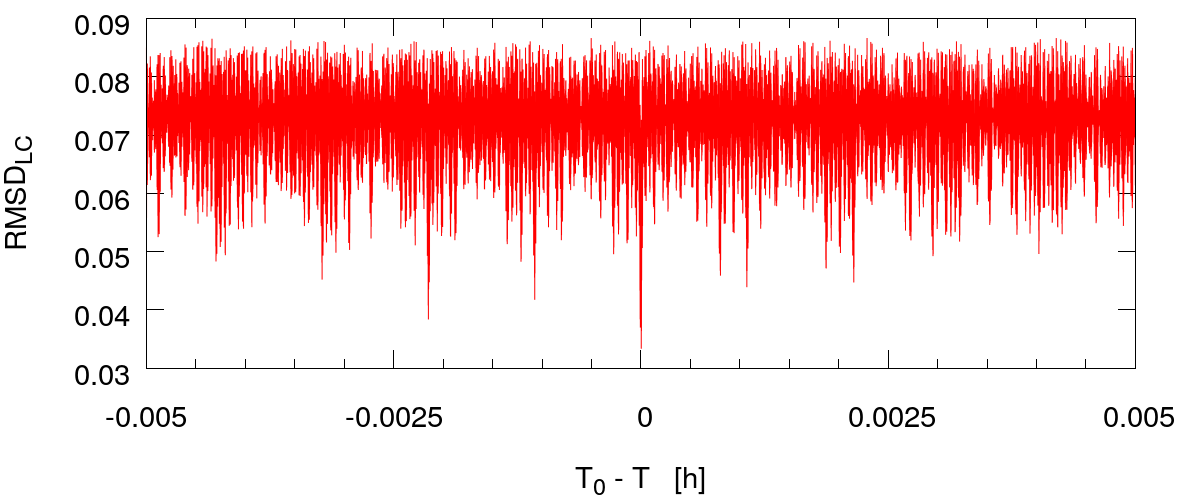}
	\caption{Periodogram for (9) Metis.}
	\label{fig:periodogram_metis}
\end{figure}

\begin{figure*}
\centering
\includegraphics[width=160mm]{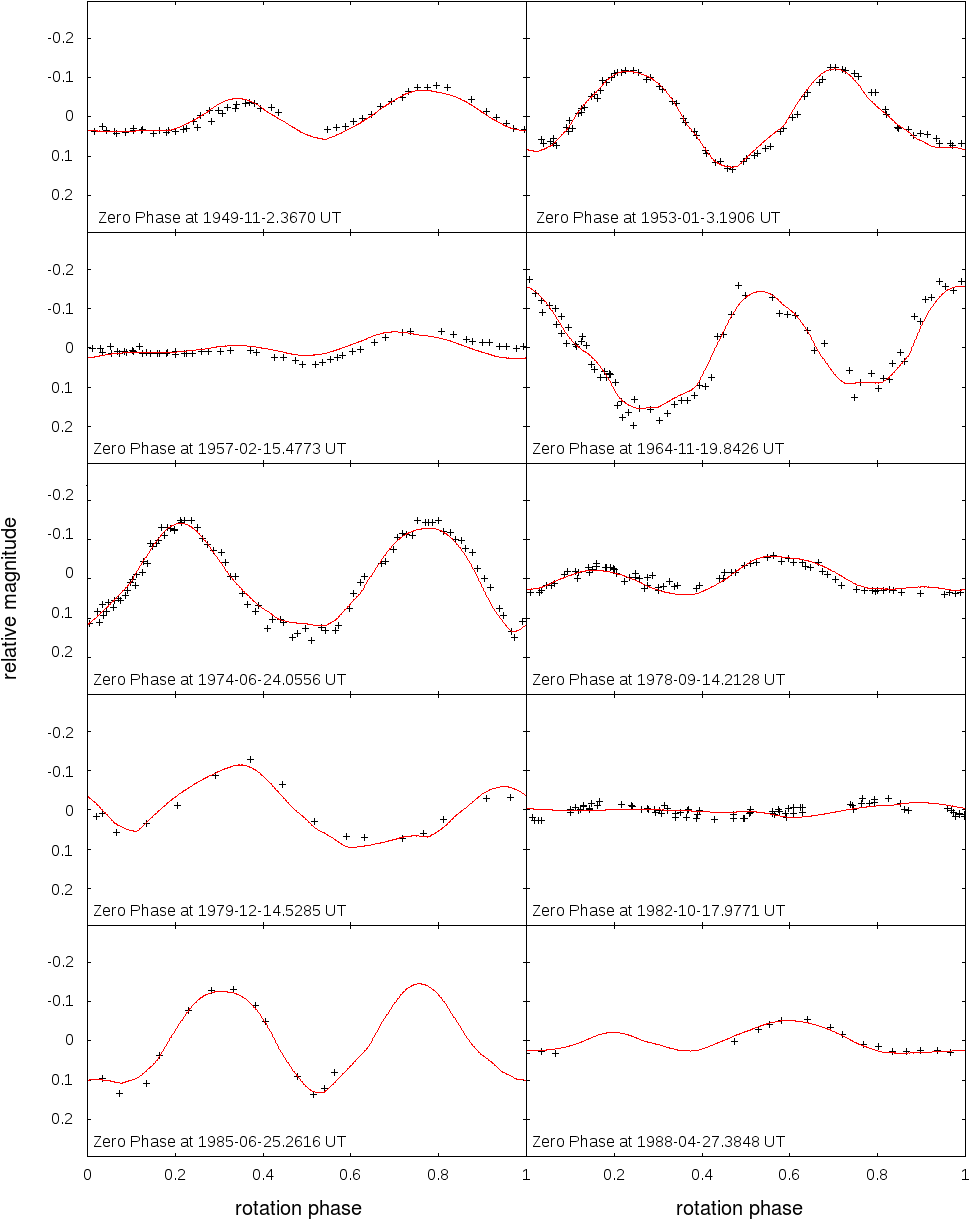}
\caption{Metis model fit to some observations. The solid line is the synthetic
	brightness associated with the model solution, while the dots correspond to
	the photometric observations.
Please notice that a different magnitude scale has been used for each
observation for clarity purposes.}
\label{fig:Metis_lc}
\end{figure*}

\begin{figure}
	\includegraphics[width=8cm]{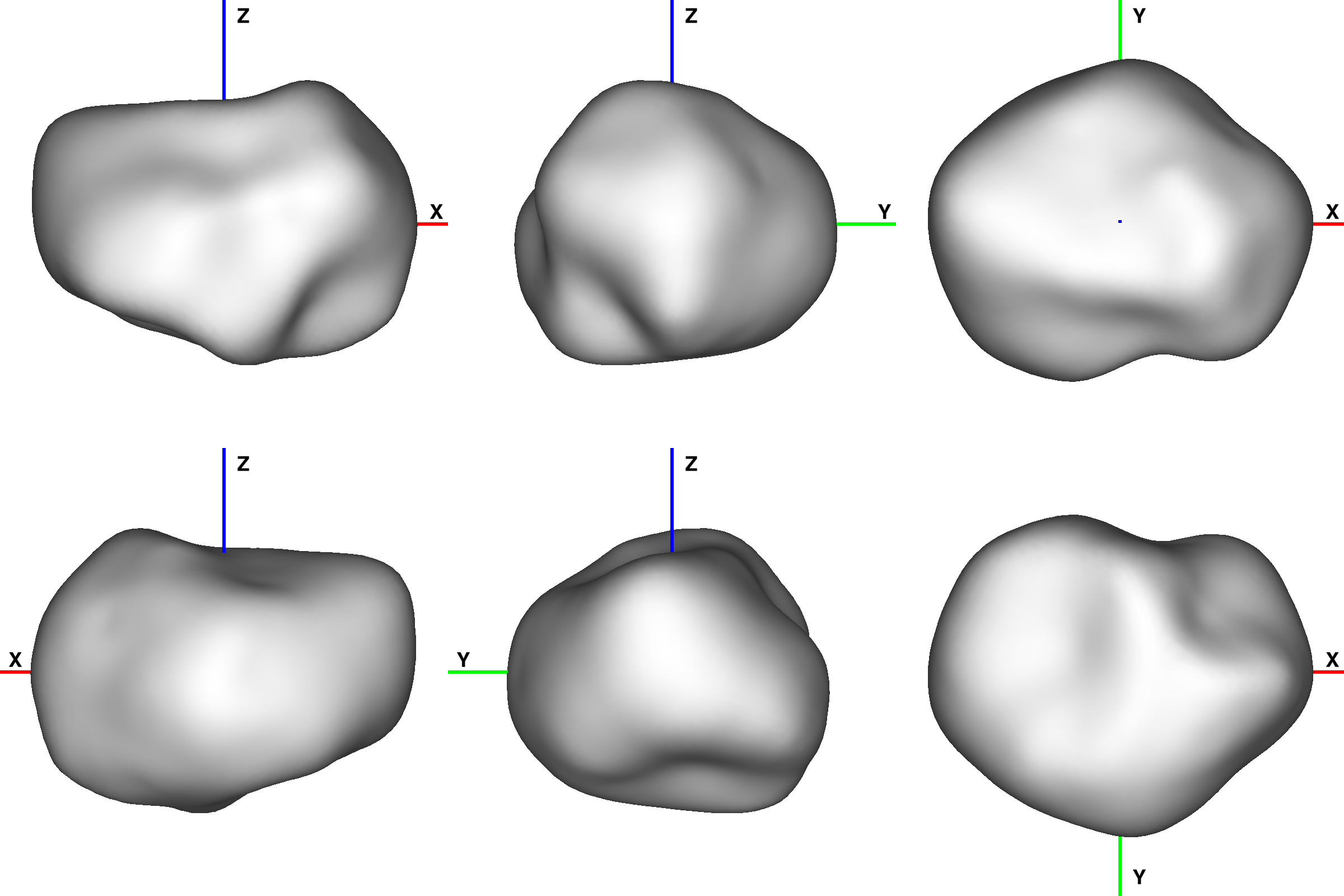}
	\caption{Projections of (9) Metis asteroid model from SAGE.}
	\label{fig:Metis_proj}
\end{figure}

The preliminary non-convex SAGE model of Metis was first introduced by
\cite{Bartczak14poster}. The model (this work, see Fig.~\ref{fig:Metis_proj} for
model's projections and Fig.~\ref{fig:periodogram_metis} for periodogram)
successfully reproduces photometric observations of Metis (see
Fig.~\ref{fig:Metis_lc}). Fig.~\ref{fig:metis_lc_comparison} shows the lightcurve
comparison with other published models of Metis by: \cite{Hanus13},
\cite{torppa2003} and \cite{ADAM15}.

SAGE model parameters:
\begin{itemize}
	\item pole coordinates:
		\begin{itemize}
			\item[] $\lambda : 182\st \pm 4\st$
			\item[] $\beta : 20\st \pm 4\st$
		\end{itemize}
	\item rotation period: $5.079177h \pm10^{-6}h$.
	\item $R_{max}$: $110$km
\end{itemize}

\begin{figure}
\centering
\includegraphics[width=8cm]{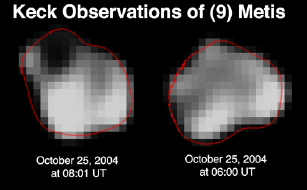}
\caption{Profile comparison of the best solution found for the (9) Metis
non-convex shape model (this work) to the Adaptive Optics observations
presented in Marchis et al. (2006) obtained with the Keck NGS AO system.}
\label{fig:metis_AO}
\end{figure}

\input{img/metis_app.tex}

\begin{table*}
	\caption{Details of the lightcurve data used for 9 Metis modelling. $\alpha$
 -- phase angle, $\lambda$ -- ecliptic longitude, $\beta$ -- ecliptic latitude.}
\label{tab:Metis_lc}
\begin{tabular}{ccccccc}
\hline
\hline
  Apparition & Year & N$_{lc}$ & $\alpha$ $[^{\circ}]$ & $\lambda$ $[^{\circ}]$ & $\beta$ $[^{\circ}]$ & reference\\
\hline
  1  & 1949 &  1  &     2.5    &  41.1  & -5.0  & \cite{groeneveld54a}\\
  2  & 1954 &  4  &  3.0 -- 9.4& 97.4  &  5.6  & \cite{groeneveld54b}\\
  3  & 1958 &  1  &     5.1    & 153.1  &  9.6  & \cite{gehrels62}\\
  4  & 1962 &  2  &  3.5--5.1    & 195.0  &  7.8  & \cite{chang62}\\
  5  & 1964 &  1  &    16.4    &  94.5  &  1.5  & \cite{yang65}\\
  6  & 1974 &  1  &     8.5    & 294.0  & -5.3  & \cite{zappala79}\\
  7  & 1978 &  3  & 4.3 -- 13.2& 320.0  & -9.0  & \cite{schober79}\\
  8  & 1979 &  3  &23.0 -- 24.2& 140.2  &  6.0  & \cite{harris89}\\
  9  & 1982/1983 & 3 & 8.5 -- 24.0 & 32.0 & -1.1 & \cite{dimartino84},\\
     & & & & & & \cite{weiden87}\\
 10  & 1984 &  8  & 3.5 -- 13.5& 178.0 & 8.9 & \cite{zeigler85},\\
     & & & & & & \cite{dimartino87},\\
     & & & & & & \cite{weiden87}\\
 11  & 1985 &  2  & 4.4 -- 5.2 & 286.0  & -4.8  & \cite{weiden87}\\
 12  & 1986 &  3  & 2.2 -- 3.8 &  67.0 & -0.8 & \cite{melillo87}\\
 13  & 1988 &  2  & 2.3 -- 2.8 & 214.0  &  5.0  & \cite{weiden90}\\
\hline
\end{tabular}
\end{table*}

\subsubsection{Stellar occultations}

Metis had multiple stellar occultations events, but only 2008 and 2014 events
were usable for shape fitting. These direct shape measurements can be used for
both model validation and scaling. We matched our Metis model silhouettes
against occultations' chords and compared with Metis models obtained using
KOALA \citep{Hanus13} (using lightcurves and adaptive optics) and ADAM
\citep{Hanus17} (using lightcurves, adaptive optics and stellar occultations)
methods and the convex model by \cite{torppa2003}.

\begin{figure*}
	\includegraphics[width=11cm]{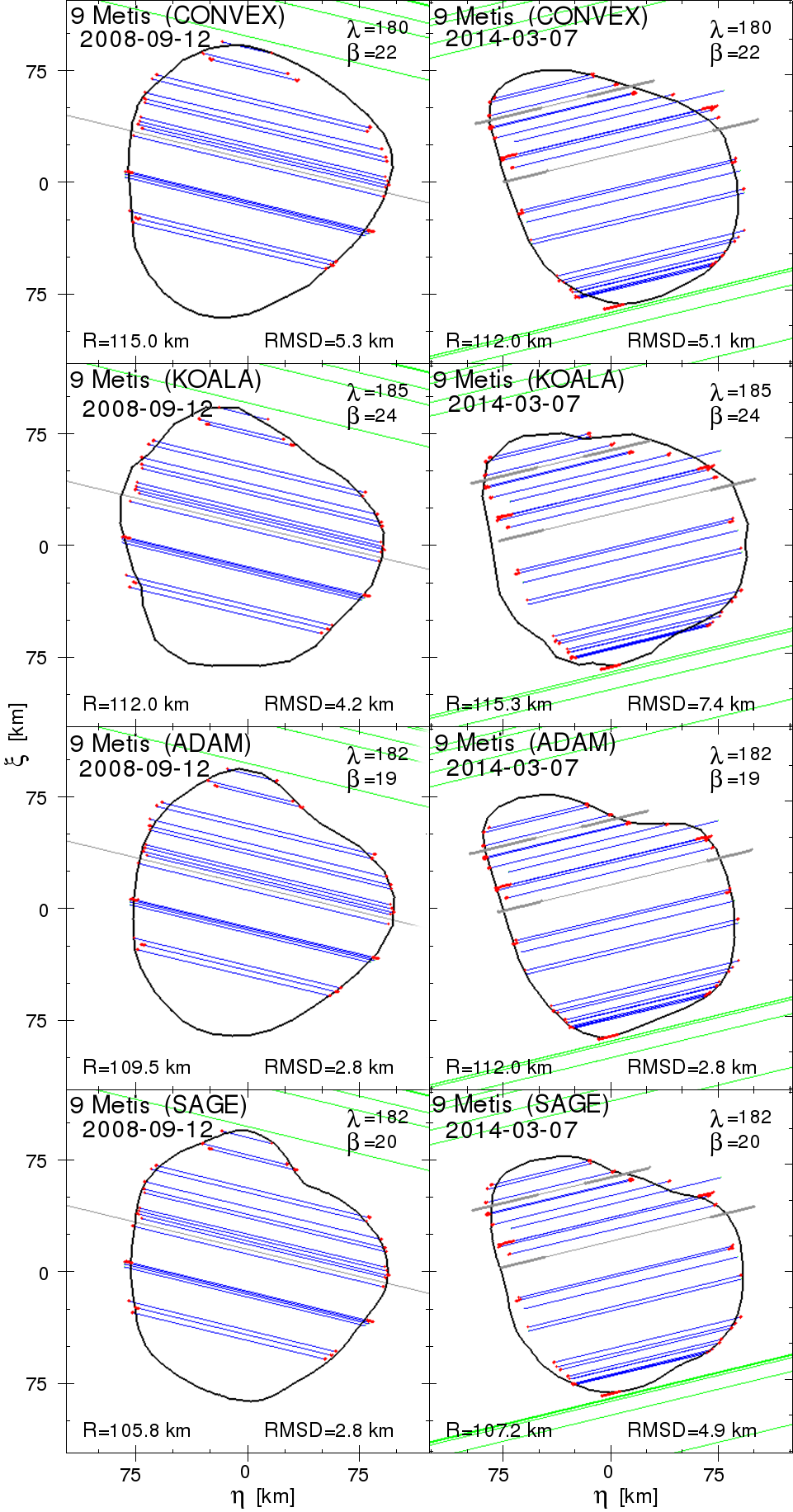}
	\caption{Metis models' silhouettes matched with 2008 and 2014 stellar
		occultations. $\chi^2$ value is calculated by summing the distances form
		the ends of the chords to the point on the silhouette along the chord
		direction. The $R$ value is the size of the model, i.e. the length of
		the longest vector in the model, based on the fit. The red colour at the
		end of the chords mark the error of the position based on the timing
		uncertainty. From the top: convex model \citep{torppa2003}, KOALA
		\citep{Hanus13}, ADAM \citep{ADAM15}, SAGE (this work).}
	\label{fig:metis_so}
\end{figure*}

To find the best fit we created the model's silhouette for a given date and
matched it against chords produced by stellar occultations' timings. The
silhouette was moved in $x$ and $y$ axes (i.e. on Earth's surface) and scaled to
provide the best fit. Figure~\ref{fig:metis_so} shows the best fit of Metis
models form various techniques.

Tab.~\ref{tab:metis_densities} shows equivalent volume sphere diameters and
densities from publications on available Metis models. Volumes and maximal radii
(and density in case of convex model) were computed for the purpose of this
work from 3D shape models for further comparison. To be able to compare models
scaled with stellar occultations we used our software on all of the models.
Diameters, volumes and densities calculated from 2008 and 2014 stellar
occultation events shown in Tab.~\ref{tab:metis_densities2} are in agreement
with the ones in Tab.~\ref{tab:metis_densities} which proves robustness of our
occultation fitting software.

We treated occultation chords from 2008 and 2014 separately to see if the models
explain both sets of observations equally well. Vertex meshes were used to find
volumes and subsequently calculate equivalent sphere diameters. Densities were
determined using Metis' mass from \cite{Carry12b} with density uncertainty
$\delta\rho$ given by
\begin{equation}
	\delta\rho = \rho\sqrt{ \left( \frac{\delta M}{M} \right)^2
								+ \left( \frac{\delta V}{V} \right)^2 }
\end{equation}
where $M$ and $V$ are mass and volume, whereas $\delta M$ and $\delta V$ are
mass and volume uncertainties.  Scalings of SAGE Metis model from separate
occultation events are consistent and in agreement with values derived for
other Metis models.

\input{img/metis/metis_tab.tex}

\subsubsection{Adaptive Optics}

\begin{figure*}
	\includegraphics[height=22.5cm]{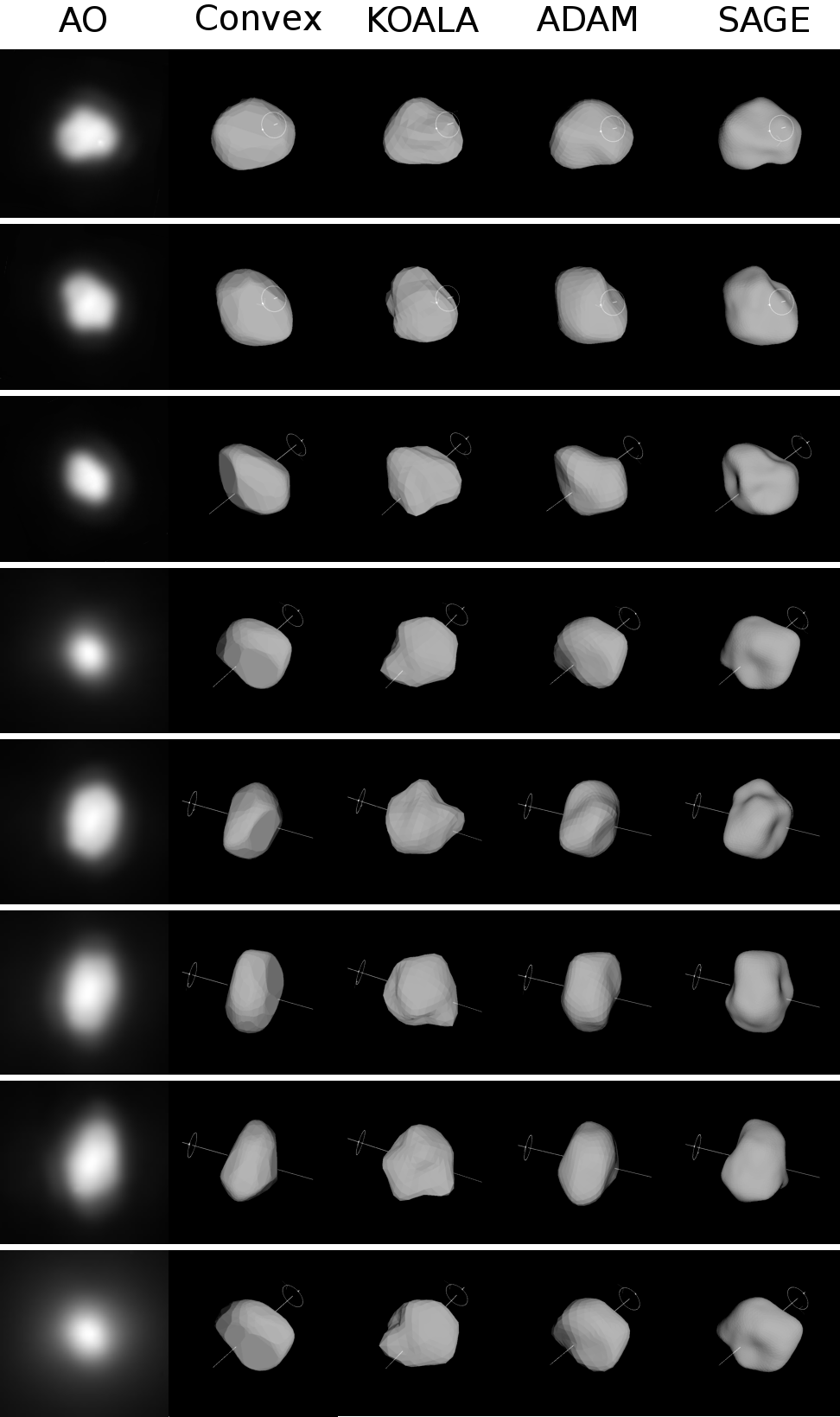}
	\caption{Comparison between various Metis models (Convex \citep{torppa2003},
	KOALA \citep{Hanus13}, ADAM \citep{ADAM15}, SAGE--this work) and Adaptive
	Optics observations \citep{Hanus17}.
	From the top, the dates and UT times of the observations are:
	2004-10-25 05:57:31; 2004-10-25
07:57:22; 2003-06-05 10:57:09; 2003-07-14 06:29:07; 2012-12-29 12:09:55;
2012-12-29 13:34:42; 2012-12-29 14:24:52; 2003-07-14 06:42:46.}
	\label{fig:metis_ao_comparison}
\end{figure*}

We have combined the AO images from \cite{Hanus17} with Metis models' sky
projections for observation dates (Fig.~\ref{fig:metis_ao_comparison}).
Visual inspection indicates SAGE model matches AO observations. There is also a
strong resemblance between SAGE and ADAM models.

\section{Conclusions}

We have developed a new modelling method -- called SAGE -- based on  photometric
lightcurves, reconstructing non-convex shapes, spin axis orientations and
rotation periods of asteroids. The method is based on a genetic algorithm that
converges to a stable solution over many iterations of random shape and spin
axis mutations.

Being computationally expensive, SAGE is run on a computer cluster of multiple
nodes equipped with graphics cards that are performing calculations when
parallelization is possible, e.g. models' lightcurve computation or search for
a rotational period.

To evaluate method's capabilities we have performed numerical tests during which
SAGE attempted to deliver a model that best fit given lightcurves.
SAGE recreated test models accurately when provided with favorable geometries:
the $\beta=45\st$ and evenly distributed apparitions with $0\st-16\st$ phase
angles producing non-flat lightcurves containing information about the whole
body. In other cases, as tests on model A indicated, the differences between
a model and a resulting model increased; the worst fit was with
$0\st$-phase-angle-only data. The general shape was recreated, but concavities
on the shape were shallow. In each case the biggest difference between test and
modelled shapes was around the south and north poles due to $z$-axis scale
uncertainty. We have tested different kinds of shapes, from random gaussian
shapes to contact binary-like body. SAGE's shape representation (fixed vectors)
was fit to describe each of them.

We picked (433) Eros and (9) Metis asteroids to test SAGE on real
observational data. Our choice was determined by the availability of the data
(detailed shape model from NEAR Shoemaker probe for Eros and stellar
occultations and adaptive optics observations for Metis) that
we were able to compare to, as well as plethora of photometric data.

The (433) Eros model reproduces general shape and major features of the
asteroid (Fig.~\ref{fig:Eros_projections}). The maximum deviation from NEAR
shape (Fig. \ref{fig:eros_topography}) is in order of $0.1$ of the $R_{max}$
with total RMSD=$0.025959$. The rotation axis orientation $\lambda=17\st\pm
5\st$, $\beta=8\st\pm 5\st$ and period $P=5.270256h \pm10^{-6}h$ are in
agreement with the values found by \cite{Miller02}.

The (9) Metis model reproduces lightcurves very well. The pole
solution found by SAGE is $\lambda=182\st\pm4\st$, $\beta=20\st\pm4\st$ with the
rotation period $P=5.079177h \pm10^{-6}h$.  By comparing the model with 2008 and
2014 stellar occultations we scaled it and obtained $R_{max}^{2008}=106\pm3km$,
$R_{max}^{2014}=107\pm5km$ which yield equivalent volume sphere diameter of
$D^{2008}=165\pm5km$, $D^{2014}=167\pm8km$; assuming the mass $M=8.39\pm1.67
\cdot 10^{18} kg$ the density is $\rho^{2008}=3.54\pm0.76\frac{g}{cm^3}$,
$\rho^{2014}=3.44\pm0.84\frac{g}{cm^3}$. As seen in
Tab.~\ref{tab:metis_densities2} these values do not deviate significantly from
the ones calculated for ADAM model, which was based on lightcurves, stellar
occultations and adaptive optics. A comparison with adaptive optics
observations (Fig.~\ref{fig:metis_ao_comparison}) also validates SAGE model.

Tests and asteroids' models described in this work demonstrate SAGE's ability to
model asteroids' physical parameters and create their non-convex shapes without
making any prior assumptions, except for uniform albedo, homogeneous mass
distribution and principal axis rotation. SAGE's software design allows to
extend the algorithm to include other types of asteroid observation techniques,
e.g. stellar occultations and adaptive optics. A merge with radar delay-Doppler
observations was already tested in \cite{Dudzinski16}. This additional data will
definitely help place more constrains during the modelling process and
produce models of better quality.

\section*{Acknowledgements}

The research leading to these results has received funding from the
  European Union's Horizon 2020 Research and Innovation Programme, under
  Grant Agreement no 687378.

This work was partialy supported by grant no. 2014/13/D/ST9/01818 from the
National Science Centre, Poland.



\bibliography{literatura}
\addcontentsline{toc}{chapter}{\bibname}
\bibliographystyle{mnras}


%% file: img/eros_app.tex
\begin{figure}
	\center
		\begin{tikzpicture}
			\draw (0,0) circle (3);

			\node(1) at (0,0) {$\Huge{\lambda}$};
			\draw (0,0) circle (0.2);

			\node(1) at (90:2.6) {$0^{\circ}$};
			\draw[color=blue] (90:2.7) -- (90:3);

			\draw [red, line width=5pt] plot [smooth, tension=1.2] coordinates {(5+90:3) (61.5+90:3) (118+90:3)};
			\node(1) at (27+90:2.2) {($5^{\circ}$ -- $119^{\circ}$)};
			\node(1) at (44+90:3.6) {1951/1952};
			\draw[color=red, line width=2pt] (30+90:2.5) -- (30+90:3);

			\draw[fill=red] (341+90:3) circle (0.2);
			\node(1) at (341+90:2.5) {$341^{\circ}$};
			\node(1) at (341+90:3.4) {1972};

			\draw [blue, line width=5pt] plot [smooth, tension=1.2] coordinates
			{(52.5+90:2.83) (105.25+90:2.83) (158+90:2.83)};
			\node(1) at (145+90:2.0) {($53^{\circ}$ -- $158^{\circ}$)};
			\node(1) at (133+90:3.6) {1974/1975};
			\draw[color=blue, line width=2pt] (140+90:2.4) -- (140+90:2.9);
			\draw [magenta, line width=5pt] plot [smooth, tension=1.2]
			coordinates {(42.4+90:2.66) (84+90:2.66) (125.6+90:2.66)};
			\node(1) at (84+90:1.2) {($42^{\circ}$ -- $126^{\circ}$)};
			\node(1) at (84+90:3.8) {1981/1982};
			\draw[color=magenta, line width=2pt] (84+90:2.66) -- (84+90:2.2);
			\draw [red, line width=5pt] plot [smooth, tension=1.2] coordinates {(294.1+90:3) (302.2+90:3)  (308.3+90:3) };
			\node(1) at (305+90:1.8) {($296^{\circ}$ -- $308^{\circ}$)};
			\node(1) at (300+90:3.5) {1993};
			\draw[color=red, line width=2pt] (302+90:2.5) -- (302+90:3);

		\end{tikzpicture}
	\caption{The distribution of Eros apparitions. $\lambda$ denotes J2000
	ecliptic longitude of the asteroid.}
	\label{fig:Eros_app}
\end{figure}

%% file: img/metis_app.tex
\begin{figure}
	\center
	\begin{tikzpicture}
		\draw (0,0) circle (3);

		\node(1) at (0,0) {$\Huge{\lambda}$};
		\draw (0,0) circle (0.2);

		\node(1) at (90:2.5) {$0^{\circ}$};
		\draw[color=blue] (90:2.7) -- (90:3);

		\fill[fill=blue] (41+90:3) circle (0.2);
		\node(1) at (41+90:2.3) {$41^{\circ}$};
		\node(1) at (41+90:3.6) {1949};

		\draw[fill=blue] (97+90:3) circle (0.2);
		\node(1) at (99+90:2.3) {$97^{\circ}$};
		\node(1) at (99+90:3.6) {1954};

		\draw[fill=blue] (153+90:3) circle (0.2);
		\node(1) at (153+90:2.3) {$153^{\circ}$};
		\node(1) at (153+90:3.6) {1958};

		\draw[fill=blue] (195+90:3) circle (0.2);
		\node(1) at (195+90:2.3) {$153^{\circ}$};
		\node(1) at (195+90:3.6) {1962};

		\draw[fill=blue] (94+90:3) circle (0.2);
		\node(1) at (92+90:2.3) {$94^{\circ}$};
		\node(1) at (92+90:3.6) {1964};

		\draw[fill=blue] (294+90:3) circle (0.2);
		\node(1) at (294+90:2.3) {$294^{\circ}$};
		\node(1) at (294+90:3.6) {1974};

		\draw[fill=blue] (320+90:3) circle (0.2);
		\node(1) at (320+90:2.3) {$320^{\circ}$};
		\node(1) at (320+90:3.6) {1978};

		\draw[fill=blue] (140+90:3) circle (0.2);
		\node(1) at (140+90:2.3) {$140^{\circ}$};
		\node(1) at (140+90:3.6) {1979};

		\draw[fill=blue] (32+90:3) circle (0.2);
		\node(1) at (32+90:2.3) {$32^{\circ}$};
		\node(1) at (32+90:3.6) {1982};

		\draw[fill=blue] (178+90:3) circle (0.2);
		\node(1) at (178+90:2.3) {$178^{\circ}$};
		\node(1) at (178+90:3.6) {1984};

		\draw[fill=blue] (286+90:3) circle (0.2);
		\node(1) at (286+90:2.3) {$286^{\circ}$};
		\node(1) at (286+90:3.6) {1985};

		\draw[fill=blue] (67+90:3) circle (0.2);
		\node(1) at (67+90:2.3) {$67^{\circ}$};
		\node(1) at (67+90:3.6) {1986};

		\draw[fill=blue] (214+90:3) circle (0.2);
		\node(1) at (214+90:2.3) {$214^{\circ}$};
		\node(1) at (214+90:3.6) {1988};
	\end{tikzpicture}
	\caption{Distribution of apparitions for Metis.}
	\label{fig:Metis_app}
\end{figure}

%% file: img/metis/metis_tab.tex
\begin{table*}
\caption{The length of the longest vector $R_{max}$, equivalent volume sphere
diameter $D$, volume $V$ and density $\rho$ of (9) Metis published
models from different modelling methods. The mass $M=8.39\pm1.67
[10^{18} kg]$ for density calculation was taken from
\protect\cite{Carry12b}.
}
\label{tab:metis_densities}
\begin{tabular}{|l|c l c l|}
\hline
     & $R_{max}$ &  D & V & $\rho$ \\
model & [km] &  [km] & [$10^6 km^3$] & [$gcm^{-3}$] \\
\hline
CONVEX & 114.5 & $169\pm20$ \citep{Durech11} & $2.52\pm0.89$& $3.33\pm 1.35$
\\
KOALA  & 102.0 & $153\pm11$ \citep{Hanus13} & $1.87\pm0.40$& $4.47\pm 1.07$
\citep{Hanus13} \\
ADAM   & 111.6 & $168\pm3 $ \citep{Hanus17}& $2.46\pm0.13$& $3.4 \pm 0.7$
\citep{Hanus17} \\ \hline
\end{tabular}
\end{table*}

\begin{table*}
\caption{(9) Metis models sizes $R_{max}$, equivalent volume sphere
diameter $D$, volume $V$ and density $\rho$ calculated based on 2008 and
2014 stellar occultations. Mass $M=8.39\pm1.67 [10^{18} kg]$
\protect\citep{Carry12b}.}
\label{tab:metis_densities2}
\begin{tabular}{|l|c c c c|c c c c|}
\hline

&\multicolumn{4}{c}{2008}&\multicolumn{4}{c}{2014}\\
\hline
     & $R_{max}$ &  D & V & $\rho$ &$R_{max}$ &  D & V & $\rho$ \\
model & [km] &  [km] & [$10^6 km^3$] & [$gcm^{-3}$] & [$km$] & [$km$] &
[$km^3$] &  [$gcm^{-3}$] \\
\hline
CONVEX & $115\pm5$ & $170\pm 7$ & $2.56\pm0.33$ & $3.27\pm 0.77$ & $112\pm5$   &
$165\pm7$ & $2.36\pm0.31$ & $3.54\pm 0.85$ \\
KOALA  & $112\pm4$ & $168\pm6$  & $2.48\pm0.26$ & $3.38\pm 0.76$ & $115\pm7$  &
$172\pm10$ & $2.68\pm0.49$ & $3.12\pm 0.84$ \\
ADAM   & $110\pm3$ & $165\pm4$   & $2.36\pm0.19$ & $3.55\pm0.76$   &
$112\pm4$   & $168\pm6$  & $2.49\pm0.26$ & $3.36\pm 0.76$ \\
SAGE   & $106\pm3$ & $165\pm 5$  & $2.36\pm0.20$ & $3.54\pm0.76$  & $107\pm5 $  &
$167\pm8$ & $2.43\pm0.34$ & $3.44\pm 0.84$ \\
\hline
\end{tabular}
\end{table*}

%% file: appendices.tex
%
\setcounter{page}{1}

\section{Topography and pole solution maps with periodograms for model A}

\begin{figure*}
	\begin{center}
		\includegraphics[width=17cm]{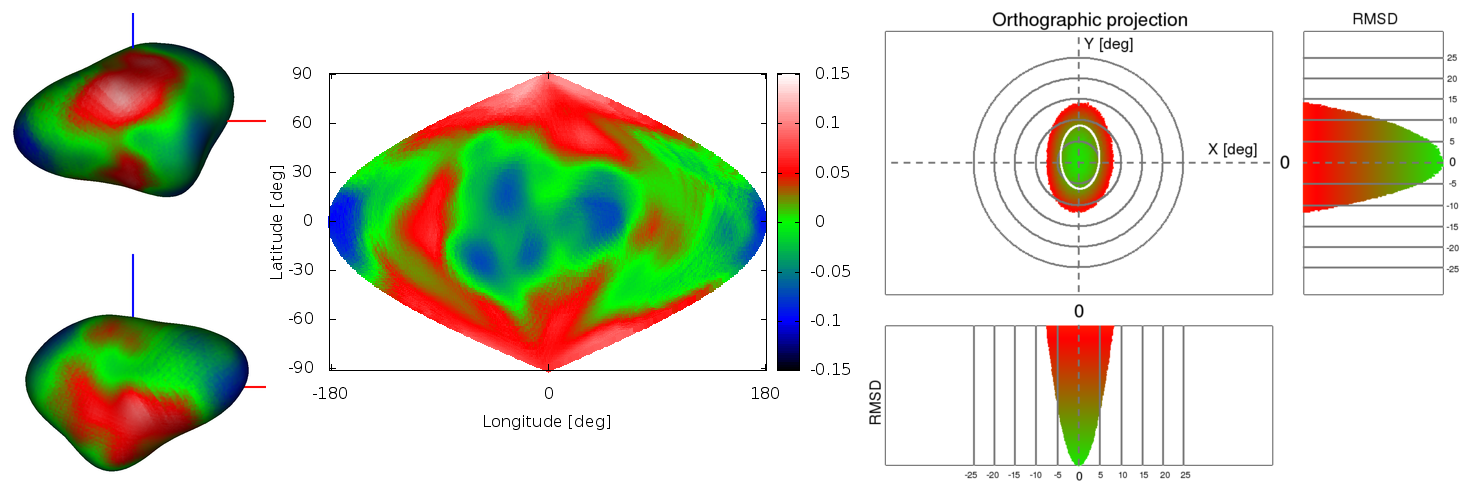}
		\includegraphics[width=8cm]{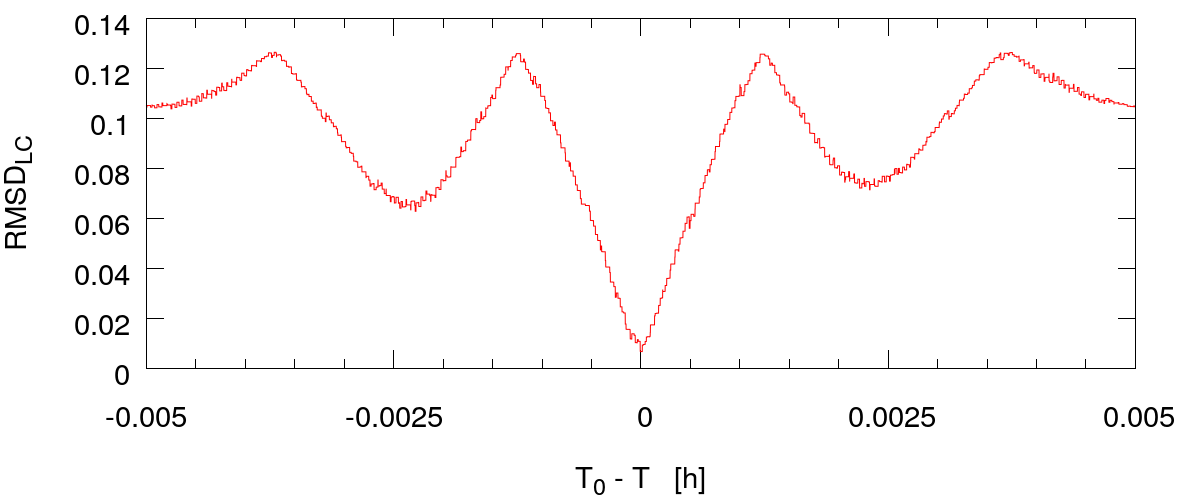}
	\end{center}
	\caption{Topography and pole solution maps with periodogram for model A, $\beta=0\st$, phase
	angles $0\st$, $14\st$, $16\st$, all apparitions. }
	\label{fig:result_A_all_all1}
\end{figure*}

\begin{figure*}
	\begin{center}
		\includegraphics[width=17cm]{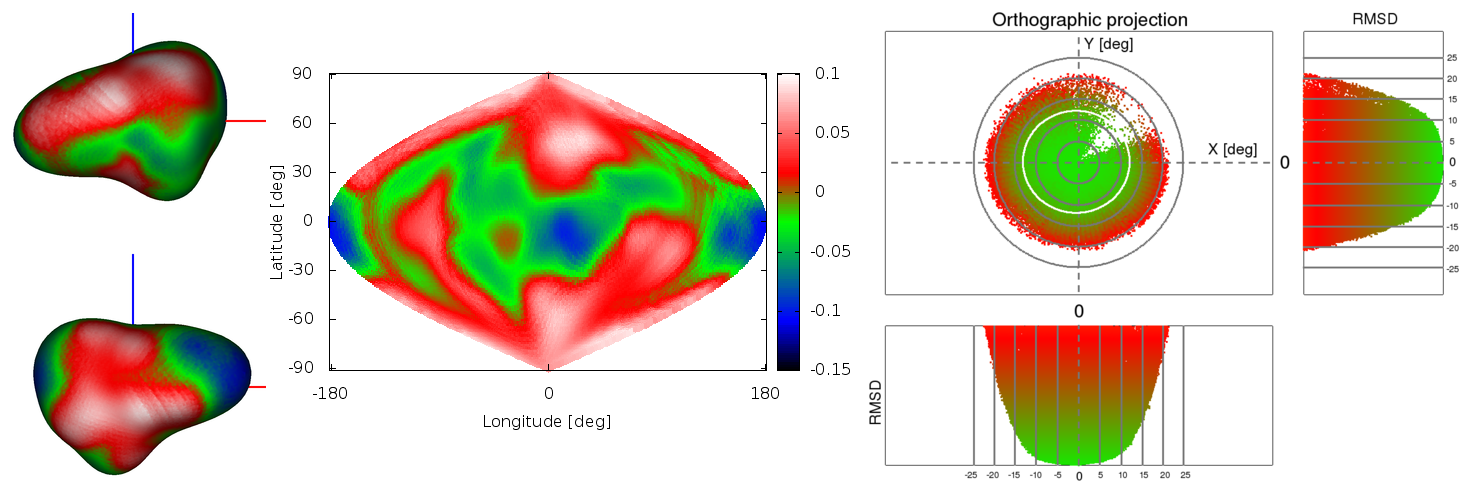}
		\includegraphics[width=8cm]{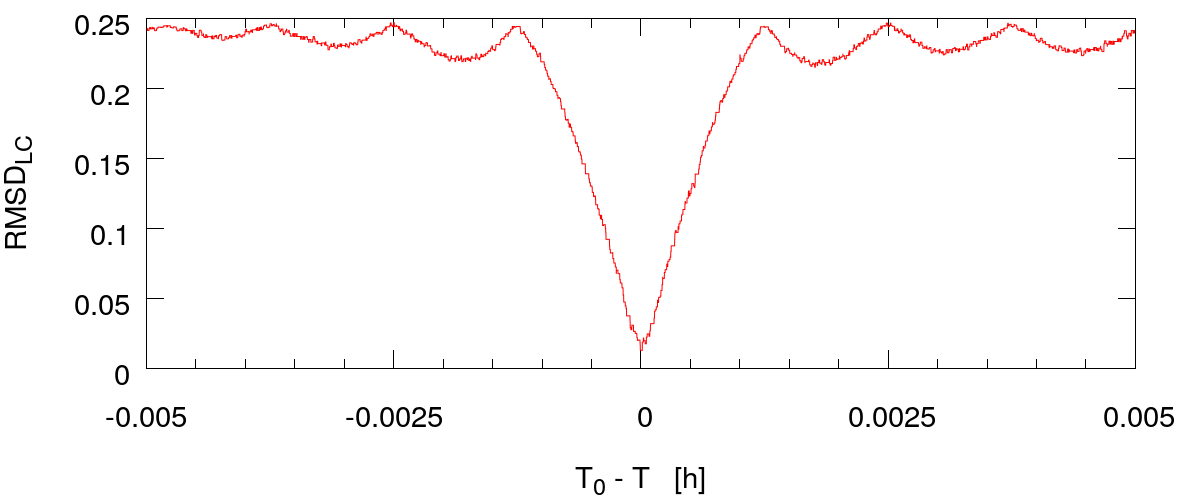}
	\end{center}
	\caption{Topography and pole solution maps with periodogram for model A, $\beta=90\st$, phase
	angles $0\st$, $14\st$, $16\st$, all apparitions. }
	\label{fig:result_A_all_all2}
\end{figure*}

\begin{figure*}
	\begin{center}
		\includegraphics[width=17cm]{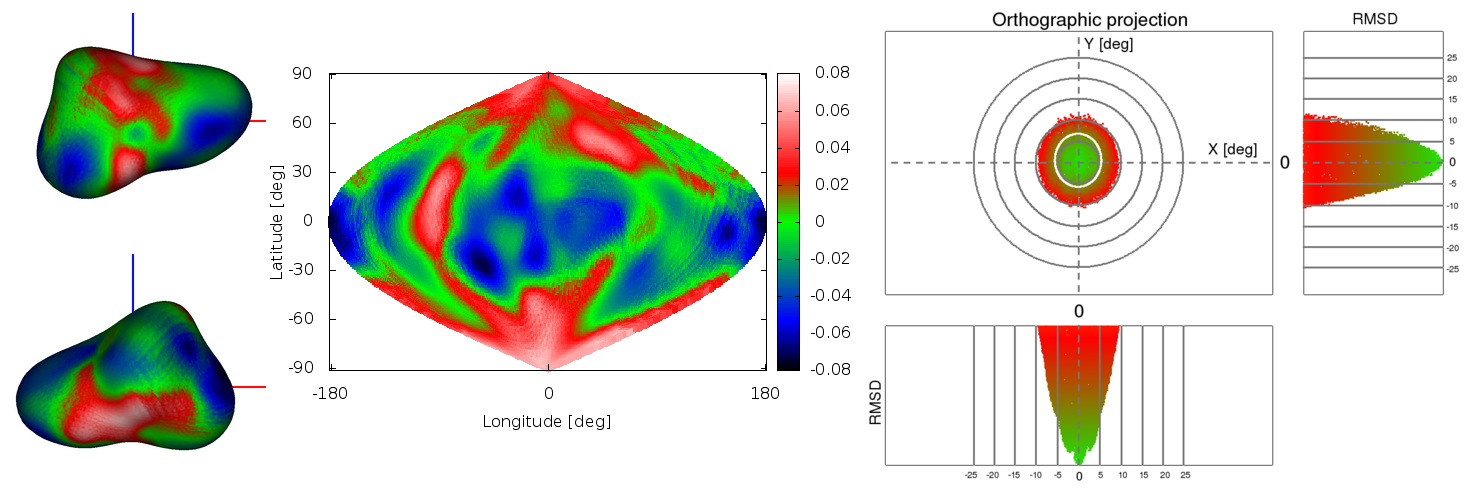}
		\includegraphics[width=8cm]{img/periodogram/periodogram_3.png}
	\end{center}
	\caption{Topography and pole solution maps with periodogram for model A, $\beta=45\st$, phase
	angles $0\st$, $14\st$, $16\st$, all apparitions. }
	\label{fig:result_A_all_all3}
\end{figure*}

\begin{figure*}
	\begin{center}
		\includegraphics[width=17cm]{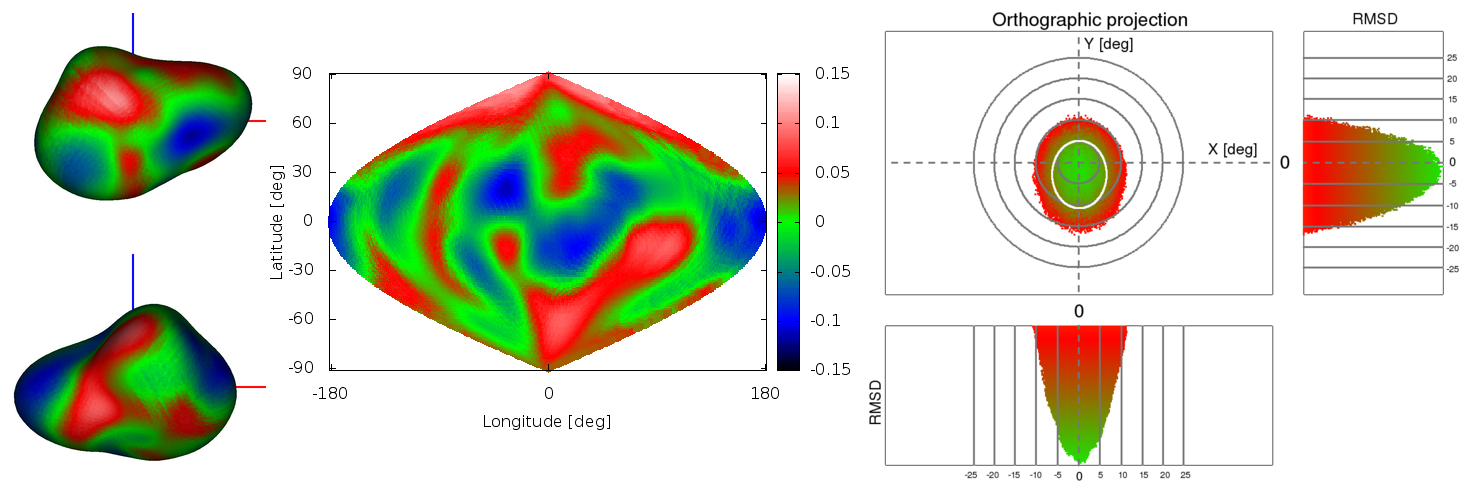}
		\includegraphics[width=8cm]{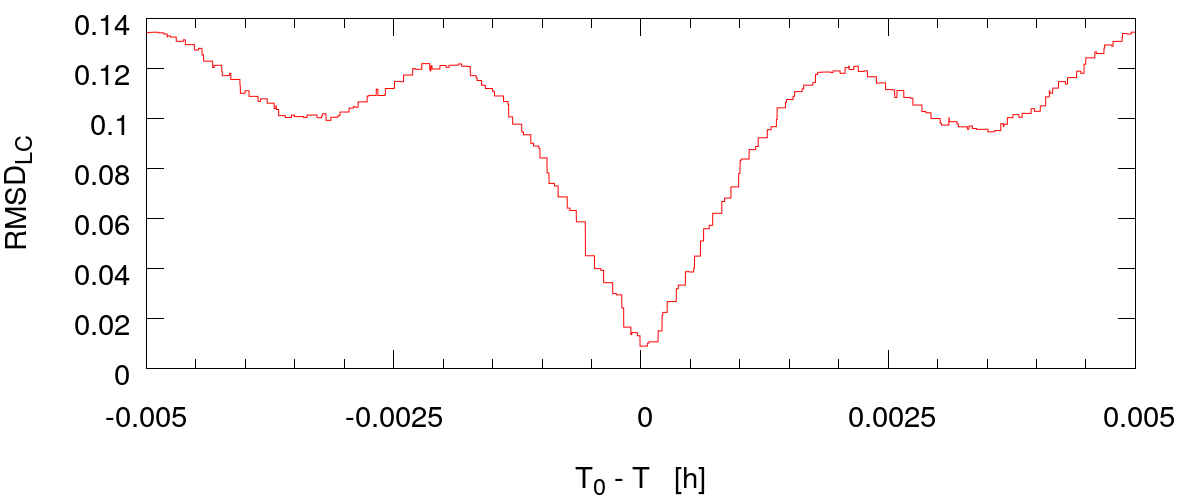}
	\end{center}
	\caption{Topography and pole solution maps with periodogram for model A, $\beta=0\st$, phase
	angles $0\st$, $14\st$, $16\st$, 1234 apparitions. }
	\label{fig:result_A_all_all4}
\end{figure*}

\begin{figure*}
	\begin{center}
		\includegraphics[width=17cm]{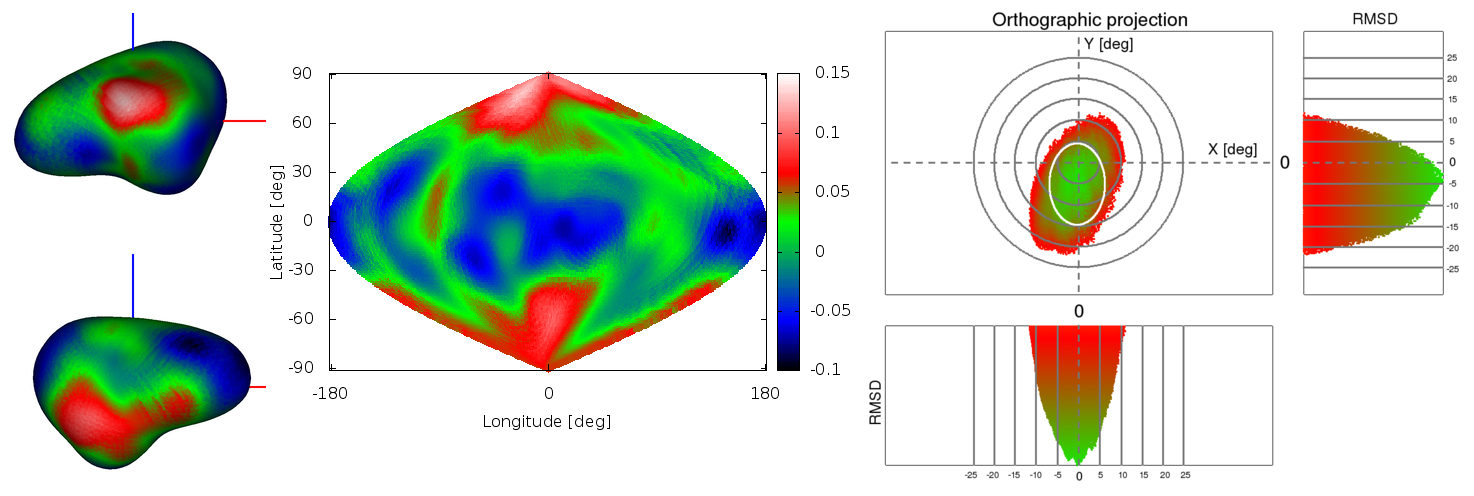}
		\includegraphics[width=8cm]{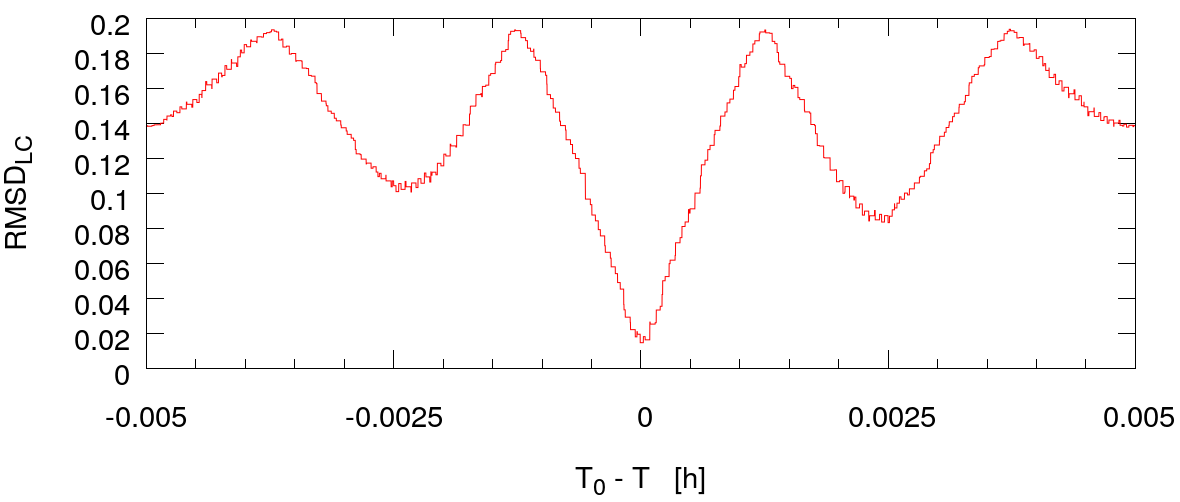}
	\end{center}
	\caption{Topography and pole solution maps with periodogram for model A, $\beta=45\st$, phase
	angles $0\st$, $14\st$, $16\st$, 1256 apparitions. }
	\label{fig:result_A_all_all5}
\end{figure*}

\begin{figure*}
	\begin{center}
		\includegraphics[width=17cm]{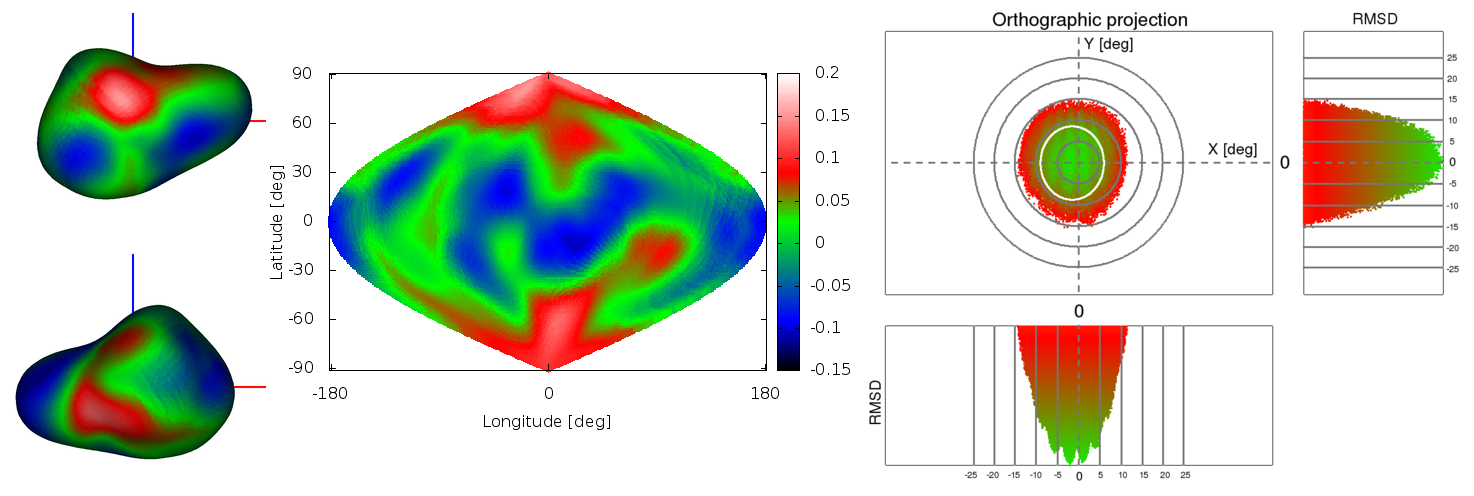}
		\includegraphics[width=8cm]{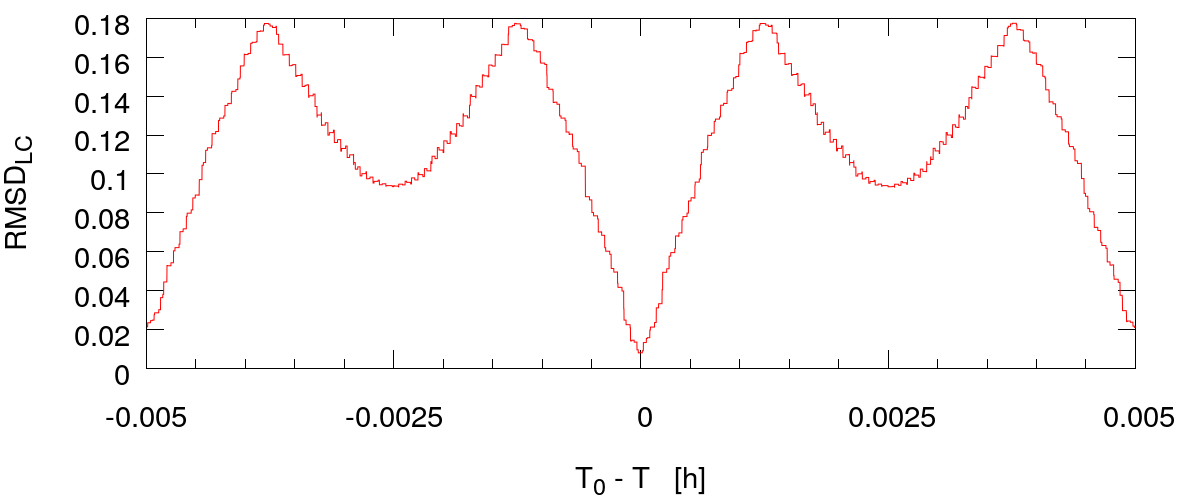}
	\end{center}
	\caption{Topography and pole solution maps with periodogram for model A, $\beta=45\st$, phase
	angles $0\st$, $14\st$, $16\st$, 1357 apparitions. }
	\label{fig:result_A_all_all5}
\end{figure*}

\begin{figure*}
	\begin{center}
		\includegraphics[width=17cm]{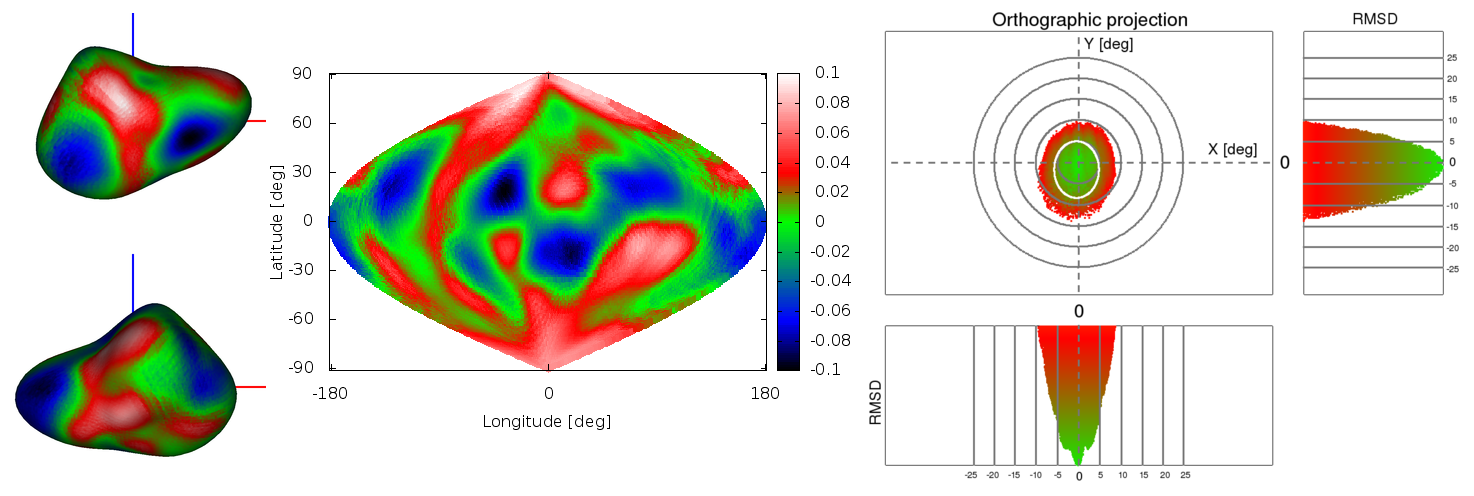}
		\includegraphics[width=8cm]{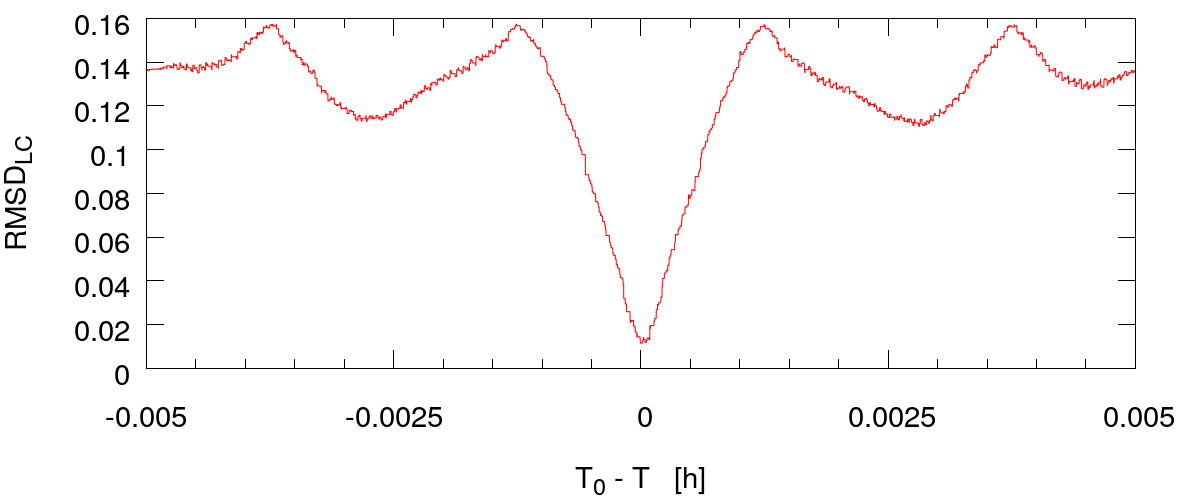}
	\end{center}
	\caption{Topography and pole solution maps with periodogram for model A, $\beta=45\st$, phase
	angles $0\st$, $14\st$, all apparitions. }
	\label{fig:result_A_all_all5}
\end{figure*}

\begin{figure*}
	\begin{center}
		\includegraphics[width=17cm]{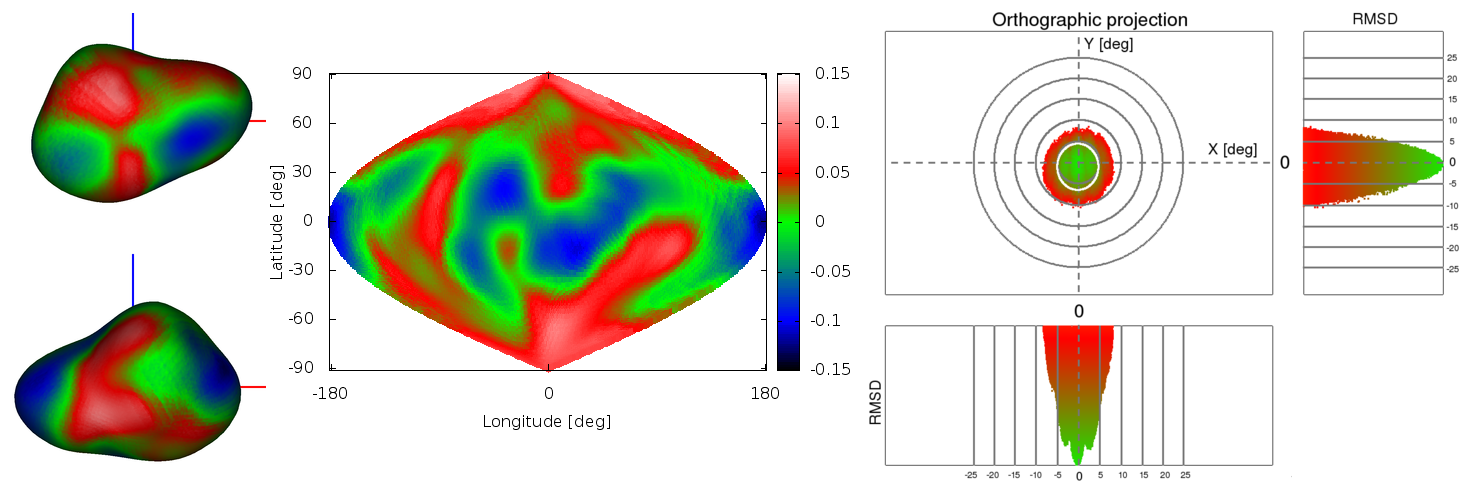}
		\includegraphics[width=8cm]{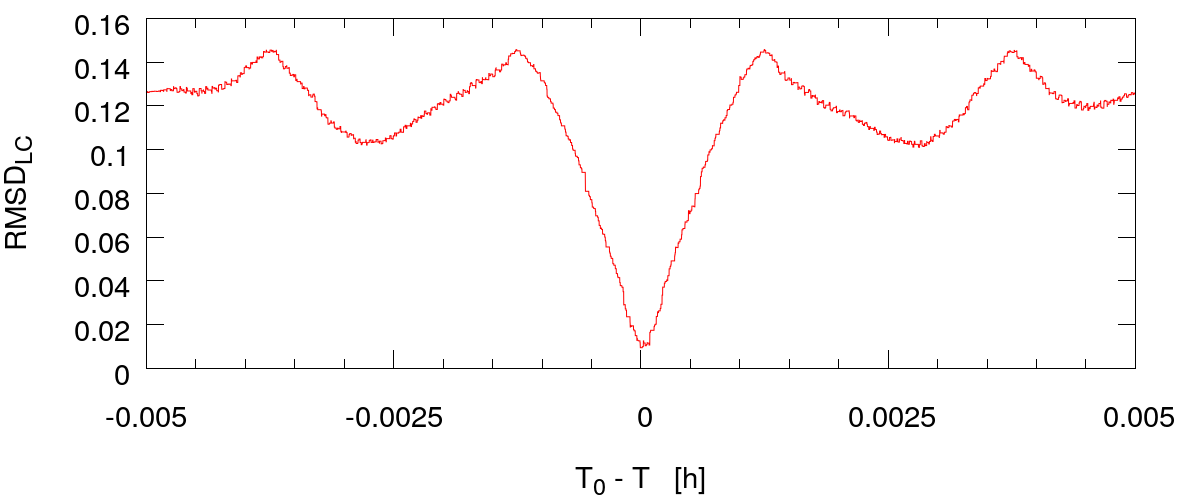}
	\end{center}
	\caption{Topography and pole solution maps with periodogram for model A, $\beta=45\st$, phase
	angles $0\st$, all apparitions. }
	\label{fig:result_A_all_all5}
\end{figure*}

\section{Results for models B, C and D}

\begin{figure*}
	\includegraphics[width=17cm]{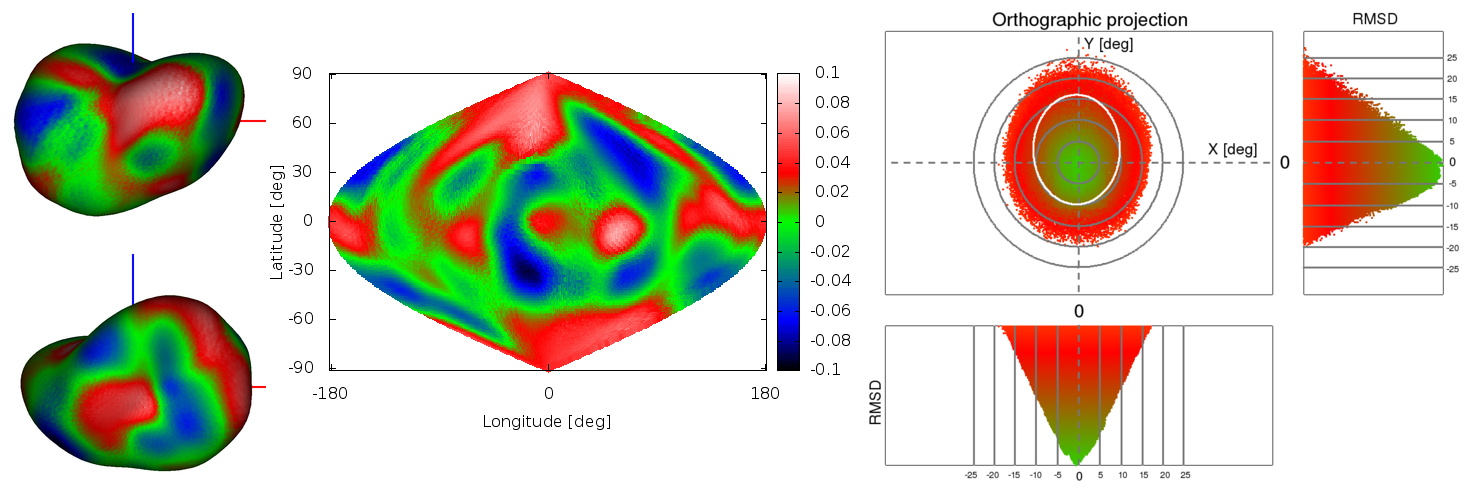}
		\includegraphics[width=8cm]{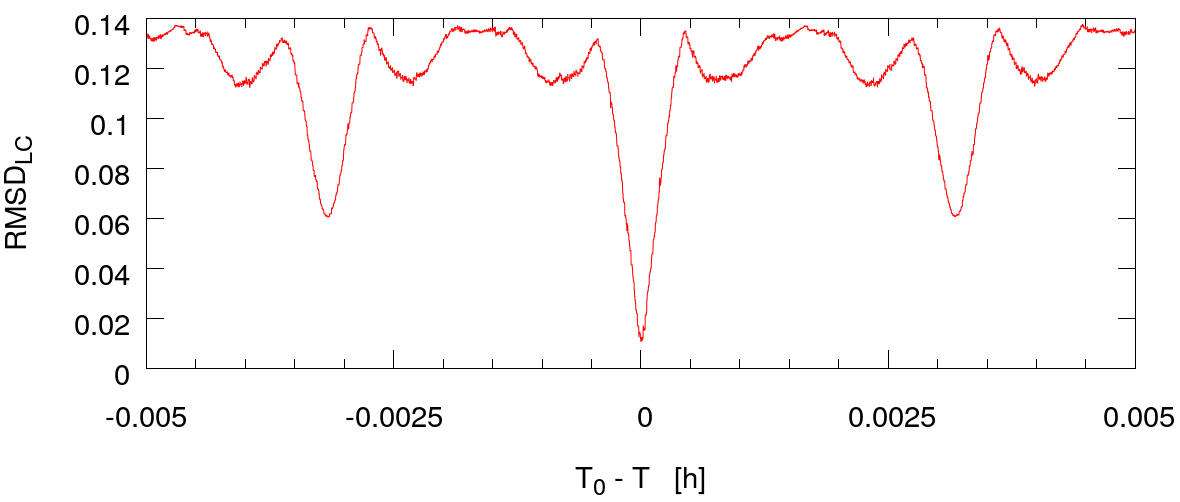}
	\caption{Topography and pole solution maps with periodogram for model B. }
	\label{fig:model_B_spin}
\end{figure*}
\begin{figure*}
	\includegraphics[width=17cm]{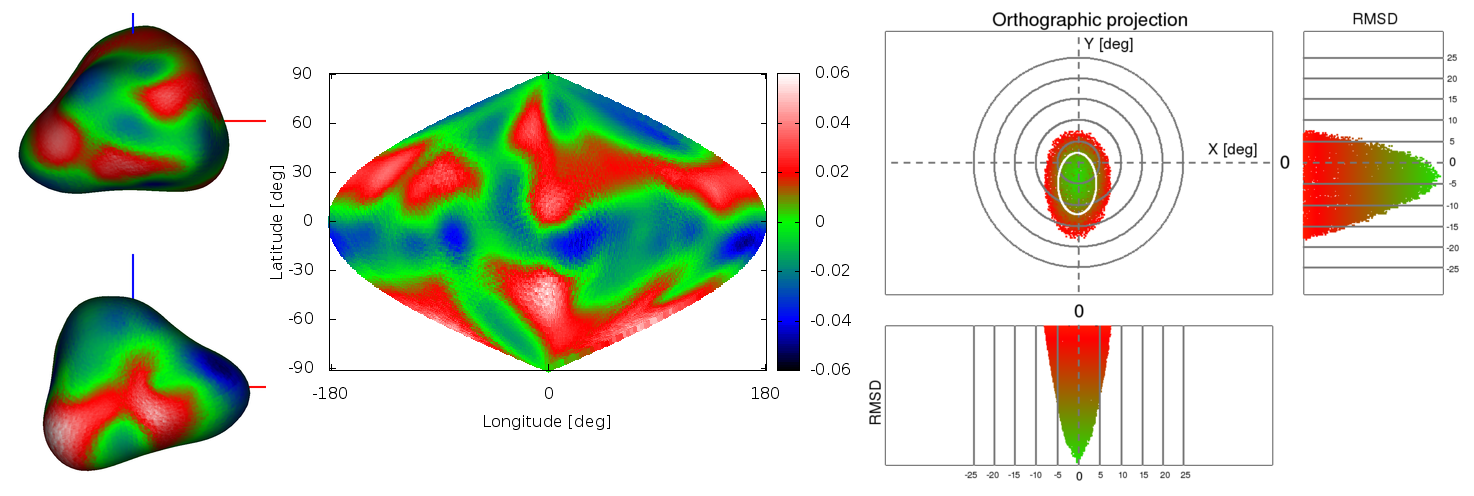}
		\includegraphics[width=8cm]{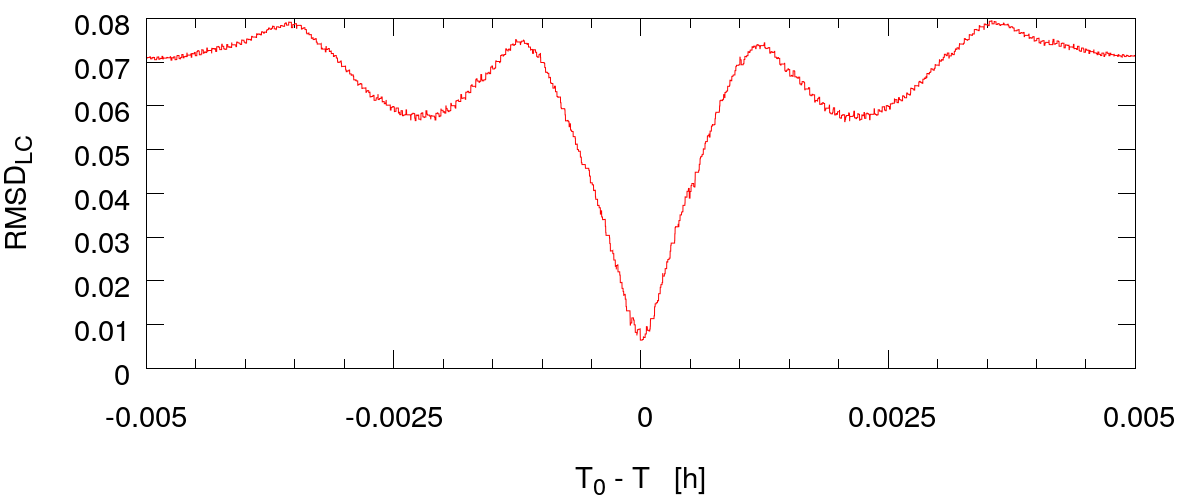}
	\caption{Topography and pole solution maps with periodogram for model C. }
	\label{fig:model_C_spin}
\end{figure*}
\begin{figure*}
	\includegraphics[width=17cm]{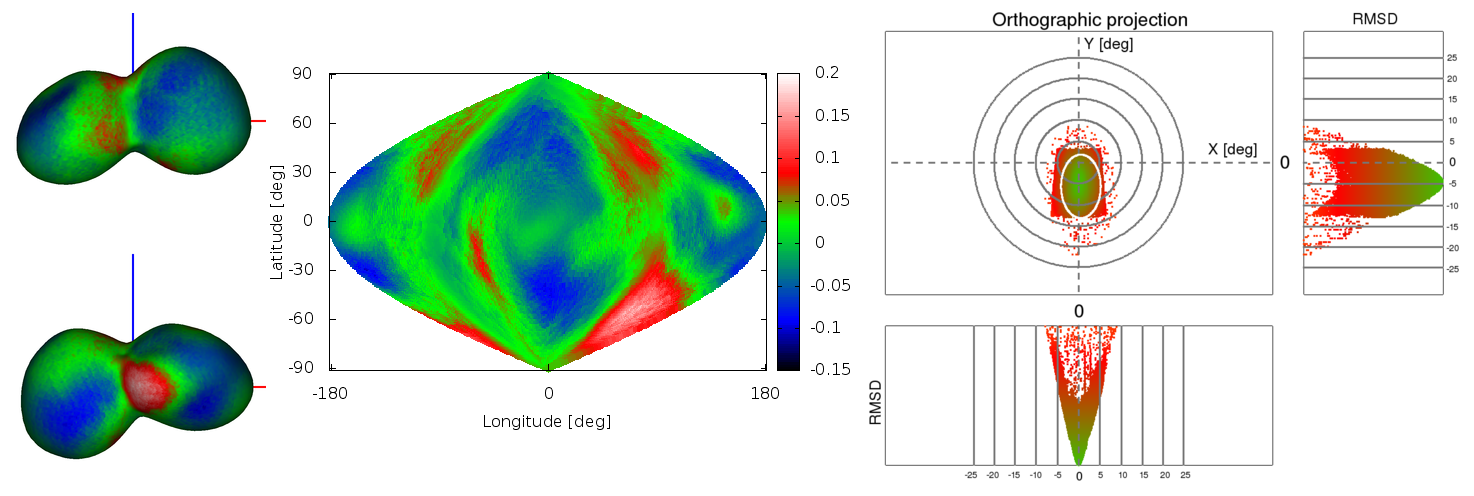}
		\includegraphics[width=8cm]{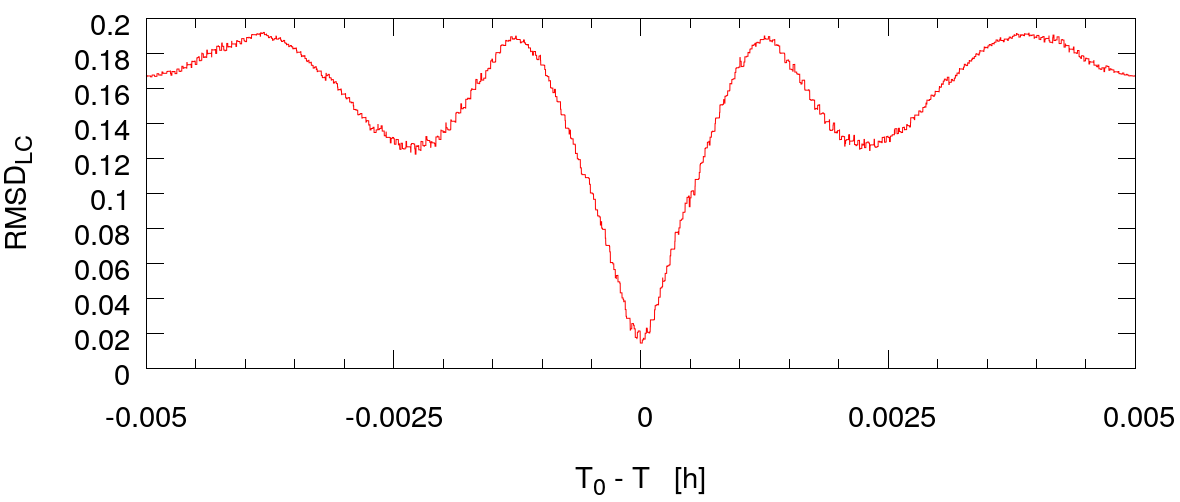}
	\caption{Topography and pole solution maps with periodogram for model D. }
	\label{fig:model_D_spin}
\end{figure*}

\section{Test models' lightcurves}

\begin{figure*}
	\includegraphics[width=17cm]{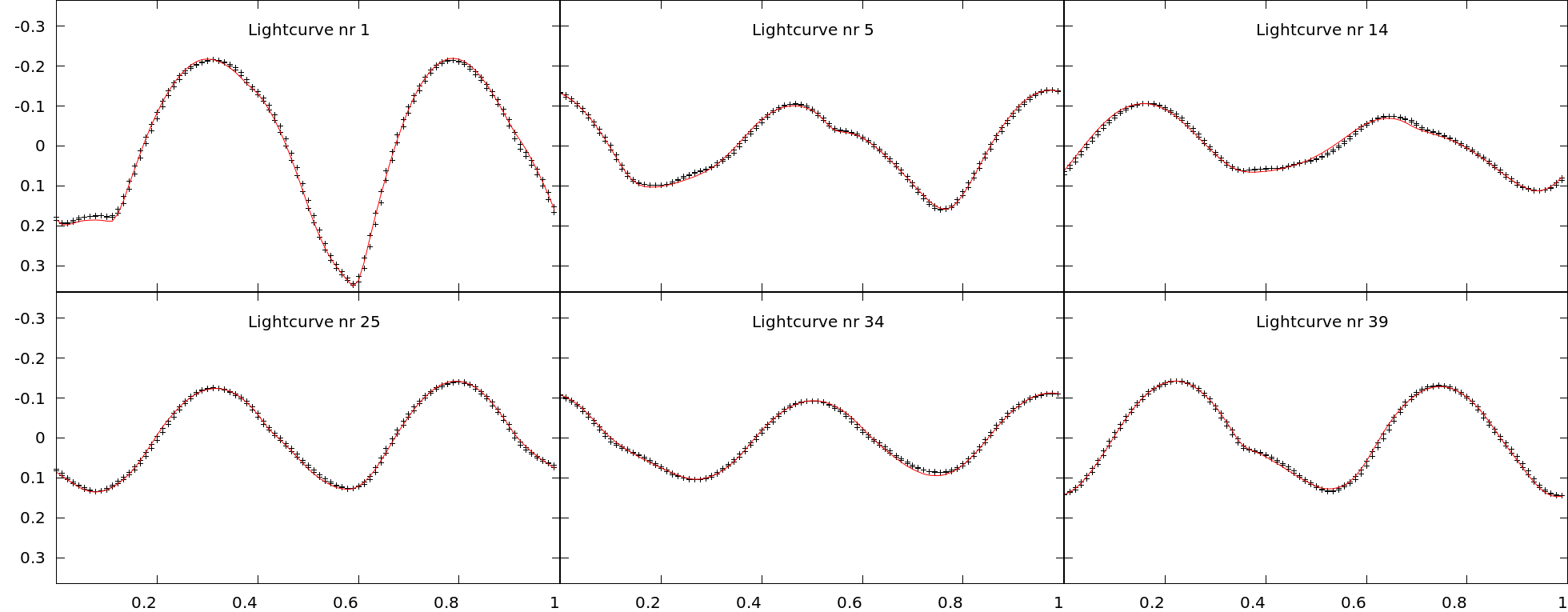}
	\caption{Some of the model A lightcurves.}
\end{figure*}

\begin{figure*}
	\includegraphics[width=17cm]{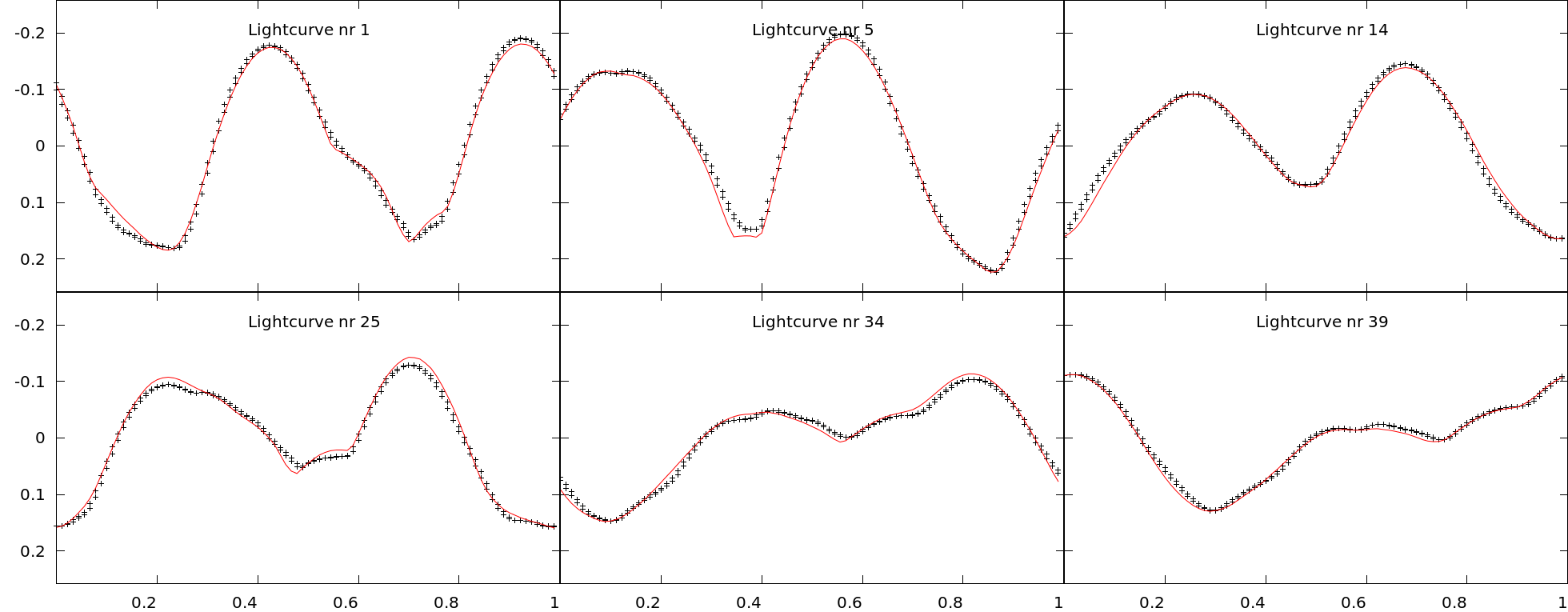}
	\caption{Some of the model B lightcurves.}
\end{figure*}

\begin{figure*}
	\includegraphics[width=17cm]{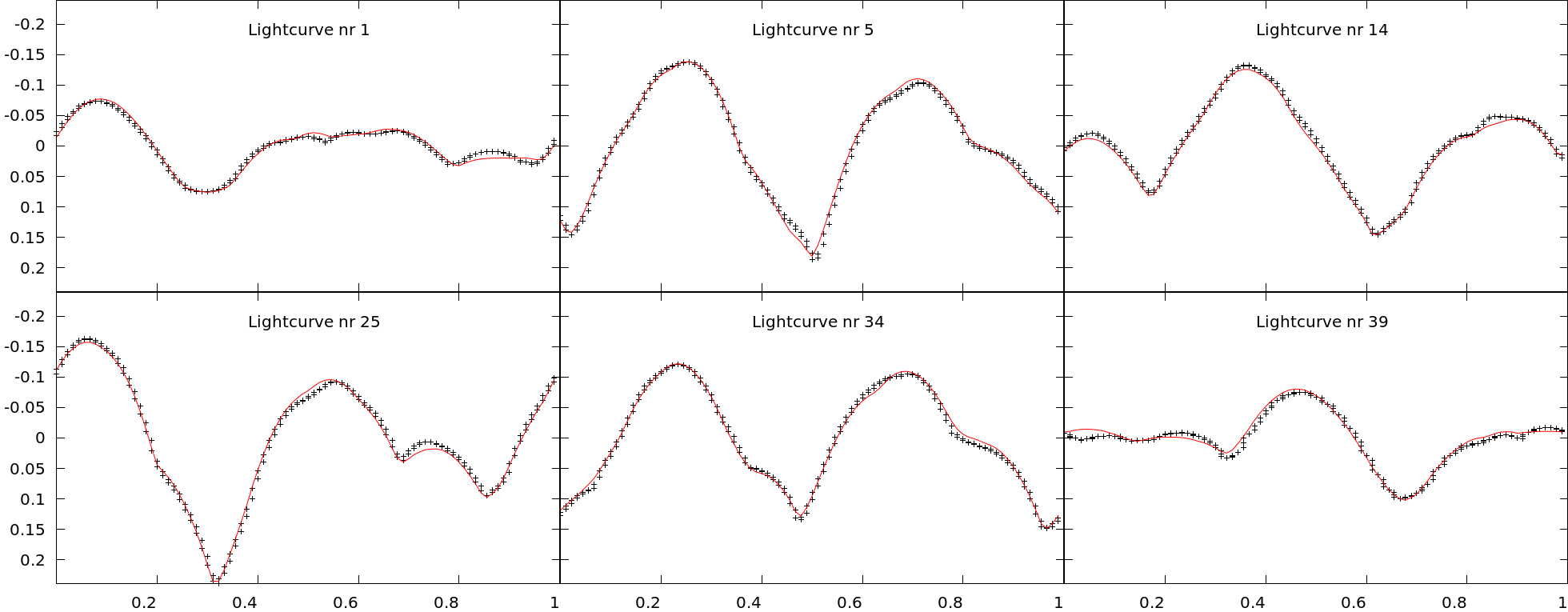}
	\caption{Some of the model C lightcurves.}
\end{figure*}

\begin{figure*}
	\includegraphics[width=17cm]{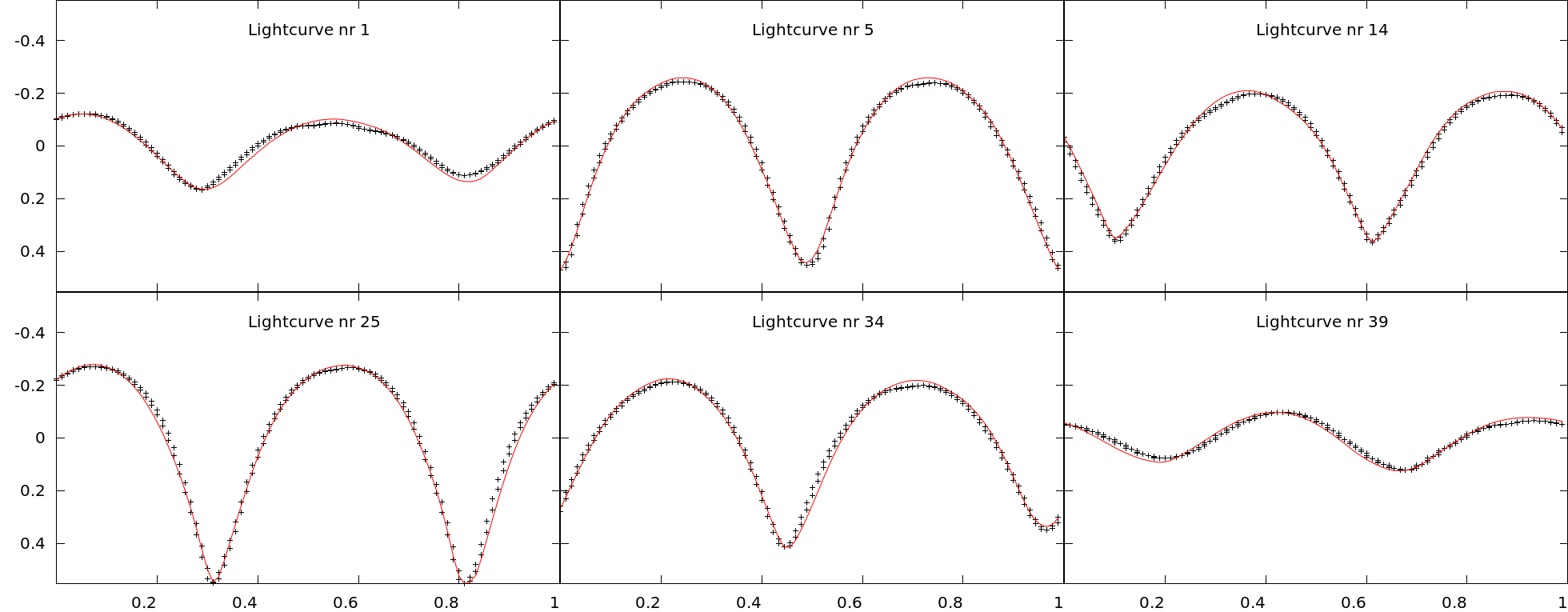}
	\caption{Some of the model D lightcurves.}
\end{figure*}

\section{(433) Eros lightcurves}
\label{app:eros_lc}
Some of the Eros lightcurves compared with model's ones.

\begin{figure*}
	\includegraphics[height=3.95cm]{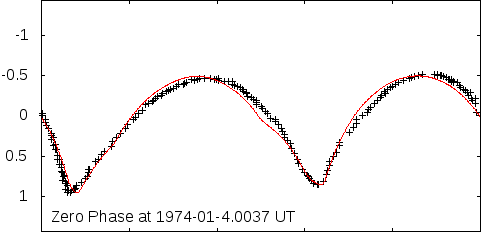}    \hspace{-7px}
	\includegraphics[height=3.95cm]{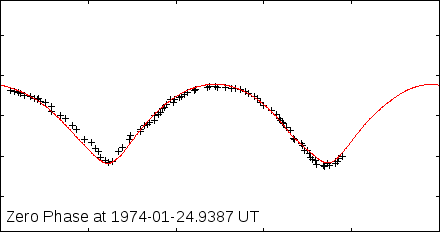}
	\includegraphics[height=3.95cm]{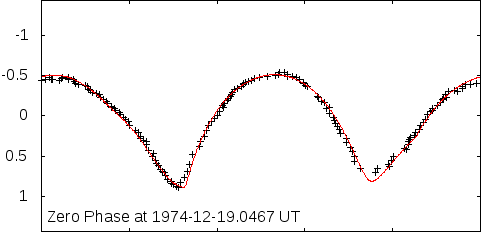}   \hspace{-7px}
	\includegraphics[height=3.95cm]{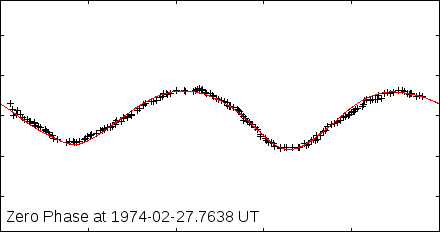}
	\includegraphics[height=3.95cm]{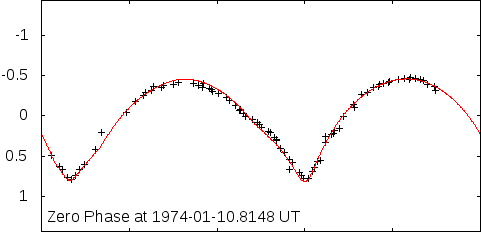}   \hspace{-7px}
	\includegraphics[height=3.95cm]{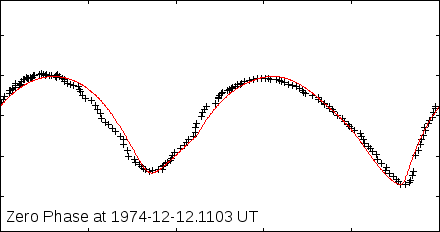}
	\includegraphics[height=3.95cm]{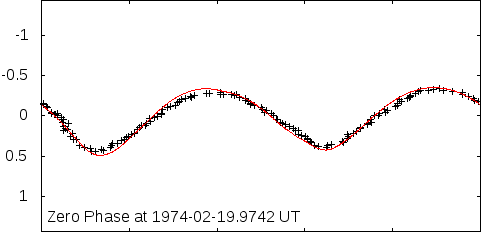}   \hspace{-7px}
	\includegraphics[height=3.95cm]{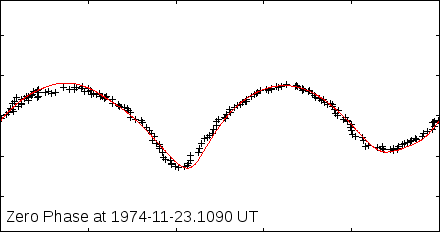}
	\includegraphics[height=3.95cm]{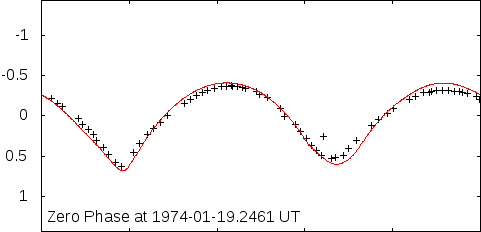}   \hspace{-7px}
	\includegraphics[height=3.95cm]{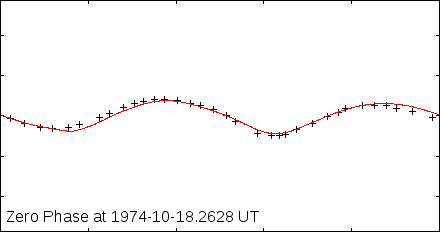}
	\includegraphics[height=4.35cm]{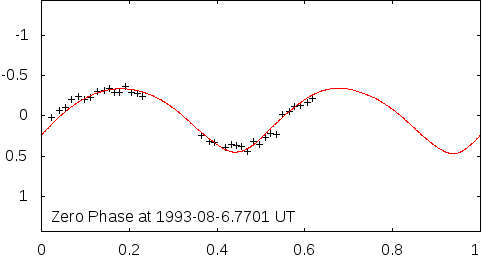}   \hspace{-7px}
	\includegraphics[height=4.35cm]{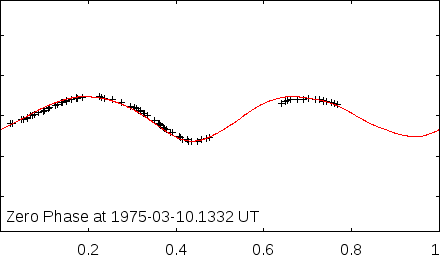}
\end{figure*}
\begin{figure*}
	\includegraphics[height=3.95cm]{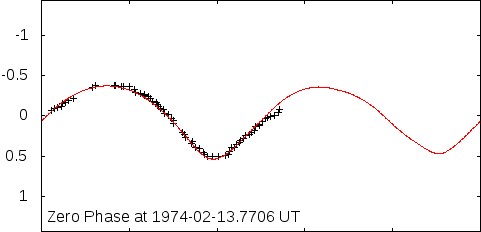}   \hspace{-7px}
	\includegraphics[height=3.95cm]{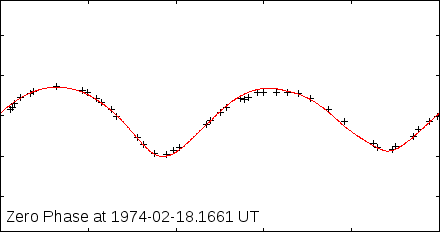}
	\includegraphics[height=3.95cm]{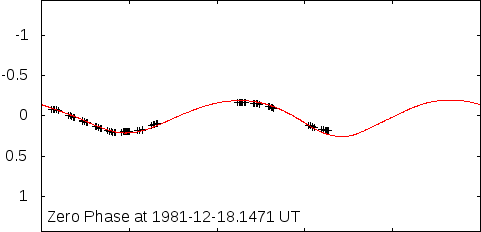}   \hspace{-7px}
	\includegraphics[height=3.95cm]{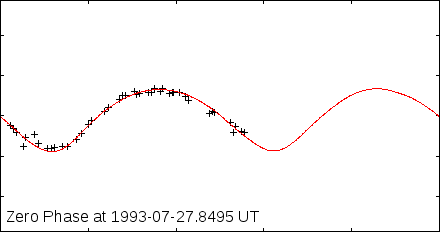}
	\includegraphics[height=3.95cm]{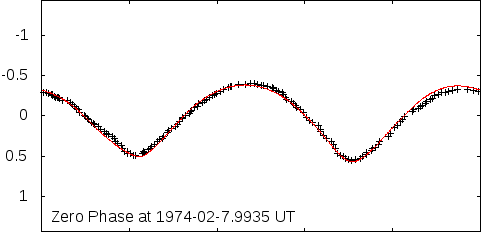}   \hspace{-7px}
	\includegraphics[height=3.95cm]{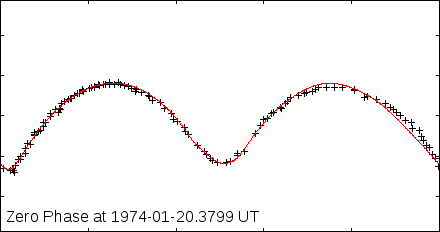}
	\includegraphics[height=3.95cm]{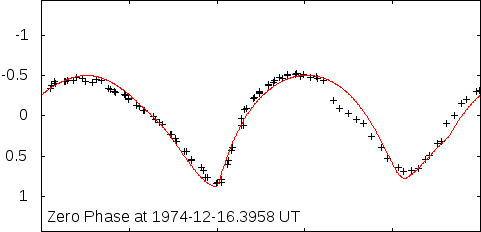}   \hspace{-7px}
	\includegraphics[height=3.95cm]{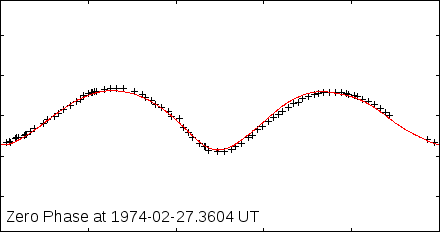}
	\includegraphics[height=4.35cm]{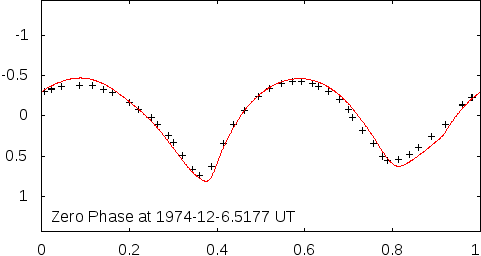}   \hspace{-7px}
	\includegraphics[height=4.35cm]{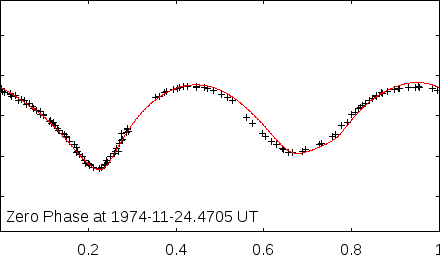}
	\caption{Some of the (433) Eros lightcurves (black points) vs. Eros
		model (red line).}
	\label{fig:metis_lc_comparison}
\end{figure*}

\section{(9) Metis lightcurves}
\label{app:metis_lc}

Comparison of lightcurves of available  Metis models from various methods.

\begin{figure*}
	\includegraphics[height=3.95cm]{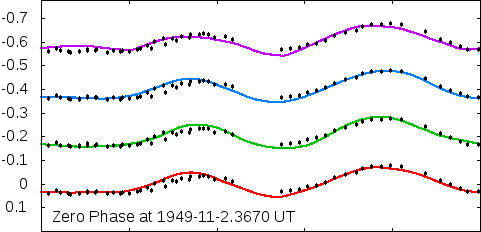}    \hspace{-7px}
	\includegraphics[height=3.95cm]{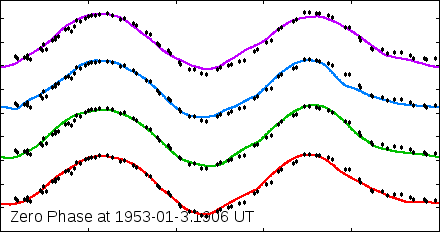}
	\includegraphics[height=3.95cm]{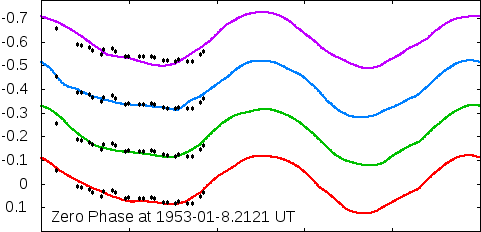}    \hspace{-7px}
	\includegraphics[height=3.95cm]{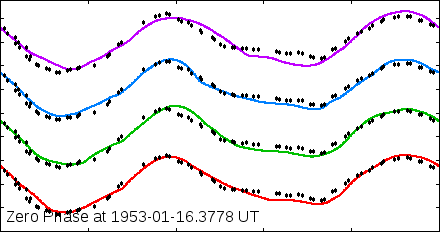}
	\includegraphics[height=3.95cm]{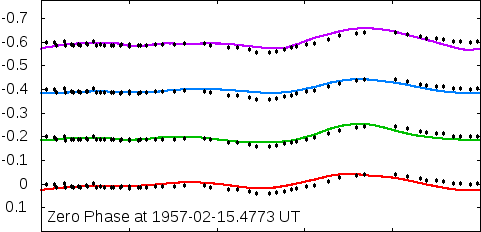}    \hspace{-7px}
	\includegraphics[height=3.95cm]{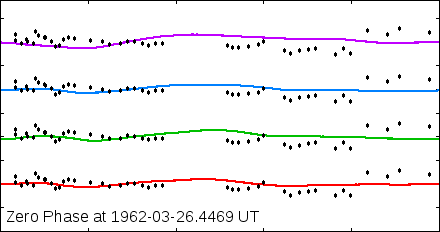}
	\includegraphics[height=3.95cm]{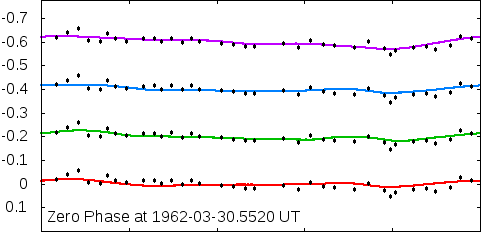}    \hspace{-7px}
	\includegraphics[height=3.95cm]{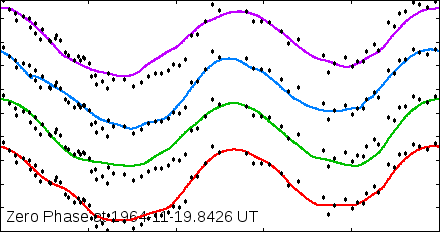}
	\includegraphics[height=3.95cm]{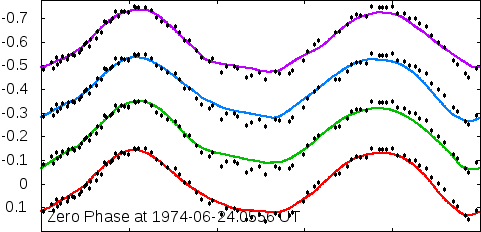}    \hspace{-7px}
	\includegraphics[height=3.95cm]{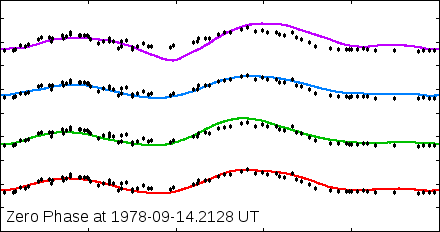}
	\includegraphics[height=4.35cm]{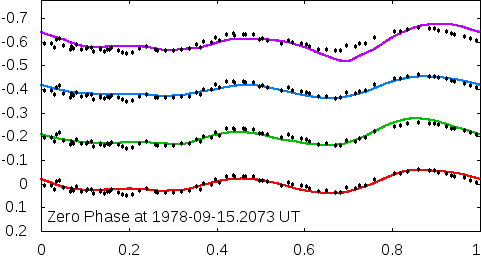}   \hspace{-7px}
	\includegraphics[height=4.35cm]{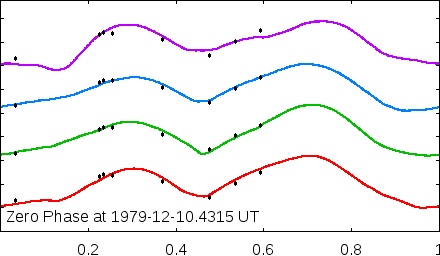}
\end{figure*}
\begin{figure*}
	\includegraphics[height=3.95cm]{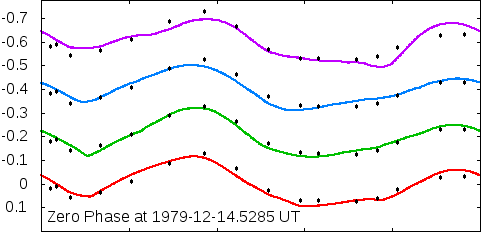}   \hspace{-7px}
	\includegraphics[height=3.95cm]{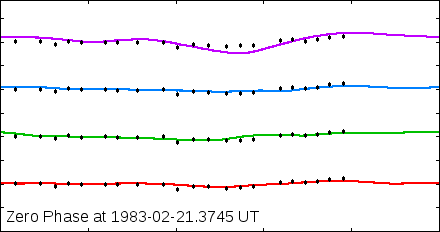}
	\includegraphics[height=3.95cm]{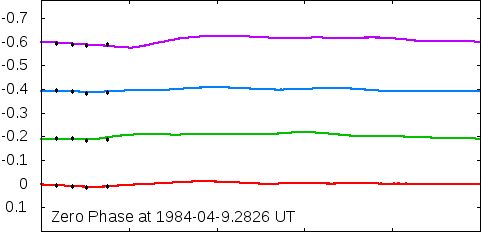}   \hspace{-7px}
	\includegraphics[height=3.95cm]{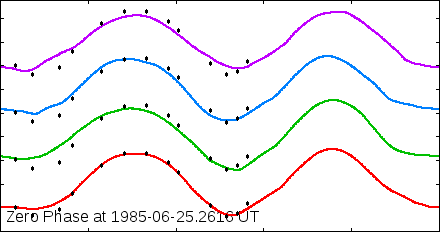}
	\includegraphics[height=3.95cm]{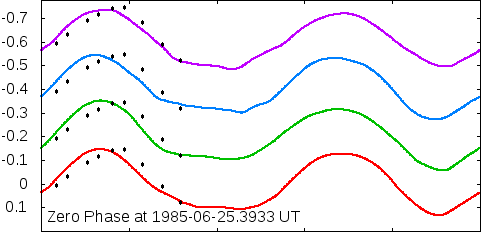}   \hspace{-7px}
	\includegraphics[height=3.95cm]{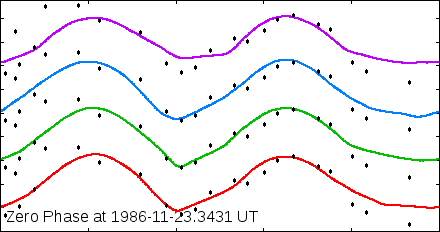}
	\includegraphics[height=3.95cm]{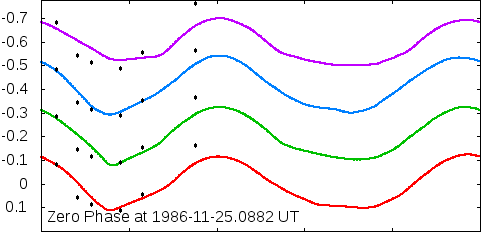}   \hspace{-7px}
	\includegraphics[height=3.95cm]{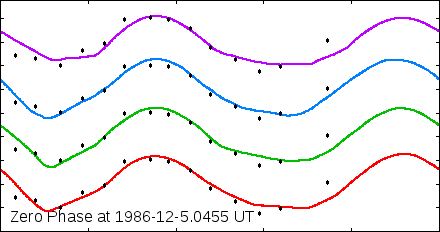}
	\includegraphics[height=4.35cm]{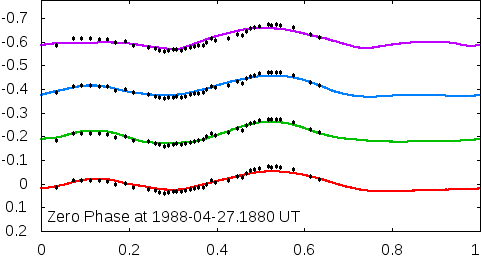}    \hspace{-7px}
	\includegraphics[height=4.35cm]{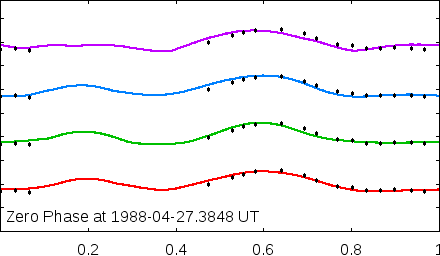}
	\caption{Some of the (9) Metis lightcurves (black points) vs. various Metis
		models.  First form the top (violet): KOALA \citep{Hanus13}, second from
		the top (blue): convex model \citep{torppa2003}, third from the top
		(green): ADAM \citep{ADAM15}, bottom (red): SAGE (this work).}
	\label{fig:metis_lc_comparison}
\end{figure*}
